\documentclass[aps, prc,
floatfix,
preprint,
tightenlines,
nofootinbib,
amsmath]{revtex4}
\usepackage{bm}
\usepackage[dvips]{color,graphicx}
\usepackage{epsfig}
\usepackage[figuresright]{rotating}
\def    \ie          {{\em i.e.\/} }
\def    \etal        {{\em et al.\/}}
\def    \eg          {{\em e.g.}}

\newcommand{\ud}     {\mathrm{d}}

\newcommand{\Mev}    {\:\mathrm{MeV}}
\newcommand{\gevsq}  {\:\mathrm{GeV}^2}

\newcommand{\tmax}   {t_{\mathrm{max}}}
\newcommand{\dirac}[1]{{\not{\!{#1}}}}
\renewcommand{\Im}{\mathop{\mathrm{Im}}}
\renewcommand{\Re}{\mathop{\mathrm{Re}}}
\newcommand{\Tr}{\mathop{\mathrm{Tr}}}
\newcommand{\eps}{\varepsilon_{\mu\nu\alpha\beta}}
\newcommand{\ceps}{\varepsilon}
\newcommand{\bra}[1]{\left\langle{#1}\left|}
\newcommand{\ket}[1]{\right|{#1}\right\rangle}
\newcommand{\average}[1]{\left\langle{#1}\right\rangle}
\newcommand{\la}{\left\langle}
\newcommand{\ra}{\right\rangle}
\newcommand{\eq}[1]{Eq.(\ref{#1})}
\newcommand{\Eqs}[2]{Eqs.~(\ref{#1}) and (\ref{#2})}
\newcommand{\eqs}[1]{Eqs.(\ref{#1})}

\newcommand{\sigmabar}{\bar{\sigma}}
\newcommand{\sigmaeff}{\bar\sigma_T}
\newcommand{\as}{\alpha_{\scriptscriptstyle S}}

\newcommand{\eth}{\varepsilon_{\mathrm{th}}}
\newcommand{\PMF}{\mathcal{P}_{\mathrm{MF}}}
\newcommand{\Pcor}{\mathcal{P}_{\mathrm{cor}}}
\newcommand{\tightspace}{\vspace{-3ex}}
\newcommand{\tunespace}{\vspace{-3ex}}

\newcommand{\figx}{0.55\textwidth}
\newcommand{\figxx}{0.69\textwidth}
\begin{document}

\title{%
Global Study of Nuclear Structure Functions
}

\author{S.~A.~Kulagin}
\email[]{kulagin@ms2.inr.ac.ru}
\affiliation{Institute for Nuclear Research, 117312 Moscow, Russia}
\author{R.~Petti}
\email[]{Roberto.Petti@cern.ch}
\affiliation{CERN, CH-1211 Gen\'eve 23, Switzerland}



\begin{abstract}
\noindent
We present the results of a phenomenological study of unpolarized nuclear 
structure functions for a wide kinematical region of $x$ and $Q^2$.
As a basis of our phenomenology we develop a model which takes into 
account a number of different nuclear effects including  nuclear 
shadowing, Fermi motion and binding, nuclear pion excess and off-shell 
correction to bound nucleon structure functions. Within this approach we 
perform a statistical analysis of available data on the ratio of the 
nuclear structure functions $F_2$ for different nuclei in the range from 
the deuteron to the lead. 
We express the off-shell effect and the effective scattering amplitude 
describing nuclear shadowing in terms of few parameters which are common 
to all nuclei and have a clear physical interpretation. The parameters are 
then extracted from statistical analysis of data. As a result, we obtain 
an excellent overall agreement between our calculations and data in the 
entire kinematical region of $x$ and $Q^2$. 
We discuss a number of applications of our model which include the 
calculation of the deuteron structure functions, nuclear valence and sea 
quark distributions and nuclear structure functions for neutrino 
charged-current scattering.
\end{abstract}


\maketitle

\section{Introduction}
\label{sec:intro}

The lepton Deep Inelastic Scattering (DIS) has been
since long time a powerful tool to probe the structure of hadrons
and nuclei at small and intermediate scales.
After the discovery of the parton structure of nucleons,
DIS remains to be the primary source of experimental information on the
distribution of quark and gluon fields in the nucleon and nuclei
and a valuable tool to test predictions of QCD.
New data from high-intensity electron (Jefferson Laboratory) and neutrino
(NuMI at Fermilab and JPARC in Japan) beams will allow in future to 
further extend our knowledge of the nucleon and nuclear structure from 
high-precision experiments.

The role of nuclei in DIS studies is dual. First, it should be noted
that the study of nuclei at small space-time scales is interesting  
by itself and it can provide valuable insights into the origin of nuclear force
and properties of hadrons in nuclear medium.
On the other hand the nuclear data often serve as the source of
information on hadrons otherwise not directly accessible. A typical
example is the extraction of the neutron structure function which is
usually obtained from deuterium and proton data in a wide kinematic
region. This procedure requires, in turn, a detailed knowledge of nuclear
effects in order to control the corresponding systematic uncertainties.
Another example is the determination of nuclear parton distribution 
functions which are universal high-momentum 
transfer characteristics of complex nuclei. 
Significant nuclear effects were discovered in charged lepton DIS
experiments~\cite{emc1,NMCdp,NMClic,NMChe,NMCpb,NMCq2,EMC,BCDMS,Gomez,E140,E665dp,E665,JLab-emc}.
These observations rule out a simple picture of a nucleus as a system
of quasi-free nucleons and indicate that the nuclear environment plays
an important role even at energies and momenta much larger than those
involved in typical nuclear ground state processes.
The study of nuclei is
therefore directly related to the interpretation of high-energy physics
from hadron colliders to fixed target experiments. The measurements of
nucleus-nucleus and proton-nucleus interactions at RHIC~\cite{RHIC} and
LHC~\cite{LHC} will help to clarify the nuclear modifications of the
parton distributions, as well as to define the initial conditions towards
the studies of new states of matter in heavy ion collisions.

The understanding of nuclear effects is particularly relevant
for neutrino physics, where the tiny cross-section with matter
requires the use of heavy nuclear targets in order to collect a
significant number of interactions. The presence of an axial-vector
component in the weak current and the quark flavour selection
differentiate neutrinos from charged leptons and imply a more
complex description of nuclear effects in neutrino scattering.
The role of nuclear corrections to neutrino structure functions
has been recently emphasized~\cite{nutev-nucl} after the NuTeV
collaboration reported a deviation from the Standard Model
prediction for the value of the weak mixing angle ($\sin^2 \theta_{W}$)
measured in neutrino DIS~\cite{nutev-sin2w}. One of the original
motivations of the present work is indeed related to the extraction
of the weak mixing angle from neutrino DIS data of the
NOMAD experiment~\cite{ichep04}.
It must be remarked that nuclear effects are important not only
in the determination of electroweak parameters, but also for
the understanding of neutrino masses and mixing.
The recent high-intensity NuMI~\cite{numi} and JPARC~\cite{JPARC}
neutrino facilities offer the possibility to perform a detailed
study of nuclear effects in neutrino interactions on a relatively
short time scale. The construction of a future neutrino
factory~\cite{Mangano:2001mj} would then allow to reach the ultimate
precision of the neutrino probe.

The main experimental information on nuclear structure functions comes
from charged-lepton scattering DIS experiments performed at
CERN~\cite{NMCdp,NMClic,NMChe,NMCpb,NMCq2,EMC,BCDMS},
SLAC~\cite{Gomez,E140}, FNAL~\cite{E665dp,E665} and recently 
at JLab~\cite{JLab-emc,JLab}.
The measurements usually refer to the
ratio $\mathcal{R}_2$ of the structure function $F_{2}$ of two nuclei
(usually a complex nucleus to deuterium).
Additional data from the Drell-Yan reaction of
protons off nuclear targets are also available~\cite{Drell-Yan}.
From the studies of data on the ratio $\mathcal{R}_2$ one can separate a few regions
of characteristic nuclear effects:
depletion of nuclear structure functions at small Bjorken $x$
($x<0.05$) known as shadowing region; a small enhancement of nuclear
structure functions for $0.1<x<0.3$ (antishadowing); depletion with a
minimum around $x=0.6\div0.7$ followed by a rise at large $x$ (known as ``EMC
effect" after the name of the experiment which discovered it).
It is interesting to note that a clear
$Q^2$ dependence has been reported only in the shadowing
region, while for $0.1<x<0.6$ $\mathcal{R}_2$ is almost $Q^2$ independent.
However, the data available on the $Q^2$ dependence of nuclear effects
are still scarce. One of the main drawbacks of all existing data
is the strong correlation between $x$ and $Q^2$ resulting from the
kinematics of fixed (stationary) target experiments. As a result,
significant regions of the $(x,Q^2)$ plane are still uncovered in
DIS experiments.

Many different theoretical models have been proposed to explain
the basic features of data (for a detailed summary of the current
understanding of nuclear corrections we refer to recent reviews and
references cited therein~%
\cite{Piller:1999wx,Thomas:1995,Arneodo:1994wf,Frankfurt:nt},
see also discussion in Sec.\:\ref{nuke-sect} of this paper).
The modelling is important to derive some insights on the underlying
physics of observed phenomena. 
However, consistent and quantitative description of nuclear effects in DIS 
in a wide kinematical region of $x$ and $Q^2$ and for a wide range of 
nuclei are clearly needed. 
In this paper we perform a quantitative study of data aiming to develop a 
model of nuclear DIS applicable in the analysis of existing data and in 
the interpretation of future experiments. 
In order to 
describe nuclear data over a wide kinematical region we take into
account many effects including nuclear shadowing, nuclear
pion excess, Fermi motion, nuclear binding and off-shell corrections
to bound nucleon structure functions.
It should be noted that if some effects, such as Fermi motion and
nuclear binding, are well constrained by other studies or data, the
remaining ones are less known. The main example is the off-shell
correction which describes the modification of structure functions of
bound nucleons in nuclear environment. We study this effect
phenomenologically by parameterizing the off-shell correction to the
nucleon structure function in terms of a few parameters which are
fixed from statistical analysis of nuclear data, together with the
corresponding uncertainties. It is worth to emphasize that these
parameters are universal, \ie  common for all nuclei, since they are
related to the nucleon structure. 
In a certain sense the off-shell correction can be
considered as a new structure function which describes the response of
the nucleon parton distributions to the variation of the nucleon
invariant mass. Even if this structure function is not accessible for
free proton and neutron, it can be probed in nuclear reactions.

It should be also emphasized that different
nuclear effects in different kinematical regions of $x$ are correlated
by DIS sum rules. For example, the light-cone momentum sum rule links
bound nucleon and pion contributions to DIS. We use this requirement 
in order to constrain mesonic contributions to nuclear
structure functions. Another example is the baryon number sum rule
which links shadowing and off-shell corrections. In our approach the
off-shell effect provides the mechanism of cancellation of a negative
nuclear-shadowing contribution to the normalization of nuclear valence
quark distributions.

After fixing the parameters of our model, we compute
predictions for a number of applications. In particular, we discuss
nuclear valence and sea quarks at high $Q^2$
and compute nuclear corrections to neutrino structure
functions.
These subjects will be treated more extensively
in future publications.

The paper is organized as follows. In Sec.\:\ref{sec:kine} we
briefly summarize the DIS kinematics for electron (muon) and neutrino
scattering and introduce notations used in this paper.
Section~\ref{nucleon} provides information on the nucleon structure
functions and parton distributions necessary for our analysis.
Section~\ref{nuke-sect} is devoted to the theoretical framework to
treat different nuclear corrections in our studies. In particular, in
Sec.\:\ref{largex-nuke} we examine the derivation of nuclear
structure functions in the approximation of incoherent scattering off
bound nucleons and nuclear pions (Sec.\:\ref{sec:pion}), 
the off-shell effects in the structure functions and quark 
distributions (Sec.\:\ref{sec:offshell}) and coherent 
nuclear effects leading to nuclear shadowing and antishadowing
(Sec.\:\ref{smallx-nuke}).  In Sec.\:\ref{sec:model} we 
discuss in detail the nuclear input which is used in our analysis 
(Sec.\:\ref{sec:deut} to \ref{sec:pion:details}), the model of 
off-shell effects (Sec.\:\ref{sec:modbn}) and effective scattering 
amplitude (Sec.\:\ref{sec:xsec}). The analysis of data is described in 
Sec.\:\ref{sec:data} and \ref{sec:fits}.  In Sec.\:\ref{sec:results} 
we present the results obtained from our fits to nuclear data. The $Q^2$ 
and $A$ dependence of nuclear effects are discussed in 
Sec.\:\ref{sec:qsq} to \ref{sec:deuterium}.  In Sec.\:\ref{sec:appl} 
we apply our approach to study nuclear parton distributions 
(Sec.\:\ref{sec:npdf}) and neutrino structure functions 
(Sec.\:\ref{sec:nuint}). 
%
In Appendix~\ref{apndx:conv} we provide the details of the
integration in nuclear convolution and in Appendix~\ref{apndx:ms} the 
multiple scattering coefficients are given.


\section{Kinematics of lepton inelastic scattering}
\label{sec:kine}

Consider the scattering of a \emph{charged lepton} (electron or muon)
off a nucleon with the four-momentum $p=(E_p, \bm{p})$ and mass
$M$. The scattering matrix element to leading order in the
electromagnetic coupling constant $\alpha=e^2/4\pi\approx 1/137$ is
determined by the standard one-photon exchange process.
In inclusive scattering, the final hadronic state is not 
detected and the differential cross section is fully given by 
hadronic tensor $W_{\mu\nu}$, which is the sum of hadronic matrix
elements of the electromagnetic current $J_\mu^{\rm em}$ over all final 
hadronic states (see, \eg, \cite{IoKhLi84})
\begin{eqnarray}
W_{\mu\nu}(p,q) = \frac{1}{4\pi}\sum_{n} 
(2\pi)^4 \delta(p+q-p_n) \bra{p}J_\mu^{\mathrm{em}}(0)\ket{n} 
                         \bra{n}J_\nu^{\mathrm{em}}(0)\ket{p}, 
\label{W}
\end{eqnarray}
where $q$ is four-momentum transfer to the target. 

We do not consider the polarization effects and assume the averaging over 
the target and beam polarizations. Then only the symmetric part of the 
hadronic tensor contributes to the cross section. Because of the 
conservation of electromagnetic current, time reversal invariance and 
parity conservation in electromagnetic interaction, the symmetric hadronic 
tensor has only 2 independent Lorentz structures (see, \eg, 
\cite{IoKhLi84}) 
\begin{eqnarray}
W_{\mu\nu}(p,q) &=& -\widetilde g_{\mu\nu}\, F_1 + \widetilde{p}_\mu 
\widetilde{p}_\nu \frac{F_2}{p\cdot q}, 
\label{SF}
\end{eqnarray}
where $F_{1,2}$ are Lorentz-invariant structure functions, and, for 
simplicity, we use the following notations: 
\begin{subequations}
\label{tilda}
\begin{eqnarray}
\widetilde g_{\mu\nu} &=& g_{\mu\nu} - \frac{q_\mu q_\nu}{q^2},\\
\widetilde{p}_\mu &=& p_\mu - q_\mu\frac{p\cdot q}{q^2}. 
\end{eqnarray}
\end{subequations}
We use the normalization of states $\langle p|p'\rangle = 
2E_{\bm{p}}(2\pi)^3 \delta(\bm{p}-\bm{p}')$ for both bosons and fermions. 
With this normalization the hadronic tensor and the structure functions 
$F_{1,2}$ are dimensionless. 

The structure functions are the functions of two independent invariant
variables. In deep inelastic regime the Bjorken
variable $x=Q^2/(2p\cdot q)$ and four-momentum transfer squared $Q^2=-q^2$ 
are usually used as the variables the structure functions depend on.

The polarization averaged differential cross section is determined by
the structure functions $F_{1,2}$. In terms of the variables $x$ and
$Q^2$ the cross section reads
\begin{eqnarray}
\label{xsec:xQ}
\frac{\ud^2\sigma}{\ud x\ud Q^2} =
\frac{4\pi\alpha^2}{xQ^4}
\left[\left(1-y-\frac{(Mxy)^2}{Q^2}\right)F_2
+ xy^2\left(1-\frac{2m_l^2}{Q^2}\right) F_1\right],
\end{eqnarray}
where $y=p\cdot q/p\cdot k$. The variable $y$ is not independent
variable and related to $x$ and $Q^2$ via the equation $xy=Q^2/(2p\cdot k)$. 
The lepton mass term is kept in \eq{xsec:xQ} for the sake of
completeness. Although it is negligible in electron deep inelastic
scattering, it might be relevant for muon scattering at
small momentum transfer or for $\tau$ lepton scattering.

The structure functions $F_{1,2}$ can be related to the virtual photon
helicity cross sections by projecting \eq{W} onto the states with
definite photon polarizations. These states are described by the
photon polation vectors.  In the reference frame, in which the
momentum transfer is along the $z$-axis, $q=(q_0,\bm{0}_\perp,q_z)$,
$q_z = -|\bm{q}|$ the photon polarization vectors are
\begin{subequations}
\label{eTL}
\begin{eqnarray}
e_\pm &=& (0,1,\pm i,0)/\sqrt2 ,\\
e_0 &=& (q_z, \bm{0}_\perp, q_0)/Q . 
\end{eqnarray}
\end{subequations}
where $Q=\sqrt{Q^2}$. The polarization vectors $e_+$ and $e_-$
describe two transversely polarized states with helicities $+1$ and $-1$.
The vector $e_0$ corresponds to the longitudinally
polarized (scalar) virtual photons.  The polarization vectors are
orthogonal to momentum transfer, $e_{\pm}\cdot q =e_0\cdot q=0$, and obey the
orthogonality and the normalization conditions, $e_\pm\cdot e_L = 0,\ 
e_\pm^*\cdot e_\pm = -1,\ e_0^2 = 1$.

The helicity structure functions are
\begin{subequations}
\label{SF:hel}
\begin{eqnarray}
W_\pm &=& {e^\mu_\pm}^* W_{\mu\nu}{e^\nu_\pm} = F_1,\\
W_0 &=& {e^\mu_0}^* W_{\mu\nu}{e^\nu_0} = \gamma^2F_2/(2x) - F_1,
\end{eqnarray}
\end{subequations}
where $\gamma=|\bm{q}|/q_0=(1+4x^2M^2/Q^2)^{1/2}$.
Instead of \eq{SF:hel}, it is more convenient to
use the transverse and the longitudinal structure functions defined as
\begin{subequations}
\label{FTL}
\begin{eqnarray}
F_T &=& x(W_+ + W_-)=2xF_1,\\
F_L &=& 2xW_0=\gamma^2 F_2 - 2xF_1.
\end{eqnarray}
\end{subequations}
%


Let us briefly consider the scattering of (anti)neutrino. In the Standard 
Model neutrino can either couple to charged $W^\pm$ bosons or to neutral 
$Z$ boson. In the former case interaction is driven by \emph{charged 
curent} (CC) $J_\mu^\pm = V_\mu^{\pm}-A_\mu^{\pm}$ with 
$V_\mu^{\pm}$ and $A_\mu^{\pm}$ the charged components of the 
vector and axial-vector current. The interaction with $Z$ boson is 
described by the \emph{neutral current} (NC) which is the superposition of 
the isovector weak left current and electromagnetic current $J_\mu^0 = 
\sqrt2 (V_\mu^3-A_\mu^3-2\sin^2\theta_W J_\mu^{\text{em}})$, where 
$\theta_W$ is the Weinberg weak mixing angle. 

The hadronic tensor for CC or NC interaction is given by \eq{W} with the
electromagnetic current replaced by the corresponding weak current.
The Lorentz decomposition of hadronic tensor is different for neutrino 
case and includes additional terms compared to \eq{SF}. For example, for 
CC neutrino interaction we have%
\footnote{The tensor $W_{\mu\lambda}^+$ corresponds to interaction mediated 
by $W^+$ boson and describes neutrino CC scattering while 
$W_{\mu\lambda}^+$ describes antineutrino. It should also be remarked 
that the neutrino and antineutrino NC structure functions are 
identical, since neutrino and antineutrino in NC scattering couple to the same 
hadronic NC. This is not the case for CC neutrino and 
antineutrino structure functions.} (see, \eg, \cite{IoKhLi84})
\begin{align}
\begin{split}
W_{\mu\lambda}^\pm &=
-\widetilde g_{\mu\lambda }\, F_1^{W^\pm}
+ \widetilde p_{\mu} \widetilde p_{\lambda} \frac{F_2^{W^\pm}}{p\cdot q}
+ i\varepsilon_{\mu\lambda}(p,q)\frac{F_3^{W^\pm}}{2p\cdot q}
\\
& \phantom{=} 
+\frac{q_\mu q_\lambda}{Q^2}F_4^{W^\pm}
+\frac{q_\mu p_\lambda + q_\lambda p_\mu}{p\cdot q}F_5^{W^\pm} ,
\end{split}
\label{CC:SF}
\end{align}
where we denote 
$\ceps_{\mu\lambda}(a,b)=\ceps_{\mu\lambda\alpha\beta}a^\alpha b^\beta$.
The first two terms with $F_1$ and $F_2$ in \eq{CC:SF} are similar to those
in charged-lepton scattering and appear due to VV and AA interactions in
Eq.(\ref{W}). The term with $F_3$ describes parity-violating VA and AV 
interference. The terms $F_4$ and $F_5$ are present because the 
axial current does not conserve.
The contributions from the structure functions $F_4$ and $F_5$ to the
neutrino production cross-section are suppessed by a small ratio
$m_l^2/(ME)$ (these terms vanish in the NC cross section). It was also 
shown that $F_4=0$ and $2xF_5=F_2$ in the leading order and in the limit 
of massless quarks (Albright--Jarlskog relations \cite{AJ75}). Recently it 
was argued that the second of these relations survives the higher order 
and the target mass corrections in massless QCD, while the relation for 
$F_4$ should be replaced by $F_4=F_2/(2x)-F_1$ \cite{KR02}.

The relations between the helicity structure functions and the
structure functions $F_{1,2,3}$ in the neutrino scattering are
\begin{subequations}\label{SF:hel:nu}
\begin{eqnarray}
W_\pm &=& F_1 \pm \gamma F_3,\\
W_0 &=& \gamma^2F_2/(2x)-F_1.
\end{eqnarray}\end{subequations}
The definition of $F_{T,L}$ by \eq{FTL} also apply in this case. One
observes from \eqs{SF:hel:nu} that the structure function $F_3$
determines the left-right asymmetry in the transverse helicity
structure functions.


\section{Nucleon structure functions}
\label{nucleon}

The structure functions remain important observables
to probe QCD structure of proton and neutron and nuclei.
In this section we briefly review the characteristics of nucleon structure 
functions necessary for our analysis.

\subsection{QCD perturbative regime}
\label{qcdpert}

In the region of $Q^2$ large compared to the nucleon scale the
structure functions can be analyzed in perturbative QCD. A working
tool of this analysis is the operator product expansion
(OPE) \cite{Wilson}. Using the OPE, the contributions from different
quark-gluon operators to hadronic tensor
can be ordered according to their \emph{twist}.
For the DIS structure functions 
this leads to the expansion in
inverse powers of $Q^2$:
\begin{eqnarray}
\label{t-expansion} F_a(x,Q^2) = F_a^{LT}(x,Q^2) +
\frac{H_a(x,Q^2)}{Q^2} + {\cal O}\left(1/Q^4 \right),
\end{eqnarray}
where $a$ labels the type of the structure function ($a=T,2,3$).
The first term is the leading twist (LT) contribution and $H_a$
are the twist-4 contributions (higher twist, HT). 

The \emph{leading twist} contribution is directly related to the
distributions of quarks and gluons inside the nucleon, the parton
distribution functions (PDFs) via the DIS factorization theorem as a
convolution with coefficient functions (for more detail see, \eg,
Ref.\cite{handbook} and references therein). 
The coefficient functions depend on the process and the type of the
structure function but are independent of the target. These functions
are computable as power series in $\as$. The parton distributions are
independent of the process but do depend on the target.

The PDFs have non-perturbative origin and cannot be calculated in
perturbative QCD. However, the $Q^2$ dependence of the PDFs can be
handled using QCD perturbation theory, and is governed by the
well-known DGLAP evolution equations with the kernel given by
the splitting functions \cite{DGLAP}.

The one-loop (NLO) coefficient and splitting functions have been
computed since long time \cite{NLO}. The two-loop (NNLO) coefficient
functions~\cite{C-NNLO} and the corresponding splitting
functions~\cite{P-NNLO} are now also available.  In our analysis of
nuclear data we use both the coefficient functions and the PDFs to
NNLO approximation calculated in $\overline{\rm MS}$ scheme using the
factorization and the renormalization scales set to $Q^2$.

The HT components involve interactions between quarks and gluons and
lack simple probabilistic interpretation.

It must be noted that the twist expansion was derived in the
massless limit. If a finite mass for the nucleon target is
considered, the new terms arise in \eq{t-expansion} that mix
operators of different spin, leading to additional power terms of
kinematical origin -- the so-called \emph{target mass corrections}
(TMC). If the parameter $x^2M^2/Q^2$ is small, the TMC series
can be absorbed in the leading twist term \cite{GeoPol76}.  Therefore,
Eq.(\ref{t-expansion}) remains valid with the LT terms replaced by
\begin{subequations}
\label{TMC}
\begin{eqnarray}
F_T^{\mathrm{TMC}}(x,Q^2) &=&
\frac{x^2}{\xi^2\gamma}F_T^{\mathrm{LT}}(\xi,Q^2) + 
	\frac{2x^3M^2}{Q^2\gamma^2}
		\int_\xi^1\frac{\ud z}{z^2}F_2^{\mathrm{LT}}(z,Q^2),\\
F_2^{\mathrm{TMC}}(x,Q^2) &=&
    \frac{x^2}{\xi^2\gamma^3}F_2^{\mathrm{LT}}(\xi,Q^2) + 
    \frac{6x^3M^2}{Q^2\gamma^4}
	\int_\xi^1\frac{\ud z}{z^2}F_2^{\mathrm{LT}}(z,Q^2),\\
xF_3^{\mathrm{TMC}}(x,Q^2) &=&
    \frac{x^2}{\xi^2\gamma^2}\xi F_3^{\mathrm{LT}}(\xi,Q^2) + 
    \frac{2x^3M^2}{Q^2\gamma^3}
        \int_\xi^1\frac{\ud z}{z^2}zF_3^{\mathrm{LT}}(z,Q^2),
\end{eqnarray}
\end{subequations}
where $\gamma=(1+4x^2M^2/Q^2)^{1/2}$ and $\xi=2x/(1+\gamma)$ 
is the Nachtmann variable \cite{Nacht}.

However, it must be remarked that the derivation of \cite{GeoPol76} was given
in the zeroth order in $\alpha_S$, assuming that the target quarks are
on-shell and neglecting the transverse degrees of freedom. 
Furthermore, Eqs.(\ref{TMC}) suffer the so called threshold
problem. Indeed, it follows from Eqs.(\ref{TMC}) that the target mass
corrected inelastic structure functions $F_2^{\mathrm{TMC}}$ remain
finite as $x\to1$ even if the LT terms vanish in this limit.
Clearly, the region $x$ close to 1 is beyond the applicability of
Eqs.(\ref{TMC}). However, in the applications to nuclear structure
functions at large $x$ it is important to meet the threshold
condition. One possible way to deal with this problem is to expand
Eqs.(\ref{TMC}) in power series in $Q^{-2}$ and keep a finite number
of terms. In particular, keeping the LT and the $1/Q^2$ term we have
\begin{subequations}
\label{TMC:exp}
\begin{align}
F_T^{\mathrm{TMC}}(x,Q^2) &= 
F_T^{\mathrm{LT}}(x,Q^2) + 
\notag\\
&
	\frac{x^3M^2}{Q^2}
\left(2\int_x^1\frac{\ud z}{z^2}F_2^{\mathrm{LT}}(z,Q^2)-
\frac{\partial}{\partial x} F_T^{\mathrm{LT}}(x,Q^2)\right),
\\
F_2^{\mathrm{TMC}}(x,Q^2) &=
\left(1-\frac{4x^2M^2}{Q^2}\right)F_2^{\mathrm{LT}}(x,Q^2) +
\notag\\
&
	\frac{x^3M^2}{Q^2}
\left(6\int_x^1\frac{\ud z}{z^2}F_2^{\mathrm{LT}}(z,Q^2)-
\frac{\partial}{\partial x} F_2^{\mathrm{LT}}(x,Q^2)\right),
\\
xF_3^{\mathrm{TMC}}(x,Q^2) &=
\left(1-\frac{2x^2M^2}{Q^2}\right)xF_3^{\mathrm{LT}}(x,Q^2) +
\notag\\
&
	\frac{x^3M^2}{Q^2}
\left(2\int_x^1\frac{\ud z}{z^2}zF_3^{\mathrm{LT}}(z,Q^2)-
\frac{\partial}{\partial x} \left(xF_3^{\mathrm{LT}}(x,Q^2)\right)\right).
\end{align}
\end{subequations}
In this approximation the structure functions have a correct threshold 
behavior and vanish in the limit of $x\to1$, provided that the LT terms 
and their derivatives vanish in this limit.

The target mass corrections should also be applied to the HT terms in
the higher order terms in the twist expansion (\ref{t-expansion}). For this
reason we do not consider $1/Q^4$ terms in the TMC formula, which are
small in the considered kinematical range. We also note, that the
extrapolation of the target mass corrections to off-shell region 
$p^2\not = M^2$ is important in the treatment
of the nuclear effects and will be discussed in 
Sec.\:\ref{off-shell-sect}.

\subsection{Structure function phenomenology}
\label{phenom-sect}

The twist expansion and PDFs as universal, process-independent 
characteristics of the target are at the basis of extensive QCD 
phenomenology of high-energy processes.
In phenomenological studies, the PDFs are extracted from QCD global fits.
A number of such analyses are available \cite{a02,cteq,mrst}. 
In our studies of nuclear data described in Sec.\:\ref{sec:data} to 
\ref{sec:deuterium} we use the results by Alekhin \cite{a02} 
who provides the set of the nucleon PDFs obtained with the coefficient and 
splitting functions calculated to the NNLO approximation.%
\footnote{In our analysis we use PDFs obtained from new fits optimized in the 
low $Q^2$ region and including additional data with respect to~\cite{a02}. 
This extraction of PDFs also takes into account the nuclear corrections 
to D data described in the present paper (see Sec.\:\ref{sec:fits}). 
Results from the new fits will be reported elsewhere.}
Furthermore, the HT terms 
and the PDF uncertainties have also been evaluated in~\cite{a02}. 

It should be also remarked that the twist expansion and perturbative QCD 
apparently breaks down at low $Q^2$. Furthermore, the conservation of 
electromagnetic current requires the structure function $F_2$ to vanish 
as $Q^2$ for $Q^2\to0$. The data seem to indicate the 
presence of a transition region between perturbative and non-perturbative 
regimes at $Q^2$ about $1\gevsq$. In our studies of nuclear effects in the 
structure functions some data points at 
small $x$ are in the low-$Q^2$ region. In order to match low-$Q$ 
and high-$Q$ regions we apply spline interpolations for the structure 
functions which obeys the current conservation requirements.


\section{Nuclear structure functions}
\label{nuke-sect}

In this Section we describe a theoretical framework which will be the
basis of phenomenological studies of nuclear DIS data discussed in
Sections~\ref{sec:model} to \ref{sec:results}.

The mechanisms of nuclear DIS appear to be different for small and
large Bjorken $x$ as viewed from the laboratory system.
The physics scale for this separation comes from the comparision
of a characteristic DIS time, which is also known as Ioffe length
$L_I=1/(Mx)$ (see, \eg, Ref.\cite{IoKhLi84}), and an average distance
between bound nucleons in nuclei which is about $1.5\,$Fm.
At large $x>0.1$ the characteristic DIS time is smaller than average
internucleon distance. This observation justifies the use of the
incoherent approximation for the nuclear Compton amplitude in this
region.
It was realized long ago that the nucleon momentum distribution 
(\emph{Fermi motion}) is important effect even in the scaling limit and 
results in the enhancement of nuclear structure functions at large Bjorken 
$x$ \cite{ref:fm}. After discovery of the EMC effect \cite{emc1} the 
calculation of nuclear DIS in impulse approximation was revisited 
\cite{ref:binding,GL92,KPW94,KMPW95,Ku98} and effect of \emph{nuclear 
binding} was emphasised which explains a significant part of the observed 
dip in the EMC ratio at $x\sim 0.6$ (for a review of the EMC effect and 
more references see \cite{Frankfurt:nt,Arneodo:1994wf,Thomas:1995}).

Effects beyond the impulse approximation are important. It should be
noted that because of binding, the nucleons do not carry all of the
light-cone momentum of the nucleus and the momentum sum rule is
violated in the impulse approximation. A natural way to correct this
problem is to explicitly consider the pion contribution to the
structure functions \cite{Sullivan:1971kd} which balances missing
momentum. Several calculations of the \emph{pion correction} to
nuclear structure functions have been performed in different
approaches and approximations
\cite{ref:pion}.
Although all calculations predict some enhancement at small $x$, the
concrete predictions are model-dependent. In this paper we calculate
nuclear pion correction following the approach of Ref.\cite{Ku89} in
which the pion contribution was constrained using the equations of
motion for interacting pion-nucleon system. By using the light-cone
momentum balance equation we effectively constrain the contribution
from all mesonic fields responsible for nuclear binding.

It should be noted that bound nucleons are off-shell particles and 
their structure functions can be different from those of free nucleons.
\emph{Off-shell effects} in nuclear DIS were discussed in a number of
papers \cite{Dunne:1985ks,GL92,KPW94,MST94,Ku98,AKL03}.  
It was shown that, because of spin 1/2, the off-shell nucleon is 
characterized by the increased number of structure functions which depend 
on the nucleon virtuality as an additional variable \cite{MST94,KPW94,KMPW95,Ku98}.
However, in the vicinity of the mass shell
(which is the relevant case for nuclei) 
the off-shell nucleon can still be described by the same number of 
structure functions as the on-shell nucleon \cite{KPW94,KMPW95,Ku98}.
Nevertheless, the off-shell dependence 
of structure functions remains an important effect through which the 
modification of the internal structure of the bound nucleon in nuclear 
environment can be assessed. It should be also emphasized that the 
off-shell effect provides the specific mechanism of balancing a negative 
contribution to the nuclear baryon number sum rule from nuclear shadowing 
effect (for details see Sections~\ref{sec:valnorm} and \ref{sec:inter}). 
In Sec.\:\ref{largex-nuke} we discuss the derivation of the nuclear 
structure functions in the presence of off-shell effects with the full 
consideration of the nucleon spin. The treatment of the off-shell effect 
in the parton distributions is discussed in more detail in 
Sections~\ref{sec:offshell} and \ref{sec:modbn}.

In the small-$x$ region the space-time picture of DIS is different.
For $x\ll 0.1$ the characteristic DIS time is large on the nuclear scale,
the nuclear DIS becomes ``stretched'' in time and in the longitudinal
direction.  The process can be viewed as the intermediate boson first
fluctuates into a quark pair which can form a complex configuration
(hadronic or quark-gluon) which then scatters off the target.
As an average time of life of such fluctuation is large compared to
average distance between bound nucleons, the photon interaction with
nuclear targets resembles hadronic properties
\cite{Gribov:1968gs,Bauer:iq}. In particular, since hadron scattering
amplitudes are almost imaginary at high energy, the double
scattering correction to the DIS cross-section is negative leading to
nuclear \emph{shadowing} effect, similar to that in hadron scattering
\cite{Glauber:1955qq}. Nuclear shadowing in DIS was subject to
intensive studies
\cite{ref:shadowing}
(for a review of nuclear shadowing and more
references see, \eg, \cite{Frankfurt:nt,Piller:1999wx}). In the present paper we
treat nuclear shadowing effect in a semi-phenomenological approach by
introducing phenomenological amplitude which describes interaction of
hadronic component of the intermediate boson with the nucleon and
consider the propagation of this state in nuclear environment using
multiple scattering theory. Details are discussed in Sec.\:\ref{smallx-nuke}.

Summarizing we write the nuclear structure functions as the sum of
incoherent and coherent contributions
\begin{equation}
\label{FA}
F_a^A = F_a^{p/A} + F_a^{n/A} + F_a^{\pi/A} + \delta F_a^A,
\end{equation}
where $F_a^{p/A}$, $F_a^{n/A}$, $F_a^{\pi/A}$ denote the contributions
to the structure function of type $a$ from bound protons, neutrons,
and nuclear pions, respectively. The last term in \eq{FA} is a
correction due to nuclear coherent interaction. The exact meaning of
all these terms will be explained in the following sections.

\subsection{Incoherent scattering approximation}
\label{largex-nuke}

The DIS hadronic tensor is given by the imaginary part of the virtual
photon Compton amplitude in the forward direction. In the incoherent
scattering regime (large $x$) taking into account the nucleon spin the
nuclear hadronic tensor can be written as (see also
\cite{Ku89,KPW94,KMPW95})
\begin{equation}
\label{WA}
W_{\mu\nu}^A(P_A,q) = \sum_{\tau=p,n}\int [\ud p]\,
\Tr\left[
\widehat{\cal W}_{\mu\nu}^\tau(p,q){\cal A}^\tau(p;A)
\right],
\end{equation}
where the sum is taken over the bound protons and neutrons, Tr is
taken in the nucleon Dirac space and the integration is performed over
the nucleon four-momentum, $[\ud p]=d^4p/(2\pi)^4$. In \eq{WA}
${\cal A}^\tau(p;A)$ is the imaginary part of the proton ($\tau=p$) or
the neuteron ($\tau=n$) propagator in the nucleus
\begin{equation}
\label{A}
{\cal A}_{\alpha\beta}^\tau(p;A)=
\int \ud t\ud^3\bm{r} e^{i p_0 t-i \bm{p}\cdot\bm{r}}
\langle A|\overline{\Psi}_\beta^\tau(t,\bm{r})\,\Psi_\alpha^\tau(0)|A\rangle
\end{equation}
with $\Psi_\alpha^\tau(t,\bm{r})$ the nucleon field operator and
$\alpha$ and $\beta$ the Dirac spinor indeces.  The off-shell nucleon
electromagnetic tensor
$\widehat{\cal W}_{\mu\nu}(p,q)$ is the matrix in the Dirac space. On the
mass shell $p^2=M^2$, averaging $\widehat{\cal
W}_{\mu\nu}(p,q)$ over the nucleon polarizations we obtain the
nucleon tensor (\ref{SF})
\begin{equation}
\label{W:on-shell}
{W}_{\mu\nu}^\tau(p,q)=\frac12\mathrm{Tr} 
\left[
(\dirac{p} + M)\widehat{\cal W}_{\mu\nu}^\tau(p,q)
\right].
\end{equation}

In off-shell region, the Lorentz tensor structure of $\widehat{{\cal
W}}_{\mu\nu}$ is more involved than the corresponding structure of
the on-shell nucleon tensor.  In order to establish the tensor
structure of $\widehat{{\cal W}}_{\mu\nu}$ we expand the latter in
terms of a complete set of Dirac matrices $ \left\{ I, \gamma^\alpha,
\sigma^{\alpha\beta}, \gamma^\alpha\gamma_5, \gamma_5\right\}$.  The
various coefficients in this expansion must be constructed from the
vectors $p$ and $q$, and from the symmetric tensor $g_{\alpha\beta}$
and the antisymmetric tensor $\epsilon_{\mu\nu\alpha\beta}$.
For the symmetric part of $\widehat {\cal W}_{\mu\nu}$ we
keep only those terms which are even under time-reversal and
parity transformations, since only such terms can contribute to
$F_{1,2}$. Keeping only current-conserving terms we have
7 independent Lorentz--Dirac structures which can be written as \cite{MST94,KPW94}
\begin{eqnarray}
\nonumber
2\, \widehat{\mathcal W}_{\mu\nu}^\mathrm{sym}(p,q) &=&
{} -\widetilde g_{\mu\nu}
	\left(\frac{f_1^{(0)}}{M} +
        \frac{f_1^{(1)}\dirac{p}}{M^2} +
        \frac{f_1^{(2)}\dirac{q}}{p\cdot q}\right)  \\
\label{W:off-shell:s}
&&{} + \frac{\widetilde p_\mu\widetilde p_\nu}{p\cdot q}
	\left(\frac{f_2^{(0)}}{M} +
        \frac{f_2^{(1)}\dirac{p}}{M^2} +
        \frac{f_2^{(2)}\dirac{q}}{p\cdot q}\right)
	+ \frac{f_2^{(3)}}{p\cdot q}\:
\widetilde p_{\{\mu}\widetilde g_{\nu\}\alpha}\gamma^\alpha ,
\end{eqnarray}
where $\widetilde g_{\mu\nu}$ and $\widetilde p_\mu$ are given by \eq{tilda}.
The curly braces in the last term denote symmetryzation over Lorentz
indeces, \ie 
$a_{\{\mu}b_{\nu\}}=\tfrac12(a_\mu b_\nu + a_\nu b_\mu)$.
The coefficients $f_i^{(j)}$ in \eq{W:off-shell:s} are the dimensionless 
Lorentz-invariant functions of $x,\ Q^2$ and the nucleon offshellness $p^2$.

Similar analysis can also be applied to the antisymmetric part
of $\widehat{\mathcal W}_{\mu\nu}$ for the neutrino scattering. This
term is described by the structure functions $F_3$ in \eq{CC:SF}. For
off-shell nucleon the result can be written as \cite{Ku98}
%
\begin{eqnarray}
\label{W:off-shell:a}
2\,\widehat\mathcal{W}_{\mu\nu}^\mathrm{asym}(p,q)
&=&
\frac{i\,\eps}{2\,p\cdot q}\, q^\alpha
\left[\left(
         \frac{f_3^{(0)}}{M}
        + \frac{f_3^{(1)}\dirac{p}}{M^2}
         + \frac{f_3^{(2)}\dirac{q}}{p\cdot q}
    \right) p^{\beta}
 + f_3^{(3)} \gamma^{\beta}
\right],
\end{eqnarray}
%
where the coefficients $f_3^{(j)}$ are dimensionless
Lorentz-invariant functions of $x$, $Q^2$ and $p^2$.

By substituting \Eqs{W:off-shell:s}{W:off-shell:a} into
Eq.(\ref{W:on-shell}) one observes that at $p^2=M^2$ \Eqs{SF}{CC:SF}
are recovered with the nucleon structure functions given by
\begin{subequations}\label{SF:on-shell}
\begin{eqnarray}
F_1&=&f_1^{(0)}+f_1^{(1)}+f_1^{(2)},\\
F_2&=&f_2^{(0)}+f_2^{(1)}+f_2^{(2)}+f_2^{(3)},\\
F_3&=&f_3^{(0)}+f_3^{(1)}+f_3^{(2)}+f_3^{(3)}.
\end{eqnarray}
\end{subequations}
It should be noted that the Dirac equation is the underlying reason of
simplification of the Lorentz structure of the hadronic tensor of the
on-shell nucleon.

One important observation which follows from this analysis is
that \eq{WA} does not factorize into completely separate nuclear and
nucleon parts. The off-shell nucleon is described by 7 independent
structure functions in the symmetric $P$-even hadronic
tensor ($f_{1}^{(i)}$ and $f_{2}^{(i)}$) and 4 independent structure functions 
$f_3^{(i)}$ in the $P$-odd antisymmetric 
hadronic tensor.  These functions depend on $p^2$ as an 
additional variable and weighted in \eq{WA} with generally different 
nuclear distributions.

Clearly, the fact that we have to deal with unknown functions
not present for the on-shell nucleon introduces additional uncertainty
in the calculation of nuclear structure functions.
However, in practice it may be quite sufficient to treat nuclei as
nonrelativistic systems. In this limit the nuclear hadronic tensor
considerably simplifies, as will be discussed in the next Section.

\subsubsection{The limit of weak nuclear binding}
\label{WNB-limit}

Let us now discuss \eq{WA} in the limit of weak nuclear binding. We
assume that the nucleus is a non-relativistic system with small
characteristic momentum and energy of bound nucleons, $|\bm{p}|\ll M,\
|p_0-M|\ll M$. The antinucleon degrees of freedom are neglected in
this approximation.
A nonrelativistic approximation to \eq{WA} is derived using the
relation between the relativistic four-component nucleon field $\Psi$
and the nonrelativistic two-component operator $\psi$ (for simplicity
we suppress the isospin index $\tau$)
\begin{eqnarray}
\label{psi:NR}
\Psi(\bm{p},t) = e^{-iMt}
\left(
	\begin{array}{r}
		(1-\bm{p}^2/8M^2)\,\psi(\bm{p},t) \\
(\bm{\sigma}\cdot\bm{p}/2M)\,\psi(\bm{p},t)
	\end{array}\right),
\end{eqnarray}
where the nucleon operators are taken in a mixed $(\bm{p},t)$
representation. The renormalization operator $1-\bm{p}^2/(8M^2)$ is
introduced to provide a correct normalization of nonrelativistic
nucleon field $\psi$, \ie  the operator $\psi^\dagger\psi$ is
normalized to the nucleon number to order $\bm{p}^2/M^2$.

In order to make the nonrelativistic reduction of \eq{WA}, we separate
the nucleon mass from the energy $p_0$ and write the four-momentum of the bound
nucleon as $p=(M+\ceps,\bm{p})$. We then substitute \eq{psi:NR} into
\eq{WA} and reduce the four-dimensional Dirac basis to the
two-dimensional spin matrices. In this way we examine all
Lorentz--Dirac structures in \Eqs{W:off-shell:s}{W:off-shell:a} and
keep the terms to order $\ceps/M$ and $\bm{p}^2/M^2$.
The result can be summarized as follows:
\begin{eqnarray}
\label{Tr-NR}
\frac1{M_A}\,
	\mathrm{Tr}\left({\cal A}(p;A)\,\widehat{\cal W}_{\mu\nu}(p,q)\right)=
\frac1{M+\varepsilon}\,{\cal P}(\varepsilon,\boldsymbol{p})\,
{W}_{\mu\nu}(p,q),
\end{eqnarray}
where $M_A$ is the mass of a nucleus $A$ and
\begin{equation}
\label{spfn}
{\mathcal P}(\varepsilon,\boldsymbol{p}) =
\int \ud t\, \exp({-i \varepsilon t})
	\langle A|
\psi^\dagger(\boldsymbol{p},t)\psi(\boldsymbol{p},0)
	|A\rangle/\langle A| A\rangle
\end{equation}
is the nonrelativistic nuclear spectral function normalized to the number
of nucleons in the corresponding isospin state
\begin{equation}
\label{norma}
\int [\ud p]\,{\cal P}^{p,n}(\varepsilon,\bm{p}) = (Z,N).
\end{equation}
Note that the factor $M_A$ in the left side of (\ref{Tr-NR}) is
absorbed in the normalization of nuclear states $\langle A| A\rangle$
in \eq{spfn}.
The hadronic tensor ${W}_{\mu\nu}(p,q)$ in \eq{Tr-NR} is given by
\eq{SF} with the structure functions
\begin{subequations}
\label{SF:off-shell}
\begin{eqnarray}
F_1(x,Q^2,p^2) &=&
	f_1^{(0)}\left(1+\frac{p^2-M^2}{2M^2}\right)+
	f_1^{(1)}\frac{p^2}{M^2}+
	f_1^{(2)},
\\
F_2(x,Q^2,p^2) &=&
	f_2^{(0)}\left(1+\frac{p^2-M^2}{2M^2}\right)+
	f_2^{(1)}\frac{p^2}{M^2}+
	f_2^{(2)}+
	f_2^{(3)} ,
\\
F_3(x,Q^2,p^2) &=&
	f_3^{(0)}\left(1+\frac{p^2-M^2}{2M^2}\right)+
	f_3^{(1)}\frac{p^2}{M^2}+
	f_3^{(2)}+
	f_3^{(3)}.
\end{eqnarray}
\end{subequations}
From \eq{Tr-NR} we obtain a nonrelativistic approximation to the
nuclear hadronic tensor (\ref{WA})
\begin{eqnarray}
\label{WA-NR}
\frac{W_{\mu\nu}^A(P_A,q)}{M_A} =
\sum_{\tau=p,n}\int \frac{[\ud p]}{M+\varepsilon}
{\mathcal P}^\tau(\varepsilon,\boldsymbol{p})\,
{W}_{\mu\nu}^\tau(p,q),
\end{eqnarray}
which is a basic equation for further analysis of nuclear DIS.

A few comments are in order.  It should be emphasized that the
nonrelativistic limit is taken with respect to the nucleon
momentum. In the derivation of Eq.(\ref{Tr-NR}) we keep terms to order
$\bm{p}^2/M^2$ and $\varepsilon/M$ and neglect the higher-order terms.
Furthermore, Eq.(\ref{Tr-NR}) is valid for arbitrary momentum transfer
$q$.
Note the factorization of the high-energy amplitude ${W}_{\mu\nu}$
from the spectral function ${\cal P}$ which describes the low-energy 
part of the problem.
In the vicinity of the mass shell the hadronic
tensor involves the same number of independent structure functions as
on the mass shell.
Therefore the problem of ``splitting'' of structure functions in the off-shell
region (i.e. the problem of additional nucleon structure functions)
can be avoided in this region.
Equations (\ref{SF:off-shell}) give
the nucleon structure functions in the off-shell
region in the vicinity of the mass-shell and it is easy to see that
Eqs.(\ref{SF:off-shell}) reduce to Eqs.(\ref{SF:on-shell}) at
$p^2=M^2$ thus assuring a correct on-shell limit.

Let us extract the relations between the nuclear and the nucleon
structure functions from \eq{WA-NR}.  Nuclear structure functions are
given by \eq{SF} with $p$ replaced by $P_A$ and $x$ by
$x_A=Q^2/(2M_A q_0)$.  However, it is convenient to consider the
nuclear structure functions as the functions of the
variable $x=Q^2/(2Mq_0)$ instead of the ``natural" nuclear scaling
variable $x_A$. We then define $F_{T,L}^A(x,Q^2)=F_{T,L}^A(x_A,Q^2)$ and
$xF_3^A(x,Q^2)=x_AF_3^A(x_A,Q^2)$. In order to separate the structure
functions we contract the both sides of \eq{WA-NR} with the virtual
photon polarization vectors (\ref{eTL}) and consider the helicity
structure functions. As a result we have
\begin{subequations}
\label{FA-TL}
\begin{align}
F_T^A(x,Q^2) &= \sum_{\tau=p,n}\int [\ud p]
	{\mathcal P}^\tau(\varepsilon,\bm{p})\,
		\left(1+\frac{\gamma p_z}{M}\right)
\left(
F_T^\tau + \frac{2{x'}^2\bm{p}_\perp^2}{Q^2}F_2^\tau 
\right),
\label{FA-T}
 \\
F_L^A(x,Q^2) &= \sum_{\tau=p,n}\int [\ud p]
	{\mathcal P}^\tau(\varepsilon,\bm{p})\,
		\left(1+\frac{\gamma p_z}{M}\right)
\left(
F_L^\tau + \frac{4{x'}^2\bm{p}_\perp^2}{Q^2}F_2^\tau
\right),
\label{FA-L}
\end{align}
\end{subequations}
where in the integrand $F_a^\tau$ with $a=T,L,2$ are the structure 
functions of bound proton ($\tau=p$) and neutron ($\tau=n$) with the 
four-momentum $p=(M+\ceps,\bm{p})$, 
$x'=Q^2/(2p{\cdot}q)=x/[1+(\varepsilon+\gamma p_z)/M]$ 
is the Bjorken variable for the bound nucleon and $\bm{p}_\perp$ is the 
transverse component of the nucleon momentum with respect to the 
momentum transfer.
The off-shell nucleon structure 
functions depend on $x'$, momentum transfer square $Q^2$ and the 
virtuality $p^2=(M+\ceps)^2-\bm{p}^2$ as additional variable.
In \eq{FA-TL} the off-shell transverse and longitudinal structure
functions are given by equations similar to (\ref{FTL}) with $M^2$
replaced by $p^2$, \ie   $F_T=2x'F_1$,  $F_L={\gamma'}^2F_2-F_T$ with
${\gamma'}^2=1+4{x'}^2p^2/Q^2$.
Using \eqs{FA-TL} we have for the nuclear structure function $F_2^A$
\begin{eqnarray}
\gamma^2 F_2^A(x,Q^2) &=&
 \sum_{\tau=p,n}\int [\ud p]
	{\mathcal P}^\tau(\varepsilon,\boldsymbol{p})\,
		\left(1+\frac{\gamma p_z}{M}\right)
\left({\gamma'}^2 +\frac{6{x'}^2 \bm{p}_\perp^2}{Q^2} \right)
F_2^\tau. 
\label{FA-2}
\end{eqnarray}

The nuclear structure function $F_3$ can be extracted from the
left-right asymmetry in the helicity amplitudes,
\eq{SF:hel:nu}. We have \cite{Ku98}
\begin{eqnarray}
xF_3^A(x,Q^2) &=&
 \sum_{\tau=p,n}\int [\ud p]
	{\mathcal P}^\tau(\varepsilon,\boldsymbol{p})\,
		\left(1+\frac{p_z}{\gamma M}\right)
        x'F_3^\tau. 
\label{FA-3}
\end{eqnarray}

Equations (\ref{FA-TL}) to (\ref{FA-3}) allow us to compute the
structure functions of a generic nucleus as a convolution of nuclear
spectral function, which describes the distribution of the bound
nucleons over momentum and separation energy, with the bound proton and neutron
structure functions.

We also comment that the transverse motion of the bound nucleon in the
target rest frame causes the mixture of different structure functions
in \Eqs{FA-T}{FA-L} to order $Q^{-2}$ (note $\bm{p}_\perp^2$ terms in 
these equations). This effect on the ratio $F_L^A/F_T^A$ was 
recently discussed in \cite{Ericson:2002ep}.

\subsubsection{Convolution representation}
\label{sec:convol}

If $Q^2$ is high enough to neglect power terms in (\ref{FA-TL}--\ref{FA-3})
then these equations can be written as two-dimensional convolution.
For example, for the structure function $F_2$ we have
\begin{align}
\label{convol:N}
F_2^A(x,Q^2) &= \int_{y>x} \ud y\ud v \left[ f_{p/A}(y,v)F_2^p(x/y,Q^2,v)+
f_{n/A}(y,v)F_2^n(x/y,Q^2,v) \right],
\end{align}
where $f_{p/A}(y,v)$ and $f_{n/A}(y,v)$ are the proton and the neutron 
distributions over the fraction of light-cone momentum $y$ and the virtuality 
$v=p^2$. The proton (neutron) distribution function is given in terms of 
the proton (neutron) spectral function as follows \cite{Ku89,KPW94}
\begin{equation}
\label{ydist:N}
f(y,v) = \int [\ud p] \mathcal{P}(\ceps,\bm{p})\left(1+\frac{p_z}{M}\right)
\delta\left(y-1-\frac{\ceps+p_z}{M}\right)\delta(v-p^2).
\end{equation}
The distribution functions are normalized to the number of bound protons 
(neutrons), as follows from \eq{norma}. Equations similar to 
(\ref{convol:N}) hold for other structure functions with the same nucleon 
distribution functions. If we further neglect off-shell effects in the 
structure functions, \eq{convol:N} reduces to the familiar one-dimensional 
convolution.

It is instructive to calculate the average nucleon light-cone momentum 
$\average{y}_N$ per one nucleon. Using \eq{ydist:N} we have 
\begin{equation}
\average{y}_N = 1+\frac{\average{\ceps}+\frac23\average{T}}{M},
\label{y:N}
\end{equation}
where $\average{\ceps}$ and $\average{T}$ are the average nucleon
separation and kinetic energies. Because of binding effect we have
$\average{y}_N<1$ (using our 
nuclear spectral function from Sec.\:\ref{sec:spfn}
we have for iron  $\average{y}_N=0.966$).  The missing
nuclear light-cone momentum apparently should be carried by fields
responsible for nuclear binding. In our approach the missing
light-cone momentum is balanced by nuclear pion field. Note that this
situation is qualitatively similar to the balance of light-cone momentum in the
nucleon in which about a half of the nucleon momentum is carried by
gluons. However, in nuclear case the fraction of pion
light-cone momentum is much smaller because the nuclei
are weakly-bound systems. 
The scattering from nuclear pions is discussed below.

\subsubsection{Pion contribution to nuclear structure functions}
\label{sec:pion}

The lepton can scatter off virtual pions which are exchanged by bound
nucleons.  The pion correction to nuclear hadronic tensor can be
written as follows (see,\eg, \cite{Ku89})
\begin{equation}
W_{\mu\nu}^{\pi/A}(P_A,q) =\frac12 \int[\ud k] 
D_{\pi/A}(k)W_{\mu\nu}^{\pi}(k,q)
\label{Wmunu:pion}
\end{equation}
where $W_{\mu\nu}^{\pi}(k,q)$ is hadronic tensor of a pion with 
four-momentum $k$ and the function $D_{\pi/A}$ describes the distribution 
of pions in a nucleus. The latter can be expressed in terms of the pion 
propagator in a nucleus as
\begin{equation}
D_{\pi/A}(k) = \int \ud^4 x \exp(ik\cdot x)
\bra{A}\bm{\varphi}(x)\bm{\varphi}(0)\ket{A},
\label{D:pion}
\end{equation}
where $\bm{\varphi}=(\varphi_1,\varphi_2,\varphi_3)$ is the pion field 
operator. The $\pi^0$ state is described by real pseudoscalar field 
$\varphi_3$, while the charged pion states are described by the complex 
pseudoscalar fields $(\varphi_1\pm\varphi_2)/\sqrt2$. The factor 1/2 in 
\eq{Wmunu:pion} is because of the chosen representation of the pion field 
operator in \eq{D:pion} in which particle and antiparticle are identical.

In the further discussion of pion effect it is convenient to consider
the normalized pion distribution, \ie independent of normalization of
the target state. We define this as follows
\begin{align}
\label{D:pion:2}
\mathcal{D}_{\pi/A}(k) &= \int \ud t \exp(ik_0 t)
        \bra{A}\varphi^*(\bm{k},t)\varphi(\bm{k},0)\ket{A}/\la A|A \ra,
\\
\varphi(\bm{k},t) &= \int\ud^3\bm{r}\, 
                     \exp({i\bm{k}\cdot\bm{r}})\varphi(\bm{r},t),
\end{align}
where $\varphi(\bm{k},t)$ is the pion field operator in momentum 
representation. Using translational invariance it is easy to verify that 
$\mathcal{D}_{\pi/A}(k)={D}_{\pi/A}(k)/(2M_A)$ in the nucleus rest frame.

In order to extract the pion contribution to nuclear structure
functions we contract both sides of \eq{Wmunu:pion} with the photon
polarization vectors.  Assuming that the hadronic tensor for off-shell
pions is given by \eq{SF} we obtain from \eq{Wmunu:pion}
\begin{subequations}
\label{SF:pion}
\begin{align}
F_T^{\pi/A}(x,Q^2) &= \int[\ud k] 
        \mathcal{D}_{\pi/A}(k)
        \left(k_0+\gamma k_z\right)
        \left(F_T^\pi + 
        \frac{2{x'}^2\bm{k}_\perp^2}{Q^2}F_2^\pi \right), \\
F_L^{\pi/A}(x,Q^2) &=  \int[\ud k] 
        \mathcal{D}_{\pi/A}(k)
        \left(k_0+\gamma k_z\right)
        \left(F_L^\pi + 
        \frac{4{x'}^2\bm{k}_\perp^2}{Q^2}F_2^\pi \right),
\\
\gamma^2 F_2^{\pi/A}(x,Q^2) &= \int[\ud k] 
        \mathcal{D}_{\pi/A}(k)
        \left(k_0+\gamma k_z\right)
        \left({\gamma'}^2+\frac{6{x'}^2\bm{k}_\perp^2}{Q^2}\right)F_2^\pi,
\\
xF_3^{\pi/A}(x,Q^2) &= \int[\ud k] 
        \mathcal{D}_{\pi/A}(k)
        \left(k_0+k_z/\gamma \right)
                x'F_3^\pi,
\end{align}
\end{subequations}
where $F_a^{\pi/A}$ denotes the pion correction to the nuclear 
structure function $F_a^A$. In the integrand $F_a^{\pi}$
are the structure functions of virtual pion with four-momentum 
$k$, $\bm{k}_\perp$ is transverse component of the pion momentum relative 
to the direction of momentum transfer and $x'=Q^2/(2k\cdot q)$ is the pion 
Bjorken variable. The pion structure functions in \eqs{SF:pion} depend on 
$x'$, $Q^2$ and pion invariant mass $k^2=k_0^2-\bm{k}^2$ as an additional 
variable. The transverse and longitudinal structure functions are related 
to $F_1$ and $F_2$ as $F_T=2x'F_1$, $F_L={\gamma'}^2F_2-F_T$ with 
${\gamma'}^2=1+4{x'}^2k^2/Q^2$. The mixture of the structure 
functions $F_T$ and $F_L$ in \eqs{SF:pion} is because of transverse motion 
of nuclear pions, similar to the corresponding effect in \eqs{FA-TL} for 
bound nucleons.

At high $Q^2$ \eqs{SF:pion} can be written in a convolution form. For 
example, pion correction to $F_2$ can be written as
\begin{align}
\label{pion:conv}
F_2^{\pi/A}(x,Q^2) &= \int_{x<y} \ud y\ud v f_{\pi/A}(y,v) 
F_2^\pi(x/y,Q^2,v), \\
f_{\pi/A}(y,v) &= 2yM \int[\ud k]\mathcal{D}_{\pi/A}(k)
                \delta\left(y-\frac{k_0+k_z}{M}\right)\delta(v-k^2),
\label{pion:f}
\end{align}
Similar equations hold for other structure functions in \eqs{SF:pion}. If 
one neglects the off-shell dependence of the pion structure functions then 
\eq{pion:conv} reduces to the standard one-dimensional convolution with the pion 
light-cone distribution which is given by \eq{pion:f} integrated over $v$.
We note that the distribution function by \eq{pion:f} is 
antisymmetric function, $f_{\pi/A}(-y)=-f_{\pi/A}(y)$. This property 
allows us to derive the sum rules for the odd moments of the pion 
distribution function which will be discussed in more detail in 
Sec.\:\ref{sec:pion:details}.


\subsubsection{Application to the deuteron}
\label{d-sect}

So far the discussion did not refer to particular nuclear
target.  In this section we apply the discussed formalism to the
\emph{deuteron}.
The deuteron is an isoscalar bound state of the proton and
the neutron.  The residual nuclear system is, therefore, the proton or
neutron and the spectral function is given in terms of the deuteron
wave function $\Psi_D(\bm{p})$
\begin{equation}
\label{spfn:D}
{\mathcal P}^{p,n}(\varepsilon,\bm{p}) = 2\pi\delta\left(
        \ceps -\ceps_D + \frac{\bm{p}^2}{2M}\right)
\left|\Psi_D(\bm{p})\right|^2,
\end{equation}
where $\ceps_D=M_D-2M$ and $\bm{p}^2/{2M}$ are the deuteron binding 
energy and the spectator nucleon recoil energy, respectively. The deuteron 
structure functions then become
\begin{subequations}
\label{D-SF}
\begin{eqnarray}
F_T^D(x,Q^2) &=& \int\frac{\ud^3\bm{p}}{(2\pi)^3}
\left|\Psi_D(\bm{p})\right|^2\left(1+\frac{\gamma p_z}{M}\right)
\left(F_T^{N}+\frac{2{x'}^2\bm{p}^2_\perp}{Q^2}F_2^{N}\right),
\\
F_L^D(x,Q^2) &=& \int\frac{\ud^3\bm{p}}{(2\pi)^3}
\left|\Psi_D(\bm{p})\right|^2\left(1+\frac{\gamma p_z}{M}\right)
\left(F_L^{N}+\frac{2{x'}^2\bm{p}^2_\perp}{Q^2}F_2^{N}\right),
\\
\gamma^2 F_2^D(x,Q^2) &=& \int\frac{\ud^3\bm{p}}{(2\pi)^3}
\left|\Psi_D(\bm{p})\right|^2\left(1+\frac{\gamma p_z}{M}\right)
\left({\gamma'}^2 +\frac{6{x'}^2 \bm{p}_\perp^2}{Q^2} \right)
F_2^{N},
\\
xF_3^D(x,Q^2) &=& \int\frac{\ud^3\bm{p}}{(2\pi)^3}
\left|\Psi_D(\bm{p})\right|^2\left(1+\frac{p_z}{\gamma M}\right)
x'F_3^{N}
\end{eqnarray}
\end{subequations}
where $F_a^N = (F_a^p + F_a^n)/2$ with $a=T,2,3$ are the structure
functions of the isoscalar nucleon. The variables of these structure
functions are similar to those of
Eqs.(\ref{FA-TL},\ref{FA-2},\ref{FA-3}) and we do not write them
explicitly.

\subsubsection{Application to complex nuclei}
\label{A-sect}

Unlike the deuteron, the spectral function of complex nuclei does not 
reduce to the ground state wave function but includes, generally infinite, 
set of excited residual states (this can be seen directly from \eq{spfn} 
by inserting the complete set of intermediate states). Furthermore, 
complex nuclei typically have different numbers of protons and neutrons 
and, in contrast to the deuteron case, the calculation of nuclear 
structure functions requires both the isoscalar and the isovector 
contributions. In order to take into account this effect we explicitly 
separate the isoscalar and the isovector contributions to 
Eqs.(\ref{FA-TL},\ref{FA-2},\ref{FA-3}). To this end we consider generic 
integrand in the convolution formulas and write
\begin{equation}
\sum_{\tau=p,n}\mathcal{P}^\tau F_{a}^\tau =
\mathcal{P}^{p+n} F_{a}^N + \mathcal{P}^{p-n}F_a^{p-n}/2,
\label{nuke:pn}
\end{equation}
where we denote $\mathcal{P}^{p\pm n}=\mathcal{P}^p\pm\mathcal{P}^n$
and $F_a^N=\tfrac12(F_a^p+F_a^n)$ and $F_a^{p-n}=F_a^p-F_a^n$ for the
structure function of type $a$.

In an isoscalar nucleus with equal number of protons and
neutrons \eq{nuke:pn} is dominated by the isoscalar contribution
and one generally assumes $\mathcal{P}^{p-n}=0$.
However, it must be remarked that this equation is violated by a
number of effects even in the isoscalar nucleus. The finite difference
between the proton and neutron spectral functions is generated by the
Coulomb interaction and isospin-dependent effects in the
nucleon--nucleon interaction. The discussion of these effects goes
beyond the scope of this paper and we leave them for future studies.
Instead we focus on the neutron excess effect for heavy nuclei.

We write the isoscalar and isovector spectral functions in terms of
reduced functions $\mathcal{P}_{0}$ and $\mathcal{P}_{1}$ as
\begin{subequations}
\label{spfn:01}
\begin{eqnarray}
\mathcal{P}^{p+n} &=& A \mathcal{P}_0,\\
\mathcal{P}^{p-n} &=& (Z-N)\mathcal{P}_1.
\end{eqnarray}
\end{subequations}
The functions $\mathcal{P}_{0}$ and $\mathcal{P}_{1}$ are normalized
to unity as follows from \eq{norma}. These spectral functions are
quite different. The function $\mathcal{P}_0$ involves the averaging
over all isoscalar intermediate states.
The function $\mathcal{P}_1$ probes the isovector component in a nucleus
and its strength is peaked about the Fermi surface as argued in 
Sec.\:\ref{sec:spfn}.
The model spectral functios $\mathcal{P}_{0}$ and $\mathcal{P}_{1}$, which 
are used in this paper, are discussed in Sec.\ref{sec:spfn}.

Using \Eqs{nuke:pn}{spfn:01} we can write each of the structure function 
$a=T,2,3$ as
\begin{equation}
\label{nuke:FA}
F_a^A = A\left\langle F_a^N \right\rangle_0 +
	\frac{Z{-}N}{2}\left\langle F_a^{p-n} \right\rangle_1,
\end{equation}
where the averaging $\left\langle F_a \right\rangle_{0,1}$ denotes the
integration in Eqs.(\ref{FA-TL},\ref{FA-2},\ref{FA-3}) with the reduced
spectral functions $\mathcal{P}_0$ and $\mathcal{P}_1$,
respectively.

We conclude this section by commenting that data are sometimes naively
corrected for the neutron excess effect neglecting Fermi motion and
binding effects (as well as any other nuclear effects) in the isovector and
the isoscalar distributions. As follows from the present discussion, the
Fermi motion and binding effects are quite different in the isoscalar
and the isovector distributions in heavy nuclei. If neglected, this effect
may cause an additional systematic uncertainty in data and a
distortion of final results.

\subsubsection{Off-shell effects}
\label{off-shell-sect}
\label{sec:offshell}

The bound proton and neutron are off-mass-shell and their structure
functions differ from those of the free proton and neutron. The off-shell
nucleon structure functions depend on the nucleon virtuality $p^2$ as an
additional variable. Therefore, the off-shell effects in the structure
functions are closely related to the target mass corrections. Target mass
effects in the off-shell nucleon can be of two different kinds. First,
similarly to the on-shell nucleon, we have to take into account the
kinematical target mass dependence due to the finite $p^2/Q^2$ ratio. We
assume that this effect is described by Eqs.(\ref{TMC}), where the nucleon
mass squared is replaced by $p^2$ (this leads in turn to the modification
of the parameter $\gamma$ and the variable $\xi$ in the off-shell region).
Furthermore, the dependence on $p^2$ appears already at leading twist (LT)
at the PDF level as was argued in \cite{GL92,KPW94,Ku98,MST94}.
Thus
off-shell effects in the LT structure functions can be viewed as a
measure of the nucleon's modification inside nuclear medium.

Since we treat nuclei as nonrelativistic systems it would be enough to
consider the off-shell effect as a correction. We expand the nucleon
LT structure functions in the vicinity of the mass shell in series in
$p^2{-}M^2$.  Keeping only the linear term we have for $F_2$
\begin{align}
\label{q:deriv}
F_2(x,Q^2,p^2) &= 
F_2(x,Q^2)\left(1+\delta f_2(x,Q^2)\,\frac{p^2{-}M^2}{M^2}\right),\\
\label{delf}
\delta f_2(x,Q^2) &= \frac{\partial\ln F_2(x,Q^2,p^2)}{\partial\ln p^2},
\end{align}
where the first term is the structure function of the on-mass-shell
nucleon and the derivative is evaluated at $p^2=M^2$. Similar expressions
can be written for the other structure functions.

The function $\delta f_2$ can be related to the corresponding
off-shell functions for the nucleon parton distributions. The
necessary relation can be obtained by writing $F_2$ in terms of a
convolution of the parton distributions with the corresponding
coefficient function according to the given order in $\as$. In order
to simplify discussion and illustrate this relation we can consider
the simple leading order expression of $F_2$
\begin{equation}
\label{F2:LT}
F_2 = x\sum e_i^2\left(q_i + \bar q_i\right),
\end{equation}
where $e_i$ and $q_i (\bar q_i)$ are the charge and the distribution
of (anti)quarks of the type $i$ and the sum is taken over different
types of quarks.  The off-shell function for the parton distribution
$q(x)$ is defined similarly to \eq{delf}, $\delta f_q = {\partial\ln
q}/{\partial\ln p^2}$. Then from \eq{F2:LT} we have a relation
\begin{equation}
\label{del:f2}
F_2(x) \delta f_2(x) =  x\sum e_q^2
\left[q(x)\delta f_q(x) + \bar q(x)\delta f_{\bar q}(x)\right].
\end{equation}
One can conclude from \eq{del:f2} that at large $x$, where the
antiquark distributions can be neglected, $\delta f_2$ is dominated by
quarks. For simplicity we neglect the isospin effect and assume
$\delta f_u=\delta f_d=\delta f_q$, then $\delta f_2=\delta f_q$ at
large $x$.  At small $x$ both, the quark and the antiquark
contributions, have to be taken into account.

Off-shell effects in nucleon structure functions were discussed in
\cite{KPW94,Ku98} using the spectral representation of the quark
distributions in the nucleon with four-momentum $p$
\begin{equation}\label{q:spec:off}
q(x,p^2) =\int\ud s\int ^{\tmax} \negthickspace
\ud t\, D_{q/N}(s,t,x,p^2).
\end{equation}
The integration in \eq{q:spec:off} is taken over
the mass spectrum of spectator states $s$ and the quark virtuality $t=k^2$
with the kinematical maximum $\tmax=x[p^2-s/(1-x)]$ for the given $s$ and
$p^2$. The invariant spectral density $D_{q/N}$ measures the probability
to find in a nucleon with momentum $p$, a quark with light-cone
momentum $x$ and virtuality $t$ and the remnant system in a state
with invariant mass $s$.

We conclude from \eq{q:spec:off} that the $p^2$ dependence of quark
distributions can have two sources: the $p^2$ term in $\tmax$
(kinematical off-shell dependence), and
the $p^2$ dependence of the quark spectral function $D_{q/N}$
(dynamical off-shell dependence).
The kinematical off-shell effect causes a negative correction to the bound
nucleon structure functions that results in an enhanced EMC effect, as
first noticed in \cite{GL92,KPW94}. However, if only the kinematical
off-shell effects are taken into account the number of valence quarks
in the nucleon would change with $p^2$. It can be seen directly from
\eq{q:spec:off} that the normalization of the quark distribution
decreases as $p^2$ decreases, provided that the spectral density is
positively defined.
This observation indicates that off-shell effect of dynamical
origin must also be present. A method to estimate the dynamical
off-shell effects minimizing the model dependence was suggested in
\cite{KPW94}, in which the conservation of the
valence quark number in off-shell nucleon was used as a
constraint.
A partial cancellation between the kinematical and dynamical off-shell
effects was found in \cite{KPW94,Ku98}. However, the off-shell effect in
the structure functions remains an important correction.
In this paper we treat the function $\delta f_2$ phenomenologically and fix
it from nuclear data as discussed in more detail in Section
\ref{sec:model}.

\subsection{Coherent nuclear effects}
\label{smallx-nuke}

Nuclear shadowing effect was extensively discussed in the literature.
A recent paper \cite{Piller:1999wx} provides a review of both data and
theoretical models of nuclear shadowing.

It appears to be a common wisdom that nuclear shadowing is a result of
coherent interaction of hadronic component of virtual photon with
target nucleus. The structure functions at small $x$ can be presented as a
superposition of contributions from different hadronic states. We
consider the helicity structure functions $W_0$ and $W_{\pm}$, as defined in
\eq{SF:hel}, that will allow us to discuss nuclear effects in
charged-lepton and neutrino interactions on the same ground. We have
\begin{equation}
W_h = \sum_v w_v \sigma_h^v(s),
\label{sum-h}
\end{equation}
where $\sigma_h^v(s)$ is the total cross section of scattering of the
hadronic state $v$ with the given helicity $h=0,\pm 1$ off the target
nucleon (or nucleus) with the center-of-mass energy
$s=Q^2(1/x{-}1)+M^2$ and the quantities $w_v$ describe the weight of
different hadronic states.

At low $Q^2$ the vector meson dominance model (VMD) was proved to be a
good tool to evaluate nuclear corrections to the structure functions
\cite{Bauer:iq}. In this model the structure functions are approximated
by contributions from a few vector-meson states.
The weights for the
electromagnetic current are $w_v=Q^2/(\pi f_v^2)(1+Q^2/m_v^2)^{-2}$
with $f_v$ the photon-meson coupling constants and $m_v$ the vector
meson mass. Usually only the lowest mass vector mesons ($\rho^0,\
\omega,\ \phi$) are important at low $Q^2\lesssim 1\gevsq$. The 
VMD structure functions have strong $Q^2$ dependence and decrease
as $Q^{-2}$ at high $Q$.
In the generalized versions of VMD, the higher-mass states including
continuum have also been considered that made it possible to apply the
model at higher $Q^2$ (see, \eg, \cite{Piller:1999wx}).

In this paper we approximate the sum over hadronic states in
\eq{sum-h} by a factorized form
\begin{equation}
W_h(x,Q^2) = w_h(x,Q^2) \sigmabar_h(s),
\label{sum-h-aprx}
\end{equation}
where $\sigmabar_h$ is an \emph{effective} cross section 
corresponding to helicity $h$ averaged over hadronic configurations and 
$w_h$ is remaining normalization factor. At low $Q^2$ the quantity 
$\sigmabar_h$ corresponds to the average over a few vector meson states.  
As $Q^2$ increases, the averaging in (\ref{sum-h-aprx}) involves the 
rising number of active hadronic configurations. Since the relative weight 
of higher-mass states increases with $Q^2$ and the cross section decreases 
with the mass, one can qualitatively conclude that $\sigmabar_h$ should 
decrease with $Q^2$. In the approach adopted in this paper we will treat 
$\sigmabar_h$ phenomenologically.

In this paper we are concerned with the relative effect of nuclear
interactions
\begin{equation}
\delta \mathcal{R}_h(x,Q^2,{A/N})={\delta W_h^A(x,Q^2)}/{W_h^N(x,Q^2)},
\label{delR}
\end{equation}
where $\delta W^A_h$ is the nuclear structure function of helicity $h$ 
subtracted incoherent contribution (cf. \eq{FA}).
Assuming that the weight factors are not affected by nuclear effects,
from \eq{sum-h-aprx} we conclude that the relative nuclear correction to
the structure functions equals the corresponding correction to the
effective cross section
\begin{equation}
\delta \mathcal{R}_h({A/N}) =
{\delta\sigmabar_h^A}/{\sigmabar_h^N},
\label{ratio:sh}
\end{equation}
where $\delta\sigmabar^A$, similar to $\delta W^A$, is the nuclear
cross section subtracted incoherent contribution. The problem of
calculation of nuclear corrections to structure functions at small $x$ thus 
reduces to the calculation of multiple scattering effects on 
effective hadronic cross section.

\subsubsection{Application to the deuteron}
\label{sec:smallx:D}

We now consider this effect in application to the {deuteron}.
In order to calculate the shadowing correction we consider hadron elastic 
scattering amplitude $a(s,k)$ with $s$ the center-of-mass energy and $k$ 
the momentum transfer. We choose the normalization of the amplitude such 
that the optical theorem reads $\Im a(s,0) = \sigma(s)/2$ and parametrize 
the scattering amplitude as $a=(i+\alpha)(\sigma/2)\exp(-B k^2/2)$, where 
the exponent describes the dependence on momentum transfer.%
\footnote{Such dependence is confirmed experimentally and for
low mass vector mesons the value of the parameter $B$ is between
4 and 10$\gevsq$ depending on $Q^2$ (see, \eg, \protect\cite{Bauer:iq}).}
The hadron-deuteron scattering 
amplitude in forward direction can be written as 
\cite{Glauber:1955qq}
\begin{align}\label{mult:D}
\begin{split}
a^D &= a^p + a^n + \delta a^D,
\\
\delta a^D &= i a^p a^n \mathcal{C}_2^D,
\end{split}
\end{align}
where $a^p$ and $a^n$ are the scattering amplitudes off the proton and
the neutron and $\delta a^D$ the double scattering 
correction. $\mathcal{C}_2^D$ can be written in terms of the deuteron wave 
function as 
\begin{align}\label{ff:D}
\mathcal{C}_2^D &= \frac{1}{(2\pi)^2}\int\ud^2\bm{k}_\perp
                   S_D(\bm{k}_\perp,k_L)e^{-B\bm{k}_\perp^2},\\
S_D(\bm{k}) &= \int\ud^3\bm{r} e^{i\bm{k}\cdot\bm{r}}
                \left|\Psi_D(\bm{r})\right|^2.
\end{align}

Note that \eq{mult:D} is the scattering amplidude of an off-shell
hadron with four-momentum $q$. For this reason there appears a finite
longitudinal momentum transfer $k_L=Mx(1+m^2/Q^2)$, which accounts
for a finite longitudinal correlation length of a virtual hadron
($k_L=0$ for the scattering of on-shell particles).

We apply \Eqs{ratio:sh}{mult:D} in order to calculate coherent nuclear 
effects for different structure functions. It should be remarked that 
helicity conserves in multiple scattering interactions and the scattering 
matrix is diagonal in helicity basis. For this reason the multiple 
scattering corrections involve the amplitudes with the same helicity. We 
also assume no isospin effect, i.e.
the effective scattering amplitudes of the given helicity are 
equal for the proton and the neutron.
Let us first discuss the transverse structure function $F_T$. The relative 
shadowing correction to the transverse structure function is 
\begin{equation}
\delta \mathcal{R}_T(x,Q^2,{D/N}) = \sigma_T(\alpha_T^2-1)\mathcal{C}_2^D/2,
\label{sh:T:D}
\end{equation}
where $\sigma_T$ and $\alpha_T$ are parameters of effective scattering 
amplitude of transversaly polarized virtual photon. A particular model of
the scattering amplitude used in our analysis is discussed in 
Section \ref{sec:xsec}.

Relation similar to \eq{sh:T:D} holds in the longitudinal channel.
It follows from \eq{sum-h-aprx} that the ratio $R=F_L/F_T$ 
equals the corresponding ratio of the cross sections 
$\sigma_L/\sigma_T$ provided that the normalization factor 
$w(x,Q^2)$ is independent of helicity. This holds in the VMD for low-mass 
mesons and we also assume that this is approximately true for 
the contribution from higher-mass states.
Thus using \eq{sh:T:D} and assuming $\alpha_L=\alpha_T$ we find that the 
relative shadowing corrections for longitudinal and transverse structure 
functions is simply determined by $R$
\begin{equation}
\frac{\delta \mathcal{R}_L(x,Q^2,{D/N})}{\delta \mathcal{R}_T(x,Q^2,{D/N})} = 
R(x,Q^2),
\label{sh:L:D}
\end{equation}
where $R(x,Q^2)$ is calculated for the nucleon.

\Eqs{sh:T:D}{sh:L:D} allow us to compute the nuclear shadowing effect
for the structure function $F_2$ in terms of the corresponding
correction to $F_T$.  Indeed, recalling \eqs{FTL} we have 
\begin{equation}
\label{sh:2:T:D}
\delta \mathcal{R}_2({D/N}) =
\frac{\delta \mathcal{R}_T({D/N}) + R \delta \mathcal{R}_L({D/N})}{1+R}.
\end{equation}
Taking into account (\ref{sh:L:D}) we find the factor $(1+R^2)/(1+R)$
difference between shadowing effect for $F_2$ and $F_T$.

Let us discuss the shadowing effect for the structure function
$xF_3$. This structure function is given by the left-right asymmetry
in helicity structure functions $W_+ - W_-$. Therefore, in this case
the problem reduces to computing the multiple scattering effect for
the difference of the corresponding scattering amplitudes. We denote 
${\Delta}a=a_+ - a_-$. The non-zero difference ${\Delta}a$ is generated 
because of vector--axial vector current transitions in the hadronic 
tensor. The double scattering correction to ${\Delta}a$ can readily be 
derived from \eq{mult:D}
\begin{equation}
\label{asym:D}
\delta {\Delta}a^D = 2i {\Delta}a\, a_T \mathcal{C}_2^D,
\end{equation}
where we denote $a_T=(a_+ + a_-)/2$.
It follows from \eq{asym:D} that the relative shadowing effect for the
cross-section asymmetry is determined by the cross-section $\sigma_T$. 
Using \Eqs{mult:D}{asym:D} we find
\begin{equation}
\frac{\delta \mathcal{R}_{\Delta}({D/N})}{\delta \mathcal{R}_T({D/N})} = 
2\frac{1-\alpha_{\Delta}\alpha_T}{1-\alpha_T^2}.
\label{sh:3:D}
\end{equation}

We observe from this equation that the shadowing effect is 
enhanced for the cross section asymmetry by the factor of 2 with respect 
to the shadowing effect for the cross section $\sigma_T$ if we neglect the 
effect of real part of the amplitudes \cite{Kulagin:1998wc}.
To clarify the origin of this enhancement we consider a somewhat 
simplified VMD model with the single vector meson ($\rho$ meson) and the 
axial-vector meson ($a_1$ meson). In this model the structure functions 
$F_L$ and $F_T$ in charged-current scattering
are determined by the diagonal vector--vector and axial 
vector--axial vector transitions $VN\to VN$ and $AN\to AN$, while the 
structure function $F_3$ is driven by the off-diagonal transitions $VN\to 
AN$ and $AN\to VN$. The cross section of the off-diagonal transitions is 
much smaller than the cross sections of the direct processes. For this 
reason, $xF_3\ll F_2$ at small $x$.
However, the nuclear multiple scattering corrections to off-diagonal 
process $V\to A$ are determined by a strong cross section of 
the diagonal processes $V\to V$ and $A\to A$. This 
becomes clear if we consider the
double scattering term for the off-diagonal nuclear amplitude. To this
order the nuclear scattering proceeds via two steps: the off-diagonal
scattering from one nucleon followed by the diagonal scattering from
the second nucleon. The off-diagonal scattering can be interchanged
with the diagonal scattering that leads to the factor of 2 enhancement,
which appears to have the combinatorial origin.%

\subsubsection{Application to complex nuclei}
\label{sec:smallx:A}

We now turn to the discussion of the shadowing effect in
{complex nuclei}.
We apply the Glauber--Gribov multiple scattering theory to calculate
the multiple scattering effect on effective cross sections. Let $a^A$
be the nuclear scattering amplitude in forward direction. We will assume 
no isospin dependence of the scattering amplitude, \ie $a^p=a^n$. Then 
$a^A$ can be written as (see, \eg, \cite{Bauer:iq} and references therein)
\begin{align}\label{mult:A}
\begin{split}
a^A &= A\, a + \delta a^A,
\\
\delta a^A &= ia^2\mathcal{C}_2^A(a),
\end{split}
\end{align}
where $a$ is the corresponding nucleon amplitude and $\mathcal{C}_2^A$
incorporates the multiple scattering effects and read as follows
\begin{equation}\label{C2A}
\mathcal{C}_2^A(a) = \int_{z_1<z_2}\hspace{-1.5em}
		\ud^2\bm{b}\ud z_1\ud z_2\,
        \rho_A(\bm{b},z_1)\rho_A(\bm{b},z_2)
\exp\left[
i\int_{z_1}^{z_2}\hspace{-1em}\ud z'
        \left(a\,\rho_A(\bm{b},z') - k_L \right)
        \right].
\end{equation}
Here $\rho_A$ is the nucleon density distribution normalized to the number 
of nucleons $A$ and the integration is performed along the collision axis, 
which is chosen to be $z$-axis, and over the transverse positions of 
nucleons (impact parameter $\bm{b}$). If only the double scattering 
approximation is considered then the exponential factor in \eq{C2A} should 
be omitted. The exponential factor in \eq{C2A} accounts for multiple 
scattering effects (see, \eg, \cite{Bauer:iq}).

We note that \Eqs{mult:A}{C2A} were derived assuming that the wave function
factorizes into the product of the single particle wave functions and
neglecting short-range correlation effects between bound nucleons 
(optical approximation). We comment in this respect that the correlations 
are relevant only if the coherence length $L_c=1/k_L$ is comparable to the 
short-range repulsive part of the nucleon--nucleon force, which is about 
0.5 Fm. This takes place at relatively large $x$, for which shadowing 
effect is small (see discussion in Ref.\cite{Piller:1999wx}).

The transverse momentum dependence of elastic scattering amplitudes
was also neglected, since the transverse size of the meson-nucleon
amplitude in the impact parameter space is of order $B^{-1/2}$,
much smaller than the radius of the nucleus.

We first discuss multiple scattering correction to the transverse 
structure function. The relative shadowing correction is determined by 
effective scattering amplitude $a_T$ of transversely polarized virtual 
photon 
\begin{equation}
\delta \mathcal{R}_T({A/N}) = 
\sigma_T\Re(i+\alpha_T)^2\mathcal{C}_2^A(a_T)/2.
\label{sh:T:A}
\end{equation}
If the real part of the amplitude is small then multiple scattering 
correction is negative because of destructive interference of forward 
scattering amplitudes on the upstream nucleons that causes 
\emph{shadowing} of virtual hadron interactions.
It should be also noted that if the real part is large then the 
interference in the double scattering term is constructive that would lead 
to antishadowing effect.

If the coherence length of hadronic fluctuation is small compared to 
average nuclear radius, $L_c\ll R_A$, then the oscillating factor in 
\eq{C2A} suppresses multiple scattering effect. The onset point of 
coherent nuclear effects can be estimated by comparing the coherence 
length of hadronic fluctuation $L_c$ with the averaged distance between 
bound nucleons in the nucleus $r_{\rm NN}$. 
The coherent nuclear effects take place if the coherence length is large 
enough $L_c > r_{\rm NN}$. Since for any mass $m^2$ of intermediate 
hadronic state $L_c < (Mx)^{-1}$ the region of coherent nuclear effects is 
limited to small $x$ for any $Q^2$, $x < (Mr_{\rm NN})^{-1}$. 
Nuclear shadowing saturates if the coherence length $L_c$ exceeds average 
nuclear radius that happens at small $x$ and the condition $L_c\sim R_A$ 
defines the transition region with strong $x$ dependence of the shadowing 
correction. 

The rate of multiple scattering interactions is controlled by mean free 
path of hadronic fluctuation in a nucleus $(\rho_A\sigma)^{-1}$. If this
is small enough compared with nuclear radius, which is the case for 
heavy nuclei, then multiple scattering effects are important.

It can be easily seen from \Eqs{mult:A}{C2A} that if the dependence of
$\mathcal{C}_2^A$ on the scattering amplitude can be neglected, then
\eq{sh:L:D} generalizes to complex nuclei. This corresponds to
the case when the double scattering saturates the multiple scattering
corrections. Generally, for heavy nuclei \eq{sh:L:D} should be replaced by
\begin{equation}
\label{sh:L:A}
\frac{\delta \mathcal{R}_L(x,Q^2,{A/N})}{\delta \mathcal{R}_T(x,Q^2,{A/N})} =
R(x,Q^2) 
        \frac{\Re\left[(i+\alpha_L)^2\mathcal{C}_2^A(a_L)\right]}
        {\Re\left[(i+\alpha_T)^2\mathcal{C}_2^A(a_T)\right]},
\end{equation}
with $a_L$ and $a_T$ the effective scattering amplitudes for 
longitudinally and transversely polarized photons. The relation between 
the nuclear shadowing effect for $F_2$ and $F_T$ in heavy nuclei can be 
derived from \eqs{sh:2:T:D}, (\ref{sh:T:A}) and (\ref{sh:L:A}).

We now discuss the multiple scattering corrections to the right-left
asymmetry in the helicity scattering amplitudes and the generalization
of \eq{sh:3:D} to heavy nuclei. The multiple scattering correction to
the difference ${\Delta}a=a_+ - a_-$, as follows from \Eqs{mult:A}{C2A}, 
can be written as 
\begin{equation}
\label{asym:A}
\delta {\Delta}a^A
  = i\left[a_+^2\mathcal{C}_2^A(a_+)
- a_-^2\mathcal{C}_2^A(a_-)\right],
\end{equation}
where $a_\pm$ are the corresponding nucleon amplitudes. We now use
the fact that $|{\Delta}a|\ll |a_T|$, where $a_T=\tfrac12 (a_+ + a_-)$ 
is the amplitude averaged over the transverse polarizations of the 
intemediate boson, and expand \eq{asym:A} in ${\Delta}a$ keeping 
only the linear term. We have
\begin{equation}
\label{asym:A:2}
\delta {\Delta}a^A = 2i {\Delta}a\, a_T\mathcal{C}_2^A(a_T) - 
                {\Delta}a\, a_T^2\mathcal{C}_3^A(a_T),
\end{equation}
where
\begin{align}\label{C3A}
\begin{split}
& \mathcal{C}_3^A(a) = -i\partial\mathcal{C}_2^A(a)/\partial a =
\\
& 
\int_{z_1<z_2<z_3}\hspace{-3em}
		\ud^2\bm{b}\ud z_1\ud z_2\ud z_3\,
        \rho_A(\bm{b},z_1)\rho_A(\bm{b},z_2)\rho_A(\bm{b},z_3)
\exp\left[
i\int_{z_1}^{z_3}\hspace{-1em}\ud z'\left(a\,\rho_A(\bm{b},z') - k_L 
        \right) \right].
\end{split}
\end{align}
The first term in the right side of \eq{asym:A:2} is similar to that in 
\eq{asym:D}. This term is driven by the double scattering term 
(the quadratic $a^2$ term in multiple scattering series). However, the 
higher order multiple scattering terms also contribute to (\ref{asym:A:2}) 
through nonlinear effects in $\mathcal{C}_2^A$ and $\mathcal{C}_3^A$. Note 
in this respect that the expantion of the term $\mathcal{C}_3^A$ in the 
multiple scattering series starts from the tripple scattering term 
$\rho_A^3$.
The analytical expressions for $\mathcal{C}_2^A$ and $\mathcal{C}_3^A$ 
calculated for uniform density distribution, which is used in our 
analysis described in Sec.\:\ref{sec:fits}, are given in 
Appendix~\ref{apndx:ms}.

Using \eq{asym:A:2} it is straightforward to compute the relative multiple 
scattering correction to the cross section asymmetry $\delta 
\mathcal{R}_{\Delta}({A/N})=\delta \Delta\sigma^A/\Delta\sigma$ that also 
determines the nuclear shadowing effect for the structure function $F_3$. 
The resulting experession for $\delta \mathcal{R}_{\Delta}({A/N})$ is somewhat 
cumbersome in general case and we do not give it explicitly here. It 
should be noted that $\delta \mathcal{R}_{\Delta}({A/N})$ does not depend on 
the cross section asymmetry $\Delta\sigma$ for the nucleon but does depend 
on $\alpha_{\Delta}=\Re {\Delta}a/\Im {\Delta}a$. If we keep only the 
double scattering term then $\delta \mathcal{R}_{\Delta}({A/N})$ is given by 
\eq{sh:3:D} also in the case of complex nuclei. However, this relation is 
violated by higher order multiple scattering terms. 


\section{Description of the model}
\label{sec:model}

In the following we discuss in detail the model which is used to describe
nuclear structure functions. The model incorporates the treatment of both the
coherent and incoherent processes as described in Sec.\:\ref{nuke-sect}.
We use the model nuclear spectral function  calculated
in a many-body approach (Section~\ref{sec:spfn}).
The model pion distribution
function is constrained by light-cone momentum conservation and equations
of motion of pion field (Section~\ref{sec:pion:details}).
The nuclear shadowing effect is described in terms
of effective scattering amplitude of intermediate hadronic states of
virtual boson off the nucleon (see Sec.\:\ref{smallx-nuke}).
In order to describe data on nuclear structure functions we 
explicitly introduce an off-shell correction to the nucleon structure 
functions, which provides a measure of the modification of the nucleon 
structure in the nuclear environment. This effect and the effective 
scattering amplitude are treated phenomenologically in terms of few 
universal parameters, common for all nuclei, which are extracted 
from nuclear DIS data in a wide kinematic range of $x$ and $Q^2$. 
The parameterizations of the off-shell 
effect and of the effective amplitude are discussed in 
Sections~\ref{sec:modbn} and \ref{sec:xsec}. Sections~\ref{sec:data} 
to~\ref{sec:syst} describe the analysis of data
and our main results.

\subsection{Deuteron wave function}
\label{sec:deut}

Nuclear spectral function $\mathcal{P}$ describes the probability to find
the nucleon with the momentum $\bm{p}$ and the (non-relativistic) energy
$\varepsilon$ in the ground state of the nucleus. We first discuss the
\emph{deuteron} for which the spectral function is determined by the wave
function (see \eq{spfn:D}).
The deuteron wave function is the superposition of $s$- and $d$-wave
states. In the momentum space it can be written as follows
\begin{equation}
\label{WF:p}
        \Psi_{D,m}(\bm{p}) = \sqrt{2\pi^2}\,
\left(\psi_0(p) - \psi_2(p)\frac{S_{12}(\widehat{\bm{p}})}{\sqrt8}
\right)
\chi_{1m},
\end{equation}
where $\psi_0$ and $\psi_2$ are respectively the $s$- and $d$-wave
function in the momentum space,%
\footnote{
In terms of the standard wave functions in the coordinate space $u(r)$
and $w(r)$ the functions $\psi_0$ and $\psi_2$ are $\psi_0(p)=(2/\pi)^{1/2}\int\ud
r\,rj_0(rp)u(r)$ and $\psi_2(p)=(2/\pi)^{1/2}\int\ud r\,rj_2(rp)w(r)$,
where $j_{0}$ and $j_{2}$ are the spherical Bessel functions. Note also a
different sign of the $d$-wave term in \eq{WF:p} with respect to the
wave function in the coordinate space.
}
$m$ is the projection of the total
angular momentum on the spin quantization axis, $\chi_{1m}$ is the
spin 1 wave function with $S_z=m$, and $S_{12}(\widehat{\bm{p}})$ is
the tensor operator
\begin{equation}
\label{S12}
S_{12}(\widehat{\bm{p}}) =
        3(\bm{\sigma}_1\cdot\widehat{\bm{p}})
        (\bm{\sigma}_2\cdot\widehat{\bm{p}})
- \bm{\sigma}_1\cdot\bm{\sigma}_2,
\end{equation}
where $\widehat{\bm{p}}=\bm{p}/|\bm{p}|$ and $\bm{\sigma}_1$ and
$\bm{\sigma}_2$ are the Pauli matrices acting on the spin variables of the
bound proton and neutron, respectively. The momentum distribution in the
deuteron is given by the wave function (\ref{WF:p}) squared
\begin{align}
\label{d:MD}
\left|\Psi_{D,m}(\bm{p})\right|^2 &=
2\pi^2\biggl[
\psi_0^2(p)+\psi_2^2(p)
{} -\chi_{1m}^\dagger S_{12}\chi_{1m}
	\left(
	\frac{\psi_0(p)\psi_2(p)}{\sqrt{2}}+\frac{\psi_2^2(p)}4
	\right)
\biggr],
\end{align}
where the last term in the right side appears due to the tensor
operator (\ref{S12}). This term vanishes after averaging over the
deuteron polarizations, which is the case for the present
paper.
The momentum space partial wave functions are normalized according to
\begin{equation}
\label{norm:p}
\int_0^\infty\!\! \ud p\, p^2
\left(\psi_0^2(p)+\psi_2^2(p)\right) = 1.
\end{equation}

In order to study the sensitivity of our result
to the choice of the deuteron wave function we used two different deuteron
wave functions: the one which corresponds to the Bonn potential
\cite{bonn:wf} and the Paris wave function \cite{paris:wf}. These two wave
functions have different high-momentum component and in this respect
represent two extreme situations.

\subsection{Nuclear spectral function}
\label{sec:spfn}

The nuclear spectral function can formally be written as a sum over the 
set of excited residual states. This can be seen directly from \eq{spfn} 
by inserting the complete set of intermediate states. For simplicity we 
suppress the explicit notations for different isospin states and write
\begin{align}\label{spfn:2}
\mathcal{P}(\varepsilon,\bm{p}) &= 2\pi \sum_{n}\left|
\langle(A-1)_n,{-}\bm{p}\left|\psi(\bm{p}) \right|A\rangle\right|^2
\delta\left(
\varepsilon + E_n^{A{-}1} + E_R - E_0^A\right) .
\end{align}
Here the sum runs over the quantum numbers of the states of $A{-}1$
nucleons, which include the bound states as well as the continuum states,
$E_n^{A{-}1}$ and $E_0^A$ are respectively the energy of the residual
nucleus (in the recoil nucleus rest frame) and the ground state energy of the
target nucleus. The residual system balances momentum of the removed nucleon
and acquires the recoil energy $E_R=\bm{p}^2/2M_{A{-}1}$.

The nuclear spectral function determines the rate of nucleon removal
reactions such as $(e,e'p)$ that makes it possible to
extract the spectral function from experimental data.
For low separation energies (for $|\ceps|<\ceps_F\sim 30-50\Mev$) the
experimentally observed spectrum is similar to that predicted by the
mean-field model \cite{Frullani:1984nn}.
The mean-field model spectral function $\PMF$ is given by the wave 
functions and energies of the occupied levels in the mean field 
\cite{deshalit}.
The mean-field picture gives a good approximation to experimentally 
observed spectrum in $(e,e'p)$ reactions in the vicinity of the Fermi 
level, where the excitation energies of the residual nucleus are small 
\cite{Frullani:1984nn}.
As nuclear
excitation energy becomes higher the mean-field model becomes 
less accurate. The peaks corresponding to the single-particle levels
acquire a finite width (fragmentation of deep-hole states).
Furthermore, the high-energy and high-momentum
components of nuclear spectrum can not be described in the mean-field
model and driven by correlation effects in nuclear ground state as
witnessed by numerous studies (for a review see \cite{Muther:2000qx}).
We denote the contribution to the spectral function which absorbs the
correlation effects as $\Pcor(\ceps,\bm{p})$.

In this paper we consider a phenomenological model of the spectral
function which incorporates both the single particle nature of the
spectrum at low energy and high-energy and high-momentum components
due to NN-correlations in the ground state. We first discuss the isoscalar
spectral function which we write as
\begin{equation}
\label{model:P0}
\mathcal{P}_0(\varepsilon,\bm{p})=\PMF(\varepsilon,\bm{p})
+\Pcor(\varepsilon,\bm{p}).
\end{equation}
The low-energy part is described by the mean-field spectral function
for which we use an approximate expression motivated by
closure, \ie the sum over occupied levels is substituted by its
average value:
\begin{equation}
\label{model:PMF}
\PMF(\varepsilon,\bm{p})=2\pi\, n_{\rm MF}(\bm{p})
\delta\left(\varepsilon+E^{(1)}+E_R(\bm{p})\right),
\end{equation}
with $E_R(\bm{p})$ the recoil energy of residual
nucleus, $E^{(1)}=E^{A{-}1}-E_0^A$ the nucleon separation energy averaged
over single-particle levels, and $n_{\rm MF}(\bm{p})$ the corresponding part
of the nucleon momentum distribution.
Note that $\PMF$ is not normalized to the number of nucleons since
a part of the strength is taken by the correlation term $\Pcor$.  

The correlated part of the spectral function $\Pcor$ is determined by
excited states in (\ref{spfn:2}) with one or more nucleons in the
continuum. Following \cite{CS96} we assume that $\Pcor$ at high
momentum and high separation energy is dominated by ground state
configurations with a correlated nucleon-nucleon pair and remaining
$A{-}2$ nucleons moving with low center-of-mass momentum
\begin{equation}
|A{-}1,-\bm{p}\rangle \approx
        \psi^\dagger(\bm{p}_1)|(A{-}2)^*,\bm{p}_2\rangle
        \delta(\bm{p}_1+\bm{p}_2+\bm{p}) .
\end{equation}
The corresponding matrix element in \eq{spfn:2} can thus be parametrized 
in terms of the wave function of the nucleon-nucleon pair embeded into 
nuclear environment. We assume factorization into relative and 
center-of-mass motion of the pair \cite{CS96}
\begin{equation}
\label{model:2NC}
\left\langle (A{-}2)^*,\bm{p}_2\left|
        \psi(\bm{p}_1)\psi(\bm{p})
\right|A\right\rangle \approx C_2
        \psi_{\rm rel}(\bm{k}) \psi_{\rm CM}^{A{-}2}(\bm{p}_\mathrm{CM})
        \delta(\bm{p}_1+\bm{p}_2+\bm{p}),
\end{equation}
where $\psi_{\rm rel}$ is the wave function of the relative
motion in the nucleon-nucleon pair with relative momentum
$\bm{k}=(\bm{p}-\bm{p}_1)/2$ and $\psi_{\rm CM}$ is the wave function
of center-of-mass (CM) motion of the pair in the field of $A{-}2$
nucleons, $\bm{p}_{\rm CM}=\bm{p}_1+\bm{p}$. The CM wave function
$\psi_{\rm CM}$ generally depends on the quantum numbers of the state
of $A{-}2$ nucleons, however the corresponding dependence of the
$\psi_{\rm rel}$ is assumed to be weak. Both the wave functions,
$\psi_{\rm rel}$ and $\psi_{\rm CM}$, are assumed to be normalized to
unity. The normalization factor $C_2$ describes the
weight of the two-nucleon correlated part in the full spectral function.

Using Eq.(\ref{model:2NC}) we sum over the spectrum of states of
$A{-}2$ nucleons and obtain an approximate expression for $\Pcor$ in 
terms of convolution of the relative and the CM momentum 
distributions
\begin{align}
\label{model:Pcor}
\Pcor(\varepsilon,\bm{p}) =& 2\pi C^2_2 \int \ud^3\bm{p}_1
\ud^3\bm{p}_2 n_{\rm rel}(\bm{k}) n_{\rm CM}(\bm{p}_2)
\delta(\bm{p}_1+\bm{p}_2+\bm{p}) \notag \\
        & \delta\left(\varepsilon+\frac{\bm{p}_1^2}{2M}+
        \frac{\bm{p}_2^2}{2M_{A{-}2}}+E^{(2)}\right).
\end{align}
Here $n_{\rm rel}$ and $n_{\rm CM}$ are the relative and the CM
momentum distributions, respectively, and $E^{(2)}=E^{A{-}2}-E_0^A$ is
the energy needed to separate two nucleons from the ground state
averaged over configurations of $A{-}2$ nucleons with low excitation
energy.  Note that the minimum two-nucleon separation energy
$E^{(2)}=E_0^{A{-}2}-E_0^A$ is of order $20\,$MeV for medium-range
nuclei like $^{56}$Fe.

We can further simplify \eq{model:Pcor} if the momentum $p$ is high enough and
$|\bm{p}|\gg |\bm{p}_2|$. This allows us to take the relative momentum
distribution out of the integral over the CM momentum at the point
$\bm{k}=\bm{p}$. Then we have
\begin{equation}
\label{model:Pcor:2}
\Pcor(\varepsilon,\bm{p}) = 2\pi C^2_2 n_{\rm rel}(\bm{p})
\left\langle \delta\left(\ceps+\frac{\bm{p}^2}{2M}
                +\frac{\bm{p}\cdot\bm{p}_2}{M}
        +\frac{\bm{p}_2^2}{2M_*}+E^{(2)}\right) \right\rangle_{\!\mathrm{CM}},
\end{equation}
where the averaging is done with respect to the CM motion of the correlated
pair and $M_*=M(A-2)/(A-1)$ is effective mass of the system of the residual
nucleus of $A-2$ nucleons and the nucleon with momentum $\bm{p}_1$. In this
approximation the high momentum part of nuclear momentum distribution is
determined by relative momentum distribution in the correlated 
nucleon-nucleon pair embedded into nuclear environment, 
$n_{\rm cor}(\bm{p})=C^2_2 n_{\rm rel}(\bm{p})$.
The energy spectrum predicted by \eq{model:Pcor:2} is a broad peak with 
the maximum at $\ceps\sim -\bm{p}^2/2M$ and effective width
$|\bm{p}| \bar p_{\rm CM}/M$ with $\bar p_{\rm CM}$ an average CM momentum. 

We perform the averaging over the CM motion of the NN pair in the field of 
other $A{-}2$ nucleons assuming that the CM momentum distribution is given 
by a Gaussian
\begin{equation}
\label{nCM}
n_{\mathrm{CM}}(\bm{p}_{\mathrm{CM}})=
        \left({\alpha}/{\pi}\right)^{3/2}
        \exp(-\alpha \bm{p}_{\mathrm{CM}}^2).
\end{equation}
The parameter $\alpha$ is related to the averaged CM momentum of the 
nucleon-nucleon pair, 
$\alpha=\frac32\langle\bm{p}^2_{\mathrm{CM}}\rangle^{-1}$.
The latter can be estimated from the balance of the overall nucleus momentum
\cite{CS96}, $\langle(\sum \bm{p}_i)^2\rangle=0$, where the sum is taken
over all bound nucleons and the expectation value is performed with respect
to the intrinsic wave function of the nucleus. This gives
$\langle\bm{p}^2_{\mathrm{CM}}\rangle= 2\langle\bm{p}^2\rangle 
(A{-}2)/(A{-}1)$, with $\langle\bm{p}^2\rangle$ the mean value of the 
squared single nucleon momentum. We consider configurations in which 
characteristic CM momenta are small. For this reason we should also 
exclude the high-momentum part in estimating $\langle\bm{p}^2\rangle$ and 
we will assume that this quantity is given by averaging over mean-field 
configurations.

Using Eq.(\ref{nCM}) we integrate over the CM momentum
in (\ref{model:Pcor:2}) and the result
reads,
\begin{equation}
\label{model:Pcor:3}
\Pcor(\varepsilon,\bm{p})=n_{\rm cor}(\bm{p})\frac{2M}{p}
\sqrt{\alpha\pi}
\left[\exp({-\alpha p_{\rm min}^2})-\exp({-\alpha p_{\rm max}^2})\right],
\end{equation}
where $p=|\bm{p}|$, $p_{\rm min}$ and $p_{\rm max}$ are respectively the
minimum and the maximum CM momenta allowed by the energy-momentum
conservation in Eq.(\ref{model:Pcor}) for the given $\varepsilon$ and
$\bm{p}$,
\begin{subequations}\label{model:limits}
\begin{align}
p_{\rm max}&={M_* p}/{M} + p_T,
\\
p_{\rm min}&=\left|{M_* p}/{M} - p_T \right|,
\end{align}
\end{subequations}
where $p_T=\left(2M_*(|\ceps|-E_{\rm th})\right)^{1/2}$ and
$E_{\rm th}=E^{(2)}+E_R(\bm{p})$. The latter is the threshold value of the nucleon
separation energy for discussed configurations. Note that in our notations
$\ceps<0$.
We also remark that $p_T$ has the meaning of the maximum CM
momentum in the correlated NN-pair in the direction transverse to $\bm{p}$
for the given $\varepsilon$ and $p$ \cite{KS00}.


In numerical evaluations we use the parameterizations for
$n_{\rm MF}(A,\bm{p})$ and $n_{\rm cor}(A,\bm{p})$ of \cite{CS96} which fit
nicely the results of many-body calculation of nuclear momentum
distribution. It follows from this calculation that low-momentum part
incorporates about 80\% of the total normalization of the spectral function
while the other 20\% are taken by the high-momentum part.
The momentum distributions are presented in \cite{CS96} for a limited
range of nuclei. In order to evaluate the momentum distributions for
other values of the nucleus mass number $A$, we interpolate
the values of the momentum distributions for each value of
momentum $|\bm{p}|$.
For the parameter $E^{(2)}$ we take the two-nucleon separation energy,
\ie  the difference $E_0^{A{-}2}-E_0^A$ between the ground state
energies (note that $E^{(2)}>0$).
The remaining parameter $E^{(1)}$ of $\PMF$ is fixed using the Koltun sum
rule \cite{KSR}, which is exact relation for nonrelativistic systems with 
two-body forces
\begin{equation}
\average{\ceps}+\average{T}=2\ceps_B,
\label{KoltunSR}
\end{equation}
where $\ceps_B=E_0^A/A$ is nuclear binding energy per bound 
nucleon and $\average{\ceps}$ and $\average{T}$ are the nucleon separation 
and kinetic energies averaged with the full spectral function
\begin{subequations}
\begin{align}
\label{av:eps}
\average{\ceps} &=A^{-1}\int[\ud p]\mathcal{P}(\ceps,\bm{p})\ceps,
\\
\average{T} &=A^{-1}\int[\ud p]\mathcal{P}(\ceps,\bm{p})\frac{\bm{p}^2}{2M}.
\label{av:T}
\end{align}
\end{subequations}

The function $\mathcal{P}_1$ describes the isovector component in a 
nucleus (see \eqs{spfn:01}).
In calculating $\mathcal{P}_1$ we assume 
that the correlation part of the spectral function $\Pcor$ is mainly 
isoscalar and cancels out in the $p-n$ difference. Then 
$\mathcal{P}^{p-n}$ is determined by the difference of the mean-field 
configurations for protons and neutrons. If we further neglect small 
differences between the energy levels of protons and neutrons then 
$\mathcal{P}^{p-n}$ is determined by the difference in the occupation 
numbers of single-particle levels for protons and neutrons. 
In a complex nucleus the deep levels are usually occupied and their 
contribution cancel out in $\mathcal{P}_1$. The Fermi level has a large 
degeneracy factor and the occupation numbers for protons and neutrons are 
different. We then conclude that the difference $\mathcal{P}^{p-n}$ is 
saturated by the Fermi level and
\begin{equation}
\label{spfn_1}
\mathcal{P}_1=|\phi_{F}(\bm{p})|^2\delta(\ceps-\ceps_F), 
\end{equation}
where $\ceps_F$ and $\phi_F$ are the energy and the wave function of the 
Fermi level (we have neglected the recoil of the $A{-}1$ nucleus).%

The isovector correction is usually relevant for heavy-weight nuclei in 
which there exists a considerable neutron excess over protons. For such 
nuclei the Fermi gas model is supposed to be a reasonable approximation 
and we use this model in numerical applications. In this model 
$|\phi_F(p)|^2 \propto \delta(p_F-p)$, where $p_F$ is the Fermi momentum 
which is determined by average nucleon density $\rho =4p_F^2/(6\pi^2)$. 
The normalization coefficient can be found from condition (\ref{norma}), 
according to which $\mathcal{P}_1$ is normalized to unity. As a result we 
have 
\begin{equation}
\mathcal{P}_1^{\mathrm{FG}} = \delta(p-p_F)\delta(\ceps-\ceps_F)/ 
                             (4\pi p_F^2). 
\label{model:P1}
\end{equation}

\subsection{Nuclear pion distribution function}
\label{sec:pion:details}

In calculating the pion effect in nuclear structure functions
the relevant quantity is the distribution of
pion excess in a nucleus since the nucleon pion cloud effect
is taken into account in the nucleon structure functions.
The inspection of Eqs.(\ref{pion:conv}) and (\ref{pion:f}) suggests that
the pion correction is located at small $x<p_F/M$, which is also
confirmed by model calculations. In this region a good approximation is
to neglect $x^2/Q^2$ terms in \eqs{SF:pion} as well as target mass
corrections to structure functions. We also assume no off-shell dependence
of pion structure function and base our discussion on convolution
approximation by \Eqs{pion:conv}{pion:f}.

Before doing model calculations it is important to realize that the pion
distribution function is constrained by a number of sum rules. The first
moment of (\ref{pion:f}) gives an average pion ligh-cone momentum
\begin{align}
\label{y:pi}
\average{y}_\pi &=\int\ud y\,yf_{\pi/A}(y)=\average{\theta_{++}^\pi}/M,
\end{align}
where $\theta^\pi_{++}=\left(\partial_0\varphi\right)^2 +
\left(\partial_z\varphi\right)^2$ is the light-cone component of pion
energy-momentum tensor. The averaging in \eq{y:pi} means
$\average{\mathcal{O}}=\int\ud^3\bm{r}\bra{A}\mathcal{O}(\bm{r})\ket{A}/
                        \left\langle A|A\right\rangle$
for any operator $\mathcal{O}$.
It is also useful to consider the average $y^{-1}$ which is proportional 
to $\varphi^2$ averaged over nuclear ground state:
\begin{equation}
\label{invy:pi}
\average{y^{-1}}_\pi =\int{\ud y}\,y^{-1}f_{\pi/A}(y)
                           =M\average{\varphi^2}.
\end{equation}

The pion and nucleon fractions of nuclear light-cone momentum are related
by the momentum balance equation
\begin{equation}
\average{y}_\pi + \average{y}_N =\frac{M_A}{AM}.
\label{balance:eq}
\end{equation}
Equation (\ref{balance:eq}), although being intuitively obvious, can
formally be derived in a meson-nucleon field-theoretic model of nuclear
Hamiltonian \cite{Ku89}.
Several constraints on nuclear pion distribution
$\mathcal{D}_{\pi/A}(k)$ can be obtained in this model using the equations
of motion for pion and nucleon operators and energy-momentum conservation
condition. In particular, for a model nuclear Hamiltonian with nucleons and
pions with pseudo-scalar interaction we obtain the following
relations \cite{Ku89}
\begin{subequations}
\begin{align}
\label{av:phi}
m_\pi^2 \average{\varphi^2} &= \ceps_B + \average{T},
\\
\label{av:phidot}
\average{\left(\partial_0\varphi\right)^2} &=
                \ceps_B-
                 \tfrac12\left(\average{\ceps} + \average{T}\right),
\\
\label{av:gradphi}
\average{\left(\nabla\varphi\right)^2} &=
                -\tfrac32\average{\ceps}
                 -\tfrac12\average{T}.
\end{align}
\end{subequations}
A few comments are in order. Pion field in nuclei is mainly generated by
nucleon sources. Time variation of the pion field describes retardation
effects in the nucleon--nucleon interaction. In a nonrelativistic system
this effect is small since typical energy variations are small compared to
the pion mass. We, therefore, take the static approximation
$\partial_0\varphi=0$. Then \eq{av:phidot} is equivalent
to the Koltun sum rule (\ref{KoltunSR}). In the static approximation for 
the pion energy-momentum tensor we have 
$\average{\theta_{++}^\pi}=\frac13\average{(\nabla\varphi)^2}$. Then using 
\eq{y:N} we conclude that \Eqs{av:gradphi}{balance:eq} are equivalent. For 
this reason only \eq{av:phi} gives independent constraint.

We use the constraints on average pion light-cone momentum $y$ and
$1/y$ which follow from \Eqs{av:phi}{balance:eq} in
order to evaluate the pion contribution to nuclear structure functions.
It should be remarked that in this approach by using momentum balance
equation (\ref{balance:eq}) we effectively take into account the
contributions from all mesons.
In order to quantitatively evaluate the pion effect in the structure functions
we use a model distribution
\begin{equation}
f_{\pi/A}(y) =C\,y(1-y)^n,
\label{fpi:model}
\end{equation}
which is motivated by the asymptotics of pion distribution function at small and
large $y$. The normalization constant $C$ and the exponent $n$
are fixed from Eqs.(\ref{balance:eq}) and (\ref{invy:pi})
using \Eqs{y:N}{av:phi}. The nucleon average separation and kinetic
energies are calculated with the spectral function described in Sec.\:\ref{sec:spfn}.


\subsection{Parameterization of off-shell effects}
\label{sec:modbn}

The off-shell effect in the structure function $F_2$ is described by \eq{delf}.
In the analysis of data, described in detail in Sec.\:\ref{sec:fits}, we 
consider a phenomenological model of the off-shell function $\delta 
f_2(x,Q^2)$. In order to choose an appropriate model we first note that 
function (\ref{delf}) describes the relative off-shell effect on the LT 
structure function and we expect that $Q^2$-independent $\delta f_2$ is a 
good approximation. We also note that off-shell effects are constrained by 
the normalization of nuclear valence quark distribution (see 
Sec.\:\ref{sec:valnorm}).
For this reason we anticipate that $\delta f_2(x)$ should have at least 
one zero. Moreover, the analysis of nuclear pion correction 
as discussed in Sec.\:\ref{sec:pion:details} suggest that $\delta f_2(x)$ 
can have two zeros. These motivate us to choose the following 
simple parameterization for the off-shell function:
\begin{equation}
\label{eq:deltaf}
\delta f_2(x) = C_N(x-x_1)(x-x_0)(h-x),
\end{equation}
where $C_{N}$ is an overall normalization constant and $0<x_1<x_0<1$ and 
$h>1$. The analysis of data indicates that the parameters $h$ and $x_0$ 
are fully correlated and suggests $h=1+x_0$. After imposing such condition 
then expression (\ref{eq:deltaf}) has only three independent 
parameters. We use this model to describe off-shell effects in the 
analysis of Sec.\:\ref{sec:fitres}.

\subsection{Effective scattering amplitude}
\label{sec:xsec}

As discussed in Sec.\:\ref{smallx-nuke},
the coherent multiple-scattering nuclear effects are determined by 
effective (averaged over hadronic configurations of the
intermediate boson) scattering amplidudes $a_h$ for different helicities 
$h=\pm 1,0$.
The amplitudes $a_h=\sigmabar_h(i+\alpha_h)/2$ are parametrized in terms 
of effective cross section $\sigmabar$ and the Re/Im ratio $\alpha$. For 
the unpolarized case, which is considered in this paper, the relevant 
amplitudes are the average transverse $a_T=(a_+ + a_-)/2$ and the 
longitudinal $a_L$ amplitudes.
One can qualitatively expect that $\sigmabar_T$ descreases with $Q^{2}$ 
since the relative weight of higher mass states increases with $Q^2$ and 
the cross-section decreases with the meson mass (see 
Sec.\:\ref{smallx-nuke}). In order to parametrize effective transverse 
cross section we use then the following expression
\begin{equation}
\label{eq:effxsec}
\sigmaeff = \sigma_{1} +
\frac{\sigma_{0}-\sigma_{1}}{\left(1 + Q^{2}/Q^{2}_{0} \right)}.
\end{equation}
The parameter $\sigma_{0}$ describes the cross section at small $Q^{2}$,
while $\sigma_{1}$ corresponds to high-$Q^{2}$ regime.
The choice of both these parameters will be discussed in detail in
Sec.\:\ref{sec:fixpars}. The free parameter $Q^{2}_{0}$ describes the
transition between low- and high-$Q^2$ regions. 
We note that in the discussed approach we only
consider relative corrections to the effective cross-section and for this
reason the analysis is not very sensitive to the detailed modelling of such
cross-section. 
The presence of non-zero real part of the amplitude is required by both
theoretical arguments and phenomenology. The choice of $\alpha_{T}$ in
our analysis will be discussed in Sec.\:\ref{sec:fixpars}.

In order to fix the effective amplitude in the longitudinal channel we use 
the relation $a_L/a_T=R=F_L/F_T$ with $R$ calculated using the PDFs and 
the structure functions of Ref.~\cite{a02} as discussed in 
Sec.\:\ref{nucleon} (see also Sec.\:\ref{smallx-nuke}).

The $C$-odd asymmetry $\Delta a=a_+ - a_-$ in the scattering amplitude of left- and 
right-polarized virtual boson does not affect the structure functions 
$F_1$ and $F_2$. However, $\Delta a$ is relevant for $F_3$ and affects the 
normalization of nuclear valence number as described in 
Sec.\:\ref{sec:valnorm}. It should be noted that the relative nuclear 
shadowing correction to $F_3$ does not depend on the cross-section 
asymmetry $\Delta\sigma$ but does depend on $\alpha_{\Delta}=\Re \Delta 
a/\Im \Delta a$ as explained in Sec.\:\ref{sec:smallx:A}. 
In order to fix $\alpha_{\Delta}$ we use the approach based on Regge 
phenomenology of high-energy hadronic amplitudes and approximate $\Delta 
a$ by the $\omega$-reggeon pole, a simple proper contribution to the 
$C$-odd amplitude. The energy dependence and Re/Im ratio of the Regge pole 
is fully determined by its intercept which is about 0.5 that leads to 
$\alpha_{\Delta}=1$ \cite{Collins:1977jy}. We use this value in the 
calculation of nuclear shadowing correction to the valence quark 
distribution in Sec.\:\ref{sec:valnorm}.

\subsection{Nuclear data}
\label{sec:data}

Table~\ref{tab:data} summarizes the list of experimental data used in
this paper. They include both muon (EMC, NMC, BCDMS, FNAL E665) and
electron (SLAC E139, E140) scattering on a variety of targets: p, D,
${}^4$He, ${}^7$Li, ${}^{9}$Be, ${}^{12}$C, ${}^{27}$Al, ${}^{40}$Ca,
${}^{56}$Fe, ${}^{63}$Cu, ${}^{108}$Ag, ${}^{119}$Sn, ${}^{197}$Au,
${}^{207}$Pb.  For each target and kinematic region, we select the
most precise and recent data and we do not use earlier results
characterized by larger uncertainties, since their contribution to the
present analysis would be negligible.%
\protect\footnote{%
Note also that the addition
of unnecessary data points with large uncertainties can produce an
artificial reduction of the $\chi^{2}$ of fits.}
Most of the data come from NMC for the small $x$ region and SLAC E139 for 
the region $x>0.1$.

We note that, since all available nuclear data are provided by
fixed-target experiments, there is always an implicit correlation
between $x$ and $Q^2$ in data points. Usually low-$x$ regions also
correspond to low-$Q^2$ values. As described in the following, this
reduces the possibility to test the $Q^2$ dependence of the model in
a complete way.
%
\begin{table}[htb]
\begin{center}
\small{
\begin{tabular}{l|c|c|c|c}
Experiment & Targets & \# of points & \multicolumn{1}{c|}{$x$ range} & \multicolumn{1}{c}{$Q^2$ range $[\gevsq]$} \\ \hline\hline 
NMC~\cite{NMCdp} & D/p  & 12  & $7.0\times 10^{-3}\div 0.70$  & $4.0$  \\ \hline 
NMC~\cite{NMChe} & ${}^4$He/D  & 18  & $3.5\times 10^{-3}\div 0.65$  & $0.77 \div 44.0$  \\ \hline 
NMC~\cite{NMClic} & ${}^7{\rm Li}$/D  & 24  & $1.4\times 10^{-4}\div 0.65$  & $0.034 \div 39.0$  \\ \hline 
NMC~\cite{NMCpb} & ${}^{9}{\rm Be}/{}^{12}$C  & 15  & $1.25\times 10^{-2}\div 0.70$  & $3.4 \div 66.7$  \\ \hline 
NMC~\cite{NMClic} & ${}^{12}$C/D & 24  & $1.5\times 10^{-4}\div 0.65$  & $0.035 \div 41.0$  \\ \hline 
NMC~\cite{NMChe} & ${}^{12}$C/${}^7{\rm Li}$  & 25  & $8.5\times 10^{-3}\div 0.60$  & $0.8 \div 17.0$  \\ \hline 
NMC~\cite{NMCpb} & ${}^{27}$Al/C  & 15  & $1.25\times 10^{-2}\div 0.70$  & $3.4 \div 63.9$  \\ \hline 
NMC~\cite{NMChe} & ${}^{40}$Ca/D  & 18  & $3.5\times 10^{-3}\div 0.65$  & $0.6 \div 41.0$  \\ \hline 
NMC~\cite{NMChe} & ${}^{40}$Ca/${}^7{\rm Li}$  & 25  & $8.5\times 10^{-3}\div 0.60$  & $0.8 \div 17.0$  \\ \hline 
NMC~\cite{NMChe} & ${}^{40}$Ca/${}^{12}$C  & 25  & $8.5\times 10^{-3}\div 0.60$  & $0.8 \div 17.0$  \\ \hline 
NMC~\cite{NMCpb} & ${}^{56}{\rm Fe}$/${}^{12}$C  & 15  & $1.25\times 10^{-2}\div 0.70$  & $3.4 \div 66.6$  \\ \hline 
EMC~\cite{EMC} & ${}^{63}{\rm Cu}$/D  & 10  & $1.5\times 10^{-2}\div 0.61$  & $3.3 \div 46.4$  \\ \hline 
NMC~\cite{NMCq2} & ${}^{119}{\rm Sn}$/${}^{12}$C  & 161  & $1.25\times 10^{-2}\div 0.70$  & $1.3 \div 110.0$  \\ \hline 
NMC~\cite{NMCpb} & ${}^{207}{\rm Pb}$/${}^{12}$C  & 15  & $1.25\times 10^{-2}\div 0.70$  & $3.4 \div 66.1$  \\ \hline \hline 
E139~\cite{Gomez} & ${}^{4}{\rm He}$/D  & 21  & $0.125\div 0.88$  & $2.0 \div 10.0$  \\ \hline 
E139~\cite{Gomez} & ${}^{9}{\rm Be}$/D  & 21  & $0.125\div 0.88$  & $2.0 \div 10.0$  \\ \hline 
E139~\cite{Gomez} & ${}^{12}{\rm C}$/D  & 17  & $0.205\div 0.80$  & $3.5 \div 10.0$  \\ \hline 
E139~\cite{Gomez} & ${}^{27}{\rm Al}$/D  & 21  & $0.125\div 0.88$  & $2.0 \div 10.0$  \\ \hline 
E139~\cite{Gomez} & ${}^{40}{\rm Ca}$/D  & 17  & $0.205\div 0.80$  & $3.5 \div 10.0$  \\ \hline 
E139~\cite{Gomez} & ${}^{56}{\rm Fe}$/D  & 23  & $0.084\div 0.88$  & $2.0 \div 10.0$  \\ \hline 
E139~\cite{Gomez} & ${}^{108}{\rm Ag}$/D  & 17  & $0.205\div 0.80$  & $3.5 \div 10.0$  \\ \hline 
E139~\cite{Gomez} & ${}^{197}{\rm Au}$/D  & 22  & $0.125\div 0.88$  & $2.0 \div 10.0$  \\ \hline \hline  
E140~\cite{E140} & ${}^{56}{\rm Fe}$/D  & 8 & $0.200\div 0.50$  & $1.0 \div 5.0$  \\ \hline  
E140~\cite{E140} & ${}^{197}{\rm Au}$/D  & 1  & $0.200$  & $1.0$  \\ \hline \hline  
BCDMS~\cite{BCDMS} & ${}^{56}{\rm Fe}$/D  & 10  & $0.07\div 0.65$  & $17.0 \div 113.0$  \\ \hline \hline 
E665~\cite{E665dp} & D/p  & 21  & $2.0\times 10^{-5}\div 0.25$  & $0.005\div 35.6.0$  \\ \hline 
E665~\cite{E665} & ${}^{207}{\rm Ca}$/${}^{12}{\rm C}$  & 10  & $1.2\times 10^{-4}\div 0.027$  & $0.15\div 7.9$  \\ \hline 
E665~\cite{E665} & ${}^{207}{\rm Pb}$/D  & 10  & $1.2\times 10^{-4}\div 0.027$  & $0.15\div 7.9$  \\ \hline 
E665~\cite{E665} & ${}^{207}{\rm Pb}$/${}^{12}{\rm C}$  & 10  & $1.2\times 10^{-4}\div 0.027$  & $0.15\div 7.9$  \\ \hline 
\end{tabular}
}
\caption{%
List of nuclear data used in the analysis. See text for details.
\label{tab:data}
}
\end{center}
\end{table}

\subsection{Extraction of parameters}
\label{sec:fits}

The numerical values of the parameters in the model are determined from
the data listed in Table~\ref{tab:data}%
\protect\footnote{%
We note that D$/p$ data were not used in our fits. 
We compare our predictions with these data in Sec.\:\ref{sec:deuterium} }
with two main steps. Initially, we verify the 
consistency of our model with $F_{2}$ data from charged-lepton scattering, 
without imposing specific constraints. We then discuss in detail the 
deconvolution of different physical effects which contribute to the 
overall nuclear modification of the structure functions.

It must be noted that the extraction of parameters responsible for nuclear 
effects is correlated with the determination of PDFs 
(see Sec.\:\ref{phenom-sect}), 
which requires both the proton and deuterium data to obtain the 
distributions of $d$ and $u$ quarks. Nuclear effects can produce 
significant distortions on parton densities and will be the subject of a 
future publication. 
In principle, using our approach it would be possible to extract 
simultaneously the proton PDFs and the parameters responsible for nuclear 
effects (such as off-shell correction and effective cross section) by 
applying QCD analysis to the extended set of data including nuclear data. 
However, in order to limit correlations we follow a different approach.

The parameters of the model are extracted only from the measured ratios
$\mathcal R_{2}({A'/A})= F_{2}^{A'}/F_{2}^{A}$, where $A'$ and $A$ are
two different nuclei (usually the denominator corresponds to
deuterium). The description of the nucleon structure functions largely
cancels in the ratios, thus effectively removing the correlation with
PDFs. In order to verify this we applied
an iterative procedure. We first extracted the parameters using PDFs
obtained without our nuclear corrections. Then we repeated the procedure after
updating the PDF extraction using the information on nuclear effects in
deuterium from the previous step (Section~\ref{phenom-sect}).  
Results indicated that the fitted parameters were stable, demonstrating the
absence of strong correlations.

Nuclear data are usually available in bins of $x$ ($\Delta x$), while
only the average $\bar{Q}^2$ in each bin is provided. We perform a 
fit to the experimental data with MINUIT~\cite{MINUIT} by minimizing 
$\chi^2 = \sum (\mathcal{R}^{\rm exp}_{2} - \mathcal{R}^{\rm th}_{2})^2 / 
\sigma^{2}(\mathcal{R}^{\rm exp}_{2})$, where $\sigma^{2}(\mathcal{R}^{\rm 
exp}_{2})$ represents the uncertainty on the measurements and the sum 
includes all data points.
For each experimental point, the model is evaluated at the given 
average $\bar{Q}^2$ and integrated over the size of the $x$ bin:%
\footnote{In a few cases, in which the
explicit $Q^2$ dependence is provided, the model is averaged over the
corresponding $Q^2$ bins.}
\begin{eqnarray}
\mathcal{R}^{\rm th}_{2}(x,\bar{Q}^2,A'/A) =
\frac{\int_{x - \Delta x/2}^{x + \Delta x/2}
  F_{2}^{A'}(x^{\prime},\bar{Q}^{2}) \ud x^{\prime}}
{\int_{x - \Delta x/2}^{x + \Delta x/2}
F_{2}^{A}(x^{\prime},\bar{Q}^{2}) \ud x^{\prime}} .
\end{eqnarray}
Both the normalization and point-to-point uncorrelated uncertainties, as
published by experiments, are taken into account.
We would like to emphasize that
the lack of knowledge of the experimental $Q^2$ distribution
in the $x$ bins can potentially result in a mismatch between
data and predictions in the regions where a significant
$Q^2$ dependence is present. As discussed in the following,
this increases the systematic uncertainties of the calculation 
from the measured parameters at $x>0.70$ and $x<0.05$. 

As explained in Sec.\:\ref{phenom-sect}, we use a phenomenological
extrapolation of free nucleon structure functions
for $Q^2<1.0$ GeV$^2$ and, in general, nuclear corrections
to structure functions can be calculated at low $Q^2$.
However, we restrict the fits to extract the free parameters
of our model to the data with $Q^2 \geq 1.0$ GeV$^2$ in order to
reduce systematic uncertainties on the parameters. We then validate
our predictions against the data points with $Q^2<1.0$ GeV$^2$, which
are included in all comparisons shown in the following.

\subsubsection{Choice of fixed parameters}
\label{sec:fixpars}

We start our fits by treating $\sigma_{0}$ and $\sigma_{1}$, the asymptotic
values of the effective transverse cross-section in \eq{eq:effxsec},
and the real part of the effective scattering amplitude
$\alpha_{T}=\Re a_{T}/\Im a_{T}$ (Section~\ref{sec:smallx:D})
as additional free parameters. This procedure allows a preliminary
estimate of their correlation with the remaining parameters and
a consistency check with the expected values.

The best fit value obtained for $\sigma_{1}$ is consistent with zero.
By setting $\sigma_{1} \neq 0$ we can still obtain an acceptable description 
of data provided $\sigma_{1}<1.0$ mb (at 90\%CL), due to the (anti)correlation 
of $\sigma_{1}$ with $Q^2_{0}$ in \eq{eq:effxsec}. After verifying that the 
correlation between $\sigma_{1}$ and the off-shell parameters in 
\eq{eq:deltaf} is negligible, we then fix $\sigma_{1}=0$ in all our fits.

We note that a non-vanishing shadowing correction at large $Q^2$
affects the normalization of the valence quark
number per nucleon (Section~\ref{sec:valnorm}). In this respect
we have two possible constraints at large $Q^2$.
The first condition is to require the conservation of the overall valence 
number in nuclei through a balance between the shadowing and the off-shell 
corrections. 
As a second choice, it is also possible to explicitly impose the off-shell 
effect to conserve the valence quark number of the off-shell nucleon. This 
implies that both the off-shell and the shadowing effects conserve 
independently the normalization of valence quark distribution. 
In our approach, initially we do not assume any specific normalization 
constraint. Instead, we verify a posteriori the magnitude of the 
renormalization introduced by the off-shell effect 
(Section~\ref{sec:valnorm}) and its balance with the shadowing correction. 
We then use the normalization condition for nuclear valence number to 
further bound some of the parameters. This procedure will be discussed in 
more detail in Sections~\ref{sec:fitres} and~\ref{sec:valnorm}.

The fits to DIS data on nuclear targets show a strong (positive)
correlation between $\sigma_{0}$ and $\alpha_{T}$. In addition,
the value of $\sigma_{0}$ is also correlated with $Q^2_{0}$ so that it is
not possible to unambiguously disentangle the three parameters from the fits.
If we fix $\alpha_{T}=0$ we obtain $\sigma_{0}=36$ mb from data. However, data
clearly prefer $\alpha_{T} \neq 0$, with a somewhat lower value of $\sigma_{0}$.
The best fit solution corresponds to $\alpha_{T} = -0.179\pm0.038({\rm stat.})$
and $\Delta \chi^2 \sim 29$ with respect to the fit with fixed
$\alpha_{T}=0$. This can be interpreted as the evidence for a sizeable
real part in the effective scattering amplitude. If we impose $\sigma_{0}=27$ mb,
as expected for electromagnetic interactions by averaging $\rho^0,\ \omega$ and $\phi$
vector mesons, we obtain $\alpha_{T} = -0.182\pm0.037({\rm stat.})$.
Note that this is in a good agreement
with the analysis of $\rho^{0}$ photoproduction experiments~\cite{Spital:ph} at low-$Q^2$.
Since we require our phenomenological model to correctly reproduce the
photoproduction limit, we fix $\sigma_{0}=27$ mb and
$\alpha_{T} = -0.20$ according to~\cite{Spital:ph}.

In our model we use the pionic parton distributions extracted from
real pion scattering data~\cite{Gluck:1999xe} to approximate the structure
functions of virtual pions in nuclei. To this end we perform fits with and
without the pionic sea distributions and we find a significantly
better description of data in the latter case. Therefore we only
consider the valence contribution to the pionic structure functions
in the following.

\subsubsection{Results}
\label{sec:fitres}

In our model we assume three main free parameters: $C_{N}, x_{0}$
and $Q^{2}_{0}$ (see Sections~\ref{sec:modbn} and \ref{sec:xsec}).
In addition, the off-shell function $\delta f_2(x)$ is
characterized by the presence of a second zero, $x_{1}$.
This specific feature has important consequences, as it is discussed
in Sec.\:\ref{sec:inter}. 
Since the parameter $x_{1}$ turns out to be strongly correlated 
with $C_{N}$ and $Q^{2}_{0}$, we perform several fits with 
different fixed values of $x_{1}$ in the range 
$0.030 \leq x_{1} \leq 0.065 $ and we evaluate the corresponding
effect on the normalization of the valence quarks at large $Q^2$.
Among the fits with comparable $\chi^2$ with respect to data, 
we choose a fixed value $x_{1} = 0.050$,
since this value provides a good cancellation between off-shell 
and shadowing corrections in the normalization for all nuclei (see 
Sec.\:\ref{sec:valnorm} for details).

In order to test our hypothesis about the universality of parameters
in \Eqs{eq:deltaf}{eq:effxsec}, we perform independent fits to
different sub-sets of nuclei (from ${}^4$He to ${}^{207}$Pb) and we
compare the corresponding values of the parameters with the ones
obtained from a combined fit to all data. As can be seen from
Table~\ref{tab:fits}, the results are compatible within
uncertainties, thus allowing a unified treatment.%
\footnote{Unfortunately it is not possible to have
data points covering both the high and low $x$ regions for all
nuclei. This can result in a small sensitivity to some of the
parameters for specific nuclei, as can be seen from the uncertainties
in Table~\ref{tab:fits}.}
The values of $\chi^{2}/\mathrm{d.o.f.}$  indicate an excellent
consistency between the model and the data points for all nuclei.

\begin{table}[htbp] 
\begin{center}
\begin{tabular}{l|c|c|c|c}
Data set & $C_{N}$ & $x_{0}$ & $Q_{0}^{2}\;[\gevsq]$ & $\chi^{2}$ / d.o.f. \\ \hline\hline 
${}^4$He/D & ~~$6.17\pm1.29$~~  & ~~$0.456\pm0.033$~~  & ~~$1.30\pm0.30$~~ & 16.0 / 35 \\ \hline
${}^7{\rm Li}/D; {}^9{\rm Be}$/D & $7.65\pm0.92$ & $0.435\pm0.022$ & $0.80\pm0.30$  & 35.1 / 35 \\ \hline
${}^{12}$C/D & $9.38\pm0.76$  & $0.449\pm0.012$  & $1.20\pm0.21$ & 23.8 / 31 \\ \hline
${}^{27}$Al/D; ${}^{27}$Al/${}^{12}$C & $7.39\pm0.86$  & $0.470\pm0.016$  & $3.60\pm2.41$  & 15.1 / 33 \\ \hline
${}^{40}$Ca/D; ${}^{40}$Ca/${}^{12}$C & $7.09\pm0.79$ & $0.482\pm0.016$  & $1.67\pm0.15$ & 56.8 / 58 \\ \hline
${}^{56}{\rm Fe}/D; {}^{63}{\rm Cu}$/D; ${}^{56}{\rm Fe}/{}^{12}$C & $8.28\pm0.53$ & $0.449\pm0.009$  & $1.39\pm0.30$  & 55.6 / 63 \\ \hline
${}^{108}{\rm Ag}/D; {}^{119}{\rm Sn}$/${}^{12}$C & $9.61\pm1.29$ & $0.448\pm0.018$  & $1.41\pm0.34$  & 21.0 / 29 \\ \hline
${}^{197}{\rm Au}/D; {}^{207}{\rm Pb}$/D; ${}^{207}{\rm Pb}$/${}^{12}$C & $8.52\pm0.87$ & $0.387\pm0.026$  & $1.31\pm0.37$ & 18.2 / 42 \\ \hline \hline
All data & $8.10\pm0.30$  & $0.448\pm0.005$  & $1.43\pm0.06$ & 458.9 / 556 \\ \hline
\end{tabular}
\caption{
Values of the parameters extracted from independent fits to different 
sub-sets of data with $Q^2>1.0$ GeV$^2$. Uncertainties are startistical 
only. The column on the right gives the $\chi^2$ from each fit and the 
corresponding number of degrees of freedom. The last row shows the result 
of the global fit. 
\label{tab:fits}
} 
\end{center}
\tunespace
\end{table}

\begin{table}[htb] 
\tunespace
\begin{center}
\begin{tabular}{l||c|c|c}
 $C_{ij}$ & $j=C_{N}$ & $j=x_{0}$ & $j=Q_{0}^{2}$ \\ \hline\hline 
$i=C_{N}$ & 1.000  & -0.067 & -0.127 \\ \hline
$i=x_{0}$ & -0.067  & 1.000  & 0.006 \\ \hline 
$i=Q_{0}^{2}$ & -0.127  & 0.006  & 1.000  \\ \hline
\end{tabular}
\caption{
Correlation coefficients between the nuclear parameters from 
the global combined fit.
\label{tab:corr}
}
\end{center}
\tunespace
\end{table}

The final values of $C_{N}$, $x_{0}$ and $Q^{2}_{0}$ obtained from a
global fit to nuclear data are given in the last line of Table~\ref{tab:fits}.
The correlation between the parameters is small and mainly related to
the normalization constant, as can be seen from Table~\ref{tab:corr}.

\begin{sidewaysfigure}[p]
\begin{center}
\epsfig{file=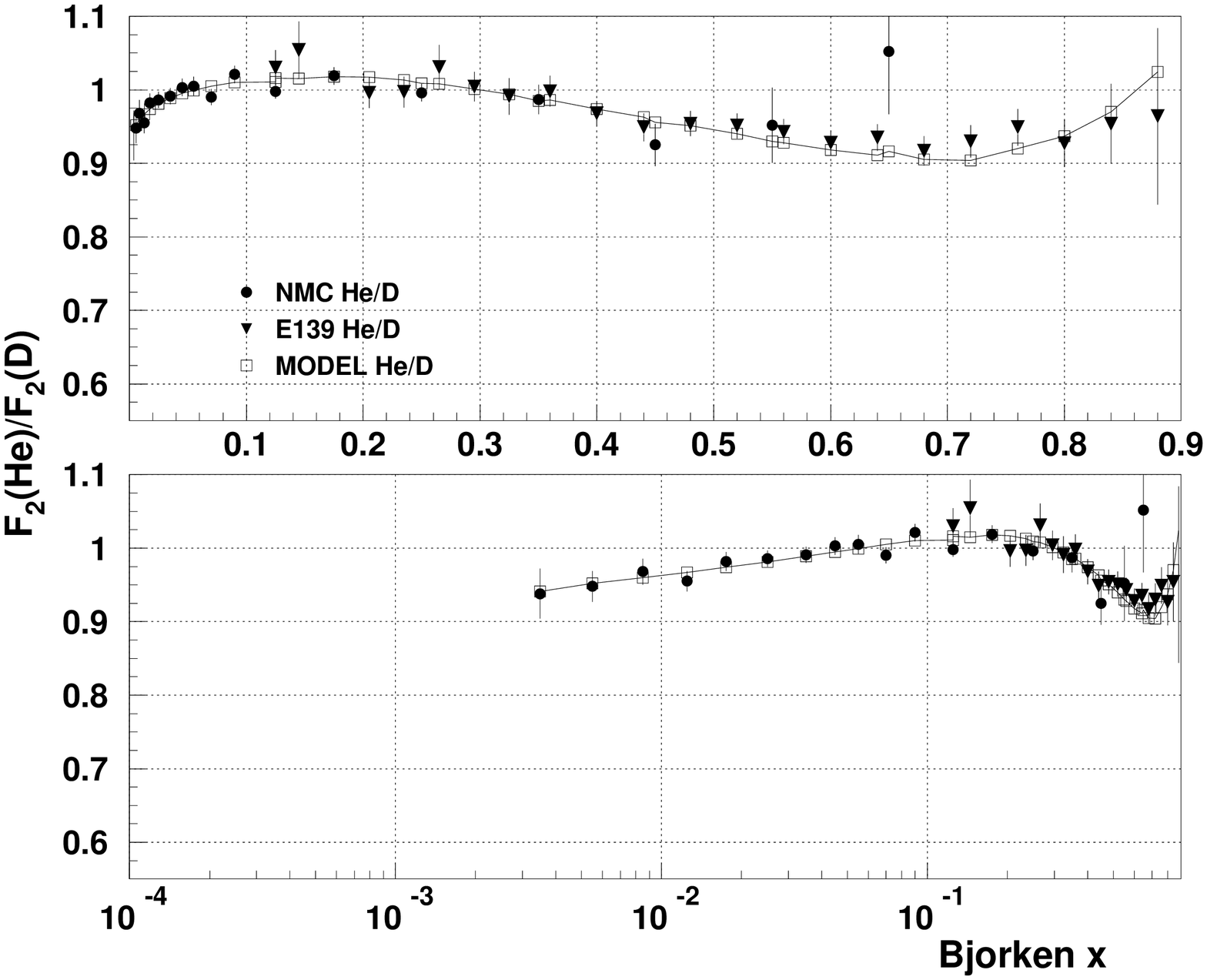,width=10.0cm}%
\epsfig{file=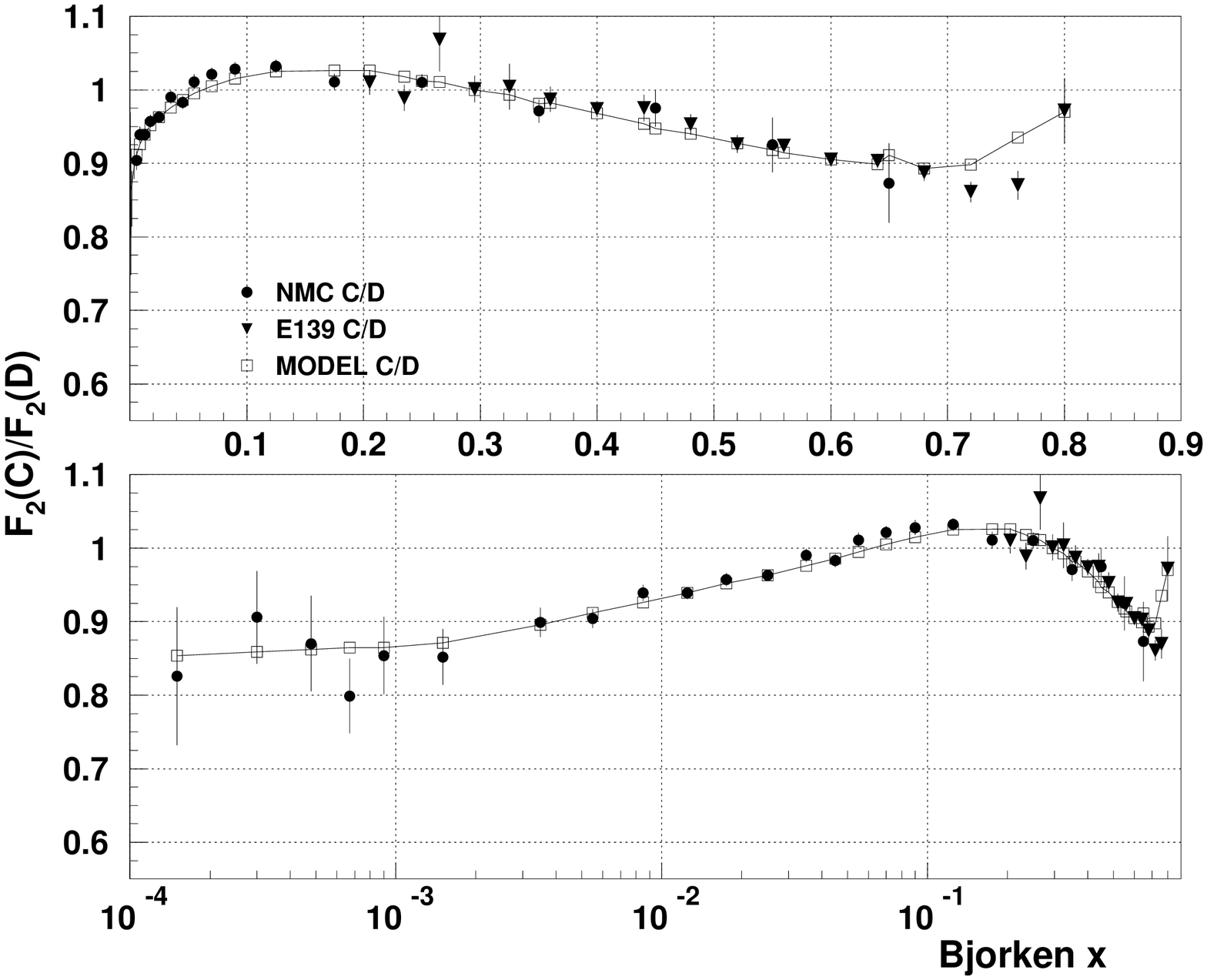,width=10.0cm}
\epsfig{file=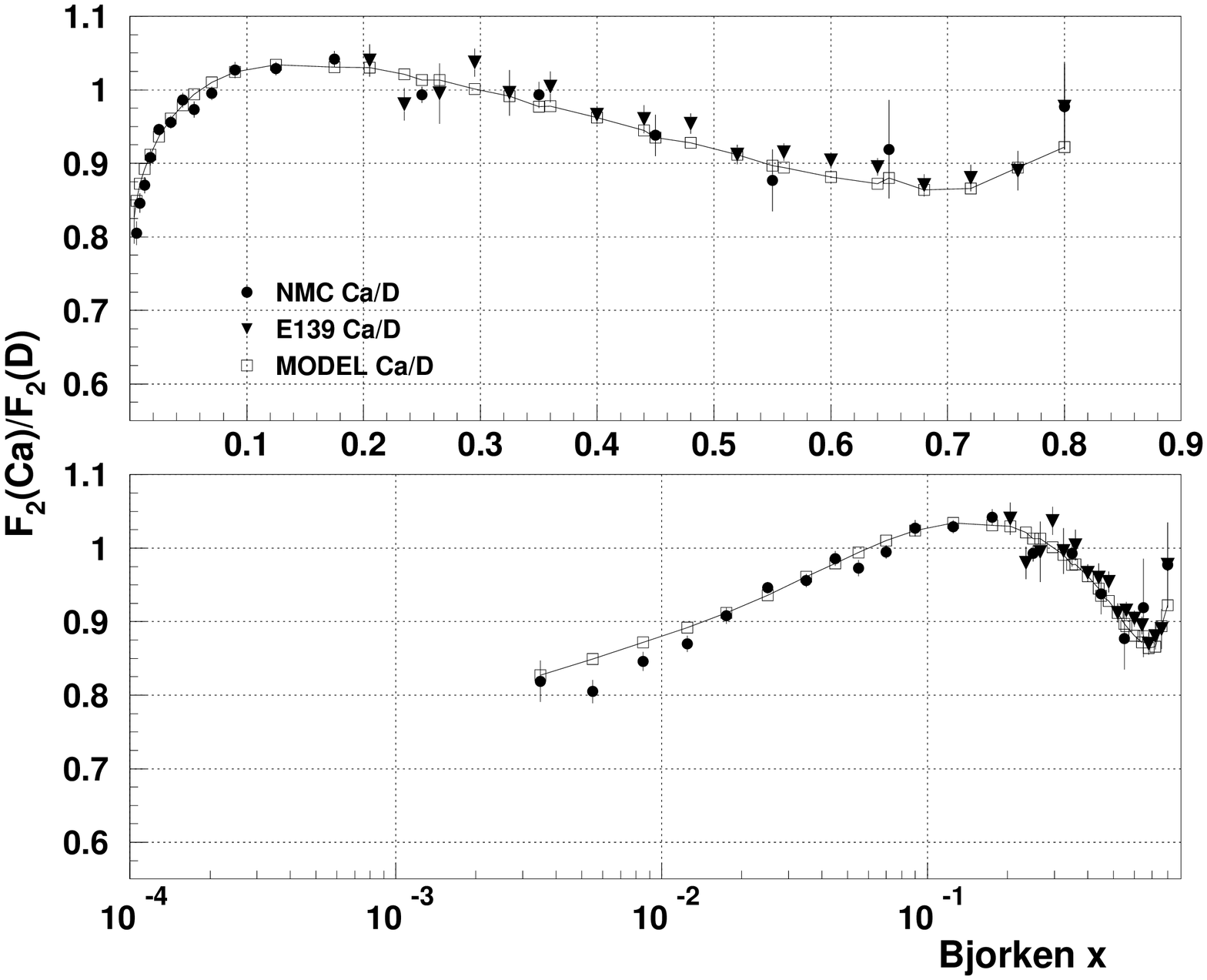,width=10.0cm}%
\epsfig{file=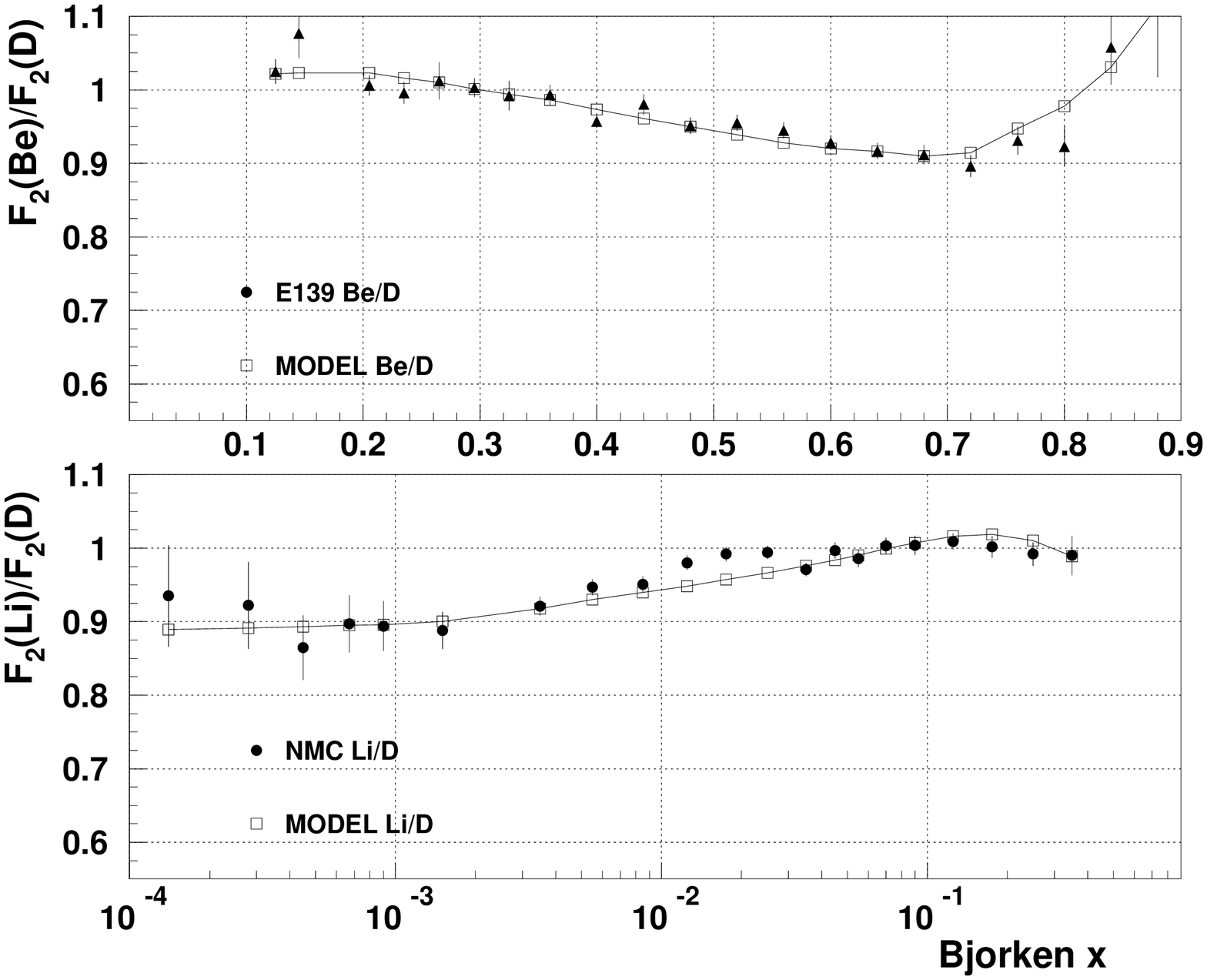,width=10.0cm}
\caption{Ratios $\mathcal{R}_{2}(x,A'/A)$ for 
${}^4$He/D, ${}^{12}$C/D, ${}^{40}$Ca/D, ${}^7$Li/D and ${}^{9}$Be/D
(left to right and top to bottom).
The curves with open squares show the corresponding model calculations 
with the parameters specified in the last line of Table~\ref{tab:fits}. For 
data points the error bars correspond to the sum in quadrature of 
statistical and systematic uncertainties, while the normalization 
uncertainty is not shown.} 
\label{fig:ratios1}
\end{center}
\end{sidewaysfigure}

\begin{sidewaysfigure}[p]
\begin{center}
\epsfig{file=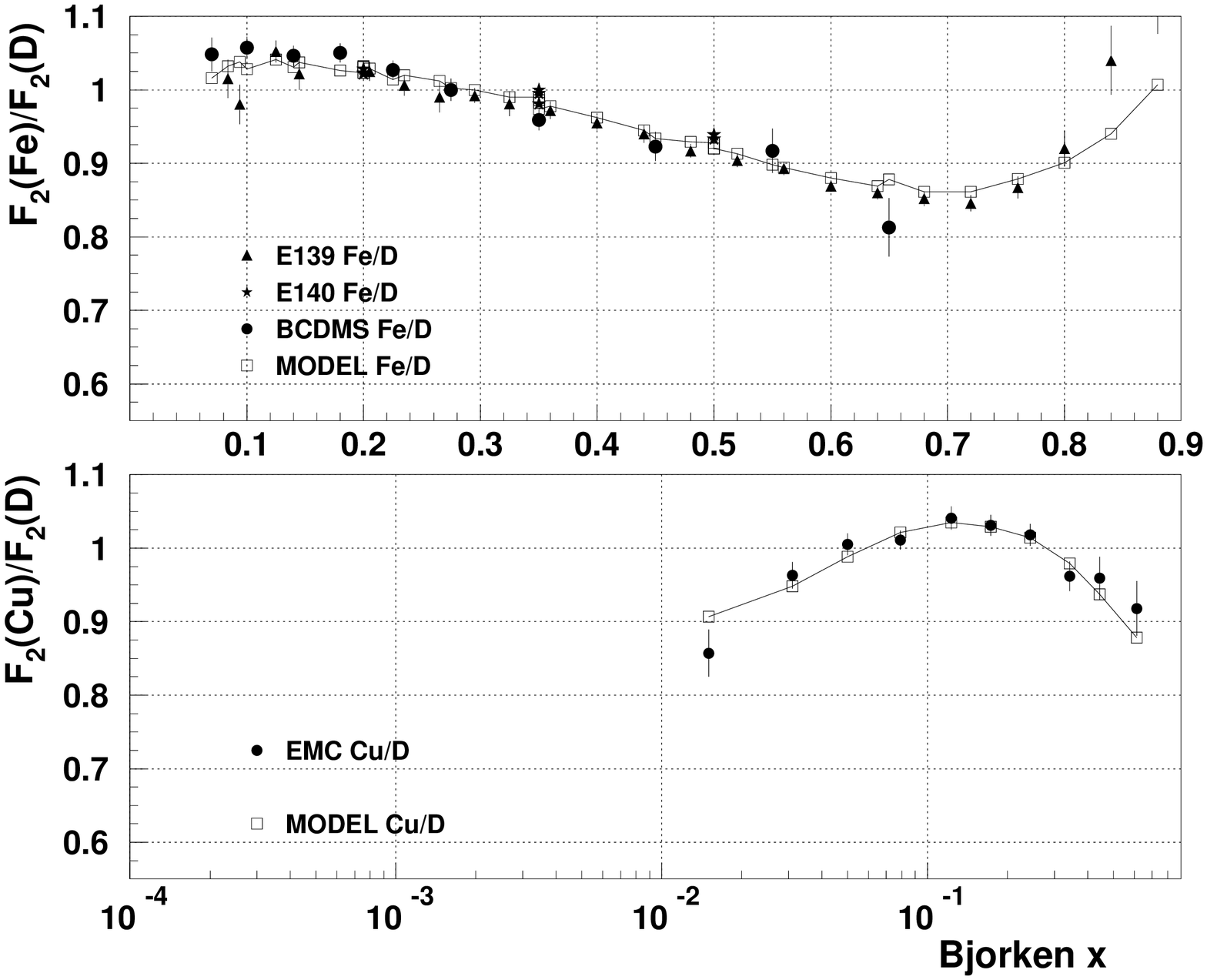,width=10.0cm}%
\epsfig{file=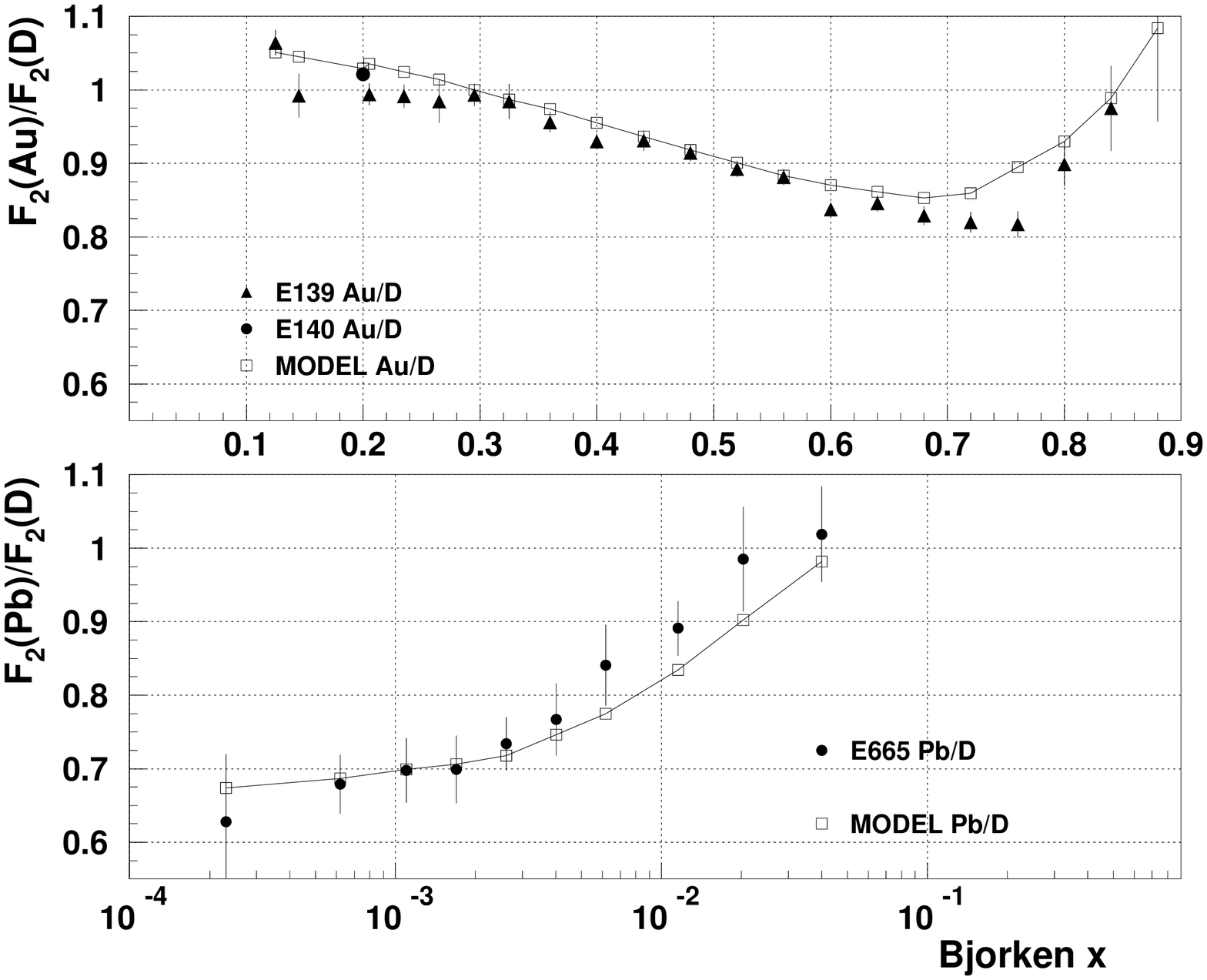,width=10.0cm}
\epsfig{file=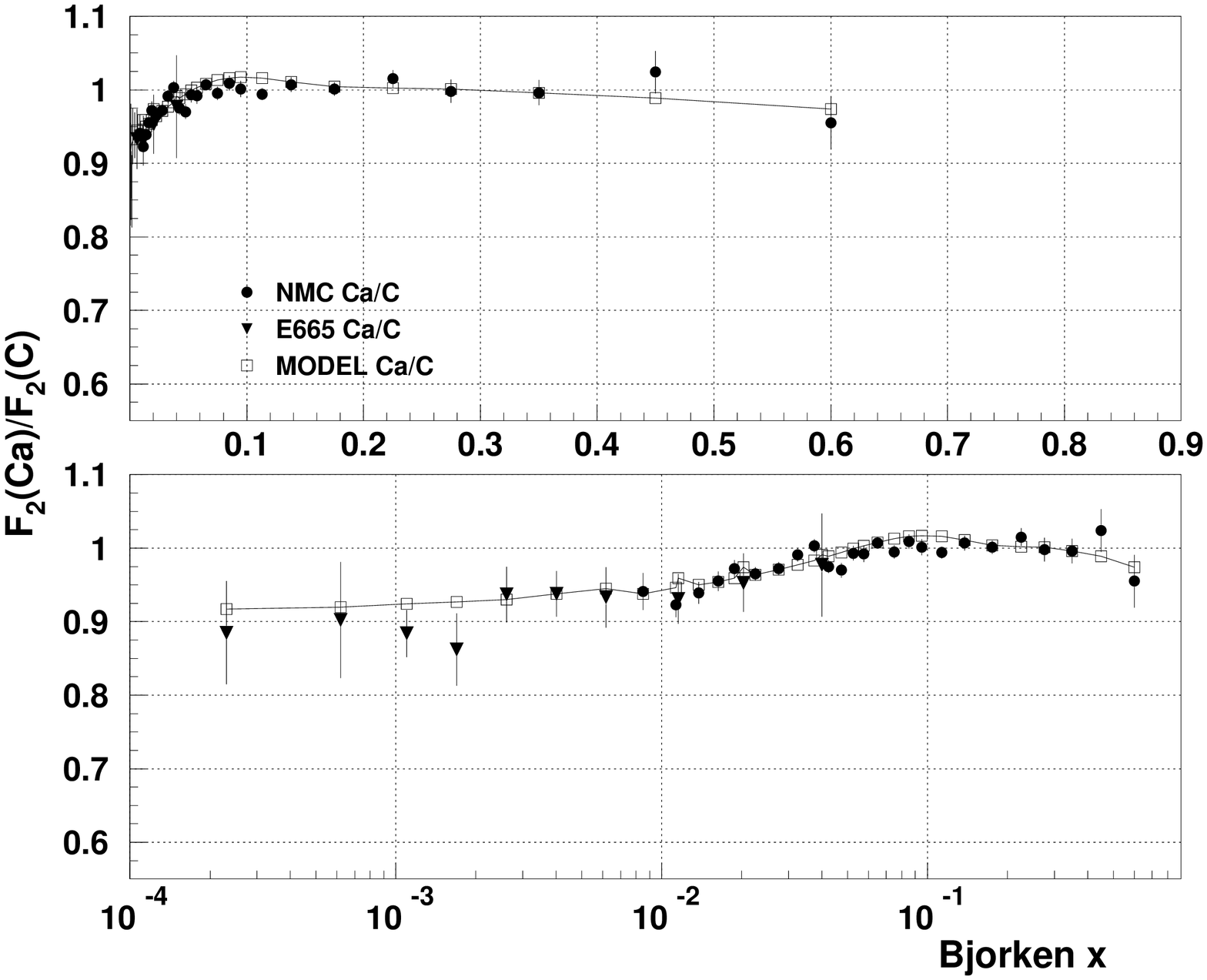,width=10.0cm}%
\epsfig{file=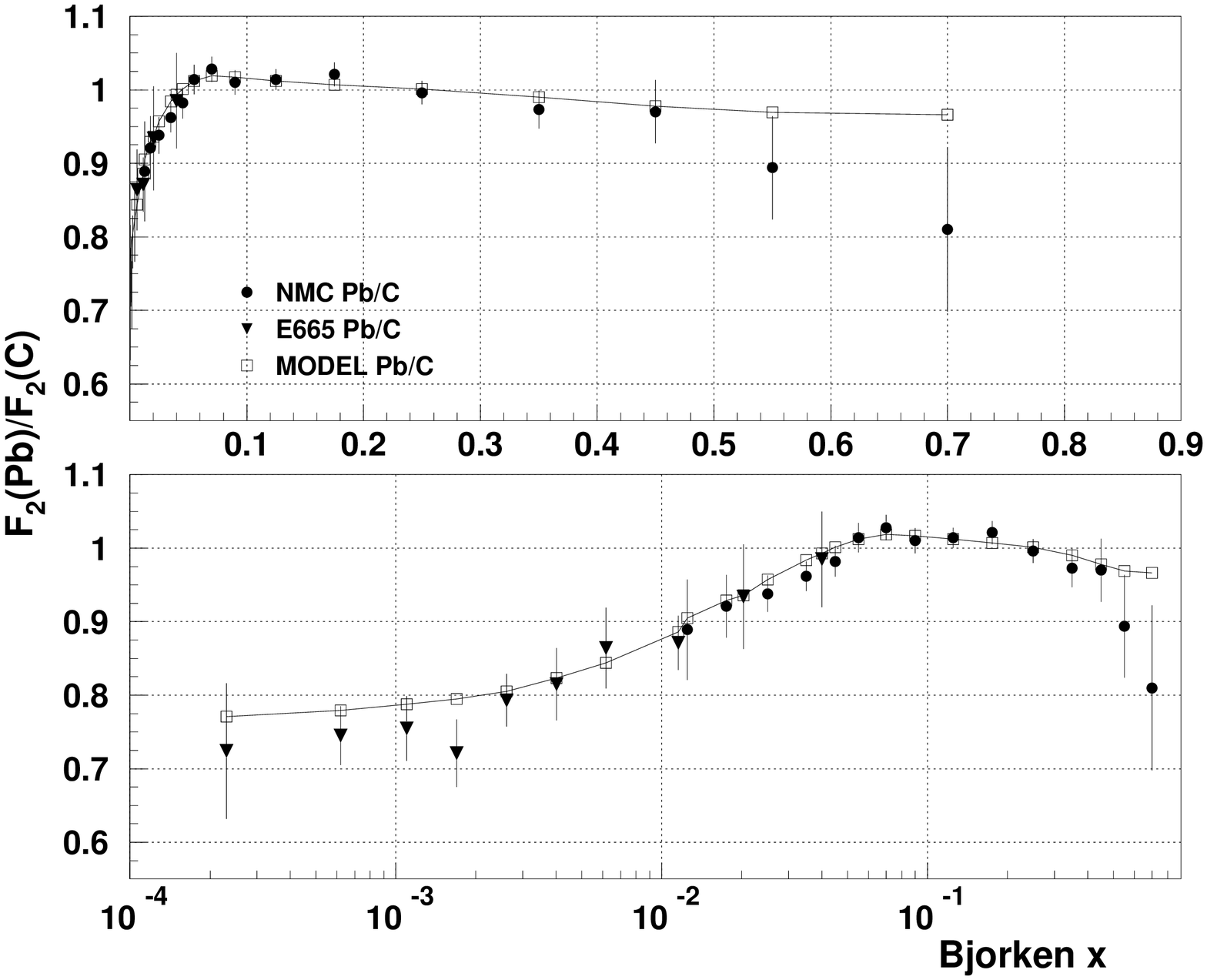,width=10.0cm}
\caption{Ratios $\mathcal{R}_{2}(x,A'/A)$ for 
${}^{56}$Fe/D, ${}^{63}$Cu/D, ${}^{197}$Au/D, ${}^{207}$Pb/D,
${}^{40}$Ca/${}^{12}$C and ${}^{207}$Pb/${}^{12}$C (left to right and top to bottom).
The curves with open squares show the corresponding model calculations
with the parameters specified in the last line of Table~\ref{tab:fits}. 
For data points the error bars correspond to the sum in quadrature of
statistical and systematic uncertainties, while the normalization uncertainty
is not shown.}
\label{fig:ratios2}
\end{center}
\end{sidewaysfigure}

Figures \ref{fig:ratios1} and \ref{fig:ratios2} show the excellent
overall agreement between the calculation and the data points for many
different nuclei. A few comments are in order. The region at $x>0.75$ is
characterized by a significant $Q^{2}$ dependence and therefore the
calculation based upon the average $\bar{Q}^{2}$ provided by the
experiments is approximate. It must also be noted that in some cases the
data points from different experiments are not fully consistent. In
particular, the data points on ${}^{12}$C/D and ${}^{40}$Ca/D ratios from
E665 experiment \cite{E665} at low $x$ seem to be systematically above the
corresponding NMC measurements, which have smaller uncertainties.
Similarly, a normalization problem could be present for ${}^{207}$Pb/D
data from E665. Assuming the effect is common to all heavy targets, in our
fits we use instead the double ratios (${}^{40}$Ca/D)/(${}^{12}$C/D) and
(${}^{207}$Pb/D)/(${}^{12}$C/D) and the E665 measurement of the ratio
${}^{207}$Pb/D. The double ratios are in good agreement with NMC data
(noticed also in~\cite{NMCpb}) as well as with our predictions, while the
${}^{207}$Pb/D points lie slightly above our calculations. Futhermore, the
ratio ${}^{7}$Li/D shown in Fig.~\ref{fig:ratios1} indicates a small
excess in the region of $x$ between $0.01$ and $0.03$, which produces
corresponding reductions in the ratios ${}^{12}{\rm C}/{}^7{\rm Li}$ and
${}^{40}{\rm Ca}/{}^7{\rm Li}$. The effect is much larger than the quoted
systematic uncertainties. For instance, the exclusion of three points at
$x=0.0125,0.0175,0.0250$ from our fit leads to the reduction of overall
$\chi^{2}$/d.o.f. for the ratio ${}^{7}$Li/D  from 1.95 to 0.72.%
\footnote{The comment is only intended to quantify the effect. We keep all
data in our fits, regardless of the inconsistencies described above.} We
also comment that the value of the ${}^{207}$Pb/${}^{12}$C ratio at
$x=0.7$ from NMC (Fig.~\ref{fig:ratios2}) is marginally
compatible with the corresponding value of ${}^{197}{\rm Au}$/D ratio from
E139 experiment.



It should be emphasized that at low $x$ there is an interplay
between the off-shell function, the pion contribution and the coherent nuclear effects.
This results in significant correlations between the corresponding parameterizations
and does not allow an unambiguous extraction of individual components
without external constraints. 
In our approach the pion (meson) excess in nuclei
is calculated as described in Sec.\:\ref{sec:pion:details}.
In order to disentangle
the actual off-shell function from the remaining coherent correction,
we use additional information from photoproduction
experiments (Section~\ref{sec:fixpars}). The agreement between our
independent extraction of the average VMD parameters and the
photoproduction limit makes us confident in the deconvolution
of different components.

We further check the interplay between nuclear pion excess and off-shell 
effects in our analysis by fitting our model without pion correction to 
$\mathcal{R}_2$ data. In this case the effective off-shell function 
$\delta f'_2$ absorbs the nuclear pion contribution to nuclear $F_2$. We 
use for this test a higher order polynomial with respect to 
\eq{eq:deltaf}, without any fixed parameter. This is intended to avoid 
biases from the functional form used to model the off-shell function. The 
results obtained for the effective $\delta f'_2$ are consistent with the 
following estimate which can be obtained by explicitly separating the 
nuclear pion contribution to nuclear $F_2$
\begin{equation}
\delta f'_2 = \delta f_2 +
                \frac{\delta F_2^{\pi/A}(x)}{\average{v}F_2^N(x)},
\label{delf:eff}
\end{equation}
where $\delta F_2^{\pi/A}$ is the nuclear pion correction calculated as 
described in Sec.\:\ref{sec:pion:details} and $\average{v}$ denotes the 
nucleon virtuality $v=(p^2-M^2)/M^2$ averaged over the nuclear spectral 
function. 
Moreover, the best fit corresponds to a value of $Q^{2}_{0}$ which is
in agreement with our fit with explicit treatment of nuclear pion 
correction.

\begin{figure}[htb]
\begin{center}
\tightspace
\epsfig{file=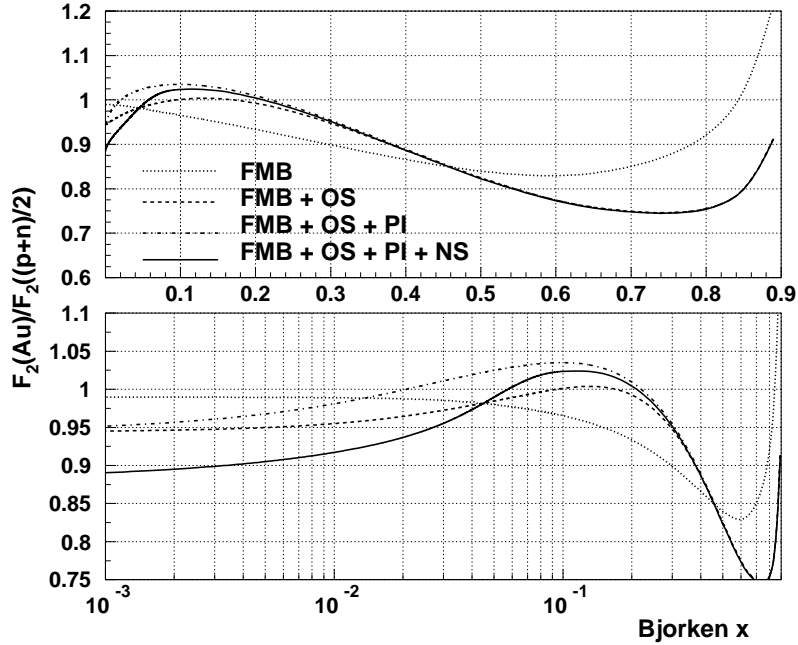,width=\figxx}
\caption{Different  nuclear effects
on the ratio of ${}^{197}$Au to
isoscalar nucleon for $F_2$ at $Q^2=10\gevsq$.
The labels on the curves correspond to effects due to Fermi motion and
nuclear binding (FMB), off-shell correction (OS), nuclear pion excess (PI) and
coherent nuclear processes (NS). Target mass and
the neutron excess corrections are included.
}
\label{fig:F2contsAu}
\tightspace
\end{center}
\end{figure}

Figure~\ref{fig:F2contsAu} illustrates different nuclear corrections
to the ratio of $F_2$ of ${}^{197}$Au to that of the isoscalar nucleon
$F_{2}^N=\tfrac12(F_2^p+F_2^n)$ calculated in our model using 
the final parameters shown in the last line of Table~\ref{tab:fits}.
As follows from comparison of Fig.~\ref{fig:F2contsAu} and the results of 
our fit displayed in Figs.~\ref{fig:ratios1} and \ref{fig:ratios2}
the standard Fermi motion and nuclear binding
effect treated in impulse approximation does not quantitatively explain 
data at large $x$. The off-shell effect is therefore an important correction which 
modifies the structure functions of bound nucleon and affects the slope and 
the magnitude of the ratio $\mathcal R_2$ at large $x$. 
As discussed above, we extract this correction from inclusive nuclear 
DIS data. The disscussion 
of off-shell correction in terms of a scale characterizing valence quark 
distribution and its modification in nuclear environment is presented in 
Sec.\:\ref{sec:radius}.


\subsubsection{Systematic uncertainties}
\label{sec:syst}

Systematic uncertainties of the model are evaluated by varying
each of the contributions from the deuterium wave function, the spectral
function, the parton distributions, the pion structure function and
the functional forms of $\delta f_2(x)$ and $\sigmabar_T(Q^2)$. New fits 
are then performed and systematics are defined from the corresponding 
variation of the nuclear parameters and from the global $\chi^{2}$ values. 
Results are listed in Table~\ref{tab:syst}.

\begin{table}[htb]
\begin{center}
\begin{tabular}{l|c|c|c}
Systematics & $\delta C_{N}$ & $\delta x_{0}$ & $\delta Q_{0}^{2}\;[\gevsq]$ \\ \hline\hline 
Deuteron wave function & ~~0.238~~ & ~~0.0016~~ & 0.026 \\ \hline 
Nuclear spectral function~~ & 0.451  & 0.0046 & 0.021 \\ \hline
Parton distributions & 0.005  & 0.0007 & 0.023 \\ \hline 
Pion structure function & 0.050 & 0.0020 & 0.065 \\ \hline 
Functional form & 0.120  & 0.0040 & 0.070 \\ \hline 
Cross-sections $\sigma_{0}$ and $\sigma_{1}$ & 0.015  & 0.0005 & 0.165 \\ \hline\hline
Total & 0.526 & 0.0067 & 0.195 \\ \hline 
\end{tabular}
\caption{
The estimate of systematic uncertainties on the extraction of 
nuclear parameters. See text for a detailed explanation.
\label{tab:syst}
}
\end{center}
\tunespace
\end{table}

Although we do not use directly deuterium data for the fits, most of the
data points come from the ratios $\mathcal{R}_{2}$ of a heavy target to
deuterium. In order to study the sensitivity of our result to the choice
of the deuteron wave function we performed independent fits with two
different choices of the deuteron wave function: the one which corresponds
to the Bonn potential \cite{bonn:wf} and the Paris wave function
\cite{paris:wf}. These two wave functions have different high-momentum
component and in this respect represent two extreme situations.

Similarly, we modify the high-momentum component of the momentum 
distribution $n_{\mathrm{cor}}(\bm{p})$ in nuclei by multiplying it by the 
ratio of the Bonn and Paris deuteron wave functions squared. This is 
motivated by the observation~\cite{ZE78,CS96} that the momentum 
distribution of finite nuclei and nuclear matter at high momenta 
are proportional to that of the deuteron. We then repeat our fits with 
modified spectral functions in order to estimate the corresponding 
variations on the parameters of the model. 

The systematic uncertainty related to the parton distributions is
estimated by varying the PDFs within their uncertainty ($\pm 1\sigma$).
In addition, we also use different sets of parton distributions,
extracted from fits to different data samples, with different
$Q^2$ boundaries and different parameterization for the low $Q^2$
extrapolation.

For the pion structure function, we repeat our fits by using
both the LO and the NLO approximations of the pionic parton
distributions from~\cite{Gluck:1999xe}. We also arbitrarily
change the parameterizations~\cite{Gluck:1999xe} within $\pm 10\%$.

We tried different functional forms in
\Eqs{eq:deltaf}{eq:effxsec} to parameterize $\delta f_2$ and
$\sigmabar_T$. In particular, for the off-shell function $\delta f_2$
we have tried a generic higher order polinomial parameterization in
\eq{eq:deltaf} and also used a parametrization with an additional $x^k$
term, with free parameter $k$. In spite of the new
parameters, all acceptable results (i.e. with the values of $\chi^2$ 
comparable to our best fit solution) extracted from fits to data were very 
similar to the ones obtained with parametrization (\ref{eq:deltaf}). We 
emphasize that the behaviour of the function $\delta f_2(x)$ for $x<0.70$ 
is well constrained by data and only small variations on both the shape 
and the position of the zero $x_{0}$ are allowed. This observation in turn 
results in reduced systematic uncertainties of the model.
%
%
\begin{figure}[htb]
\begin{center}
\tightspace 
\epsfig{file=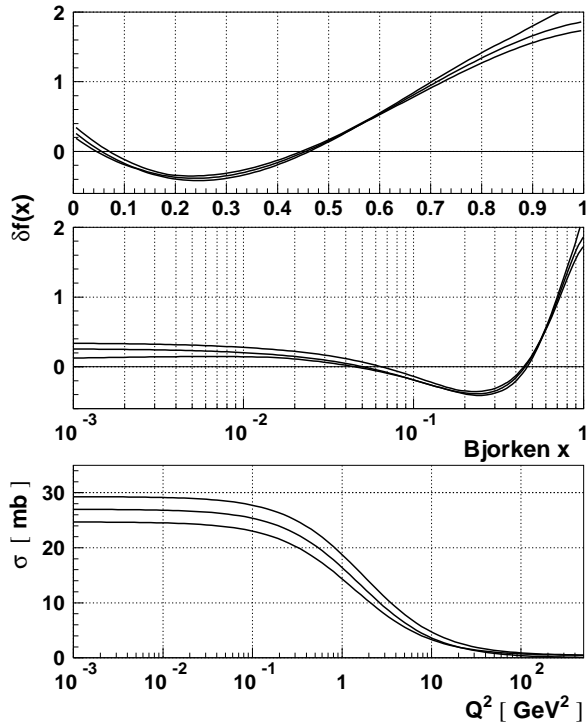,width=\figx} 
\caption{Off-shell function $\delta f_2(x)$ and effective cross-section 
$\sigmabar_T(Q^{2})$ corresponding to the parameters from 
Table~\ref{tab:fits}. The curves show the size of the uncertainty bands 
($\pm 1\sigma$ ), including both statistical and systematic 
(Table~\ref{tab:syst}) uncertainties. The effect of different functional 
forms is also included, as explained in Sec.\:\ref{sec:syst}.} 
\label{fig:functions}
\tightspace 
\end{center}
\end{figure}

For the coherent nuclear effects, we varied $\sigma_{0}$ within
the uncertainty estimated by averaging over $\rho^0,\ \omega$ and $\phi$
mesons, $\pm 3$ mb (Section~\ref{sec:xsec}). As explained in
Sec.\:\ref{sec:fixpars}, this parameter is strongly correlated with
$\alpha_{T}$ and $Q^2_{0}$. Similarly, we varied $\sigma_{1}$
within the $1\sigma$ allowed range (Section~\ref{sec:fixpars}).
We also tried to change the exponent controlling
$Q^2$ dependence in \eq{eq:effxsec}. We obtained
almost equally good fits with the dipole and monopole forms in
\eq{eq:effxsec}. The monopole form of \eq{eq:effxsec} had lower
$\chi^2/\mathrm{d.o.f.}$ for the overall data set.

Figure~\ref{fig:functions} shows the off-shell function 
$\delta f_2(x)$ and the effective cross-section $\bar{\sigma}_T(Q^2)$ 
obtained in Sec.\:\ref{sec:fitres}, together with the corresponding 
total uncertainty bands (including both statistical and systematic 
uncertainties).
We comment that in our analysis the off-shell correction is treated as the 
first order correction in the parameter $v=(p^2-M^2)/M^2$. 
At large $x>0.7$ the off-shell correction can be as large as 25\% for 
heavy nuclei (see also Fig.~\ref{fig:F2contsAu}) indicating that 
higher-order terms in $v$ might not be negligible. This can also be a 
source of systematic uncertainty. However, going beyond the first order in 
$v$ requires the consideration of higher-order relativistic corrections to 
nuclear wave and spectral functions (see discussion in 
Sec.\:\ref{WNB-limit}) that would also affect the treatment of 
``standard'' FMB effect. This goes beyond the scope of the present 
analysis. 
We also note that our phenomenological function $\delta f_2$ is 
extracted from nuclear data and hence it effectively incorporates 
additional contributions from missing terms.  

The final results are dominated by systematic uncertainties,
as can be seen from Tables~\ref{tab:fits} and~\ref{tab:syst}.
However, we note that the magnitude of systematic uncertainties is 
constrained by the value of $\chi^2$ of a global fit to data, as explained 
above. Therefore, the availability of more accurate (or with wider 
coverage of kinematics and nuclei) experimental measurements would 
significantly improve our results.


\section{Discussion}
\label{sec:results}

We now discuss the results obtained from our fit to nuclear data in 
Sec.\:\ref{sec:fits}. In Sec.\:\ref{sec:valnorm} we address the 
problem of the normalization of the nuclear valence quark distribution and 
Sec.\:\ref{sec:inter} is focused on the implication of this constraint 
for our analysis. Section~\ref{sec:qsq} is devoted to the $Q^2$ and $A$ 
dependence of nuclear effects predicted by our model. In Section 
\ref{sec:deuterium} we discuss nuclear effects on the deuteron structure 
functions.

\subsection{Nuclear valence quark number}
\label{sec:valnorm}

It is instructive to study the contributions due to different nuclear
effects to the normalization of valence quark distribution in a nucleus.
Common wisdom is that this quantity should not be corrected by nuclear effects
since it counts the baryon number of the system. Therefore, it is
important to verify if different nuclear effects cancel out in the
normalization. In the impulse approximation,
\ie  if no shadowing and off-shell effects are taken into account, the
cancellation of nuclear binding and Fermi motion effects in the
normalization is explicit and it is guaranteed by the normalization
of nucleon distribution function (\ref{ydist:N}) to the number of nucleons.
It should be also noted that nuclear pions do not contribute to nuclear
valence distribution. In the presence of off-shell (OS) and nuclear
shadowing (NS) effects different contributions to the valence quark
normalization per one nucleon can be written as
\begin{align}
\label{nval:A}
N_{\text{val}/A} &= \int_0^A\ud x\,q_{\mathrm{val}/A}(x) =
        N_{\text{val}/N} + 
                \delta N_{\text{val}}^{\text{OS}} +
                        \delta N_{\text{val}}^{\text{NS}},
\end{align}
where $N_{\text{val}/N}=3$ is the number of valence quarks in the nucleon and
\begin{align}
\label{del:val:off}
\delta N_{\text{val}}^{\text{OS}} &= \average{v}\int_0^1\ud x \sum_{i=u,d}
\left(
q_{i/N}(x)\delta f_q(x) - \bar q_{i/N}(x)\delta f_{\bar q}(x)
\right),
\\
\label{del:val:sh}
\delta N_{\text{val}}^{\text{NS}} &=
  \int_0^1\ud x\,q_{\mathrm{val}/N}(x)\delta \mathcal{R}_{\text{val}}(x).
\end{align}
Here $q_{\mathrm{val}/N}=u-\bar u+d-\bar d$ is the nucleon valence
quark distribution, $\delta f_q$ and $\delta f_{\bar q}$ are off-shell
correction functions for quark and antiquark distributions, $\delta
\mathcal{R}_{\text{val}}(x)$ is the shadowing correction to the
valence quark distribution and $\average{v}=\average{p^2{-}M^2}/M^2$
is the bound nucleon off-shellness averaged over nuclear spectral
function (for more details see Sec.\:\ref{sec:npdf}).%
\footnote{Note that the normalization of valence quark distribution is
not affected by the order of perturbation theory analysis and
therefore \protect\eq{nval:A} holds to any order in $\as$, unlike
the Gross--Llewellyn-Smith sum rule
\protect\cite{Gross:1969jf} which is corrected by both perturbative
\protect\cite{gls:rad} and nonperturbative effects.  }

In general, the off-shell corrections $\delta f_q$ and $\delta f_{\bar q}$ 
could be different. Since we phenomenologically extract the off-shell correction 
$\delta f_2$ from a study of $\mathcal{R}_2$ data it is difficult to 
unambiguously disentangle off-shell effects for quark and antiquark distributions. 
The analysis of additional data from either Drell-Yan production 
or neutrino scattering would be therefore important. 
While we defer a detailed analysis of the existing Drell-Yan data from nuclear 
targets~\cite{Drell-Yan} to a future publication, no sensitive neutrino data 
about nuclear effects on structure functions are currently available 
(see also the discussion in Sec.\:\ref{sec:nuint}).  
In this paper we rather try to use simple considerations on the 
nuclear valence quark number in order to test the hypothesis of a single 
universal off-shell correction for all partons in the bound nucleon 
against the case of different corrections $\delta f_q$ and 
$\delta f_{\bar q}$.
For this purpose it is enough to focus on the high $Q^2$ region, where 
we can use \eq{del:f2}.%

\begin{figure}[htb]
\begin{center} 
\epsfig{file=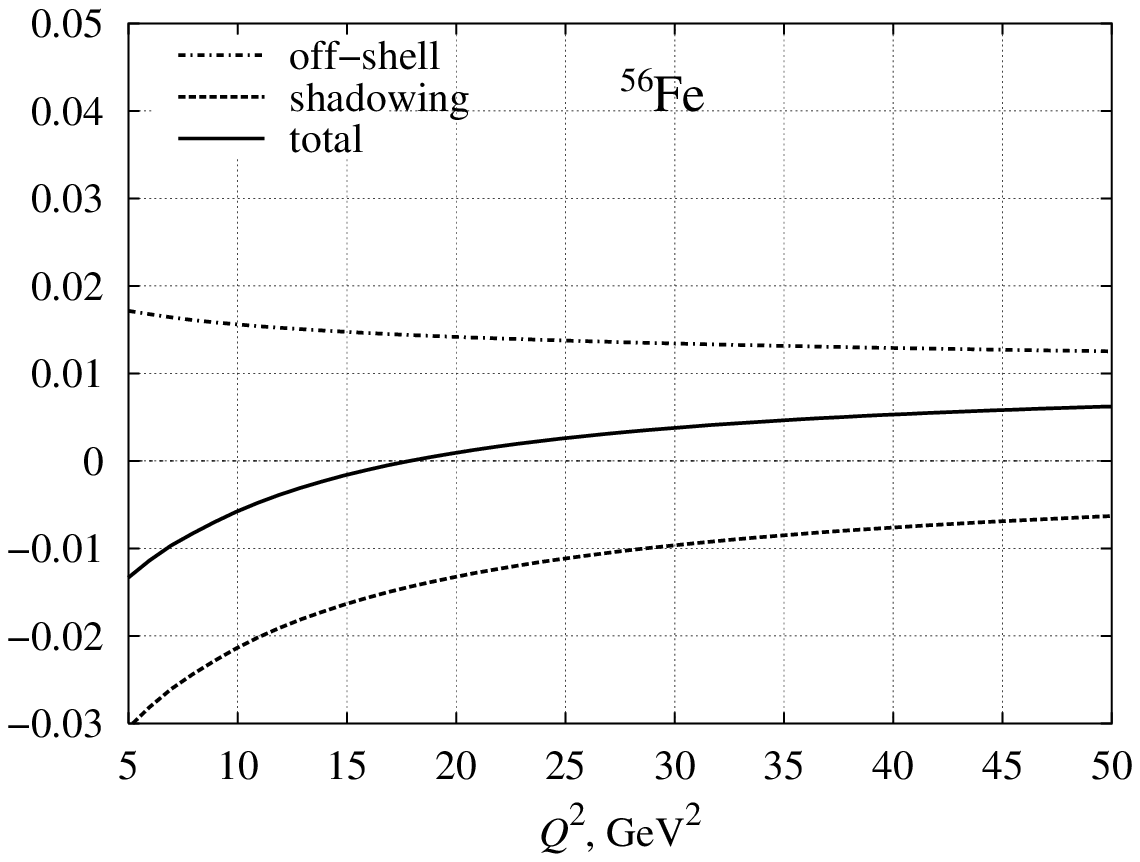,width=0.49\textwidth}%
\epsfig{file=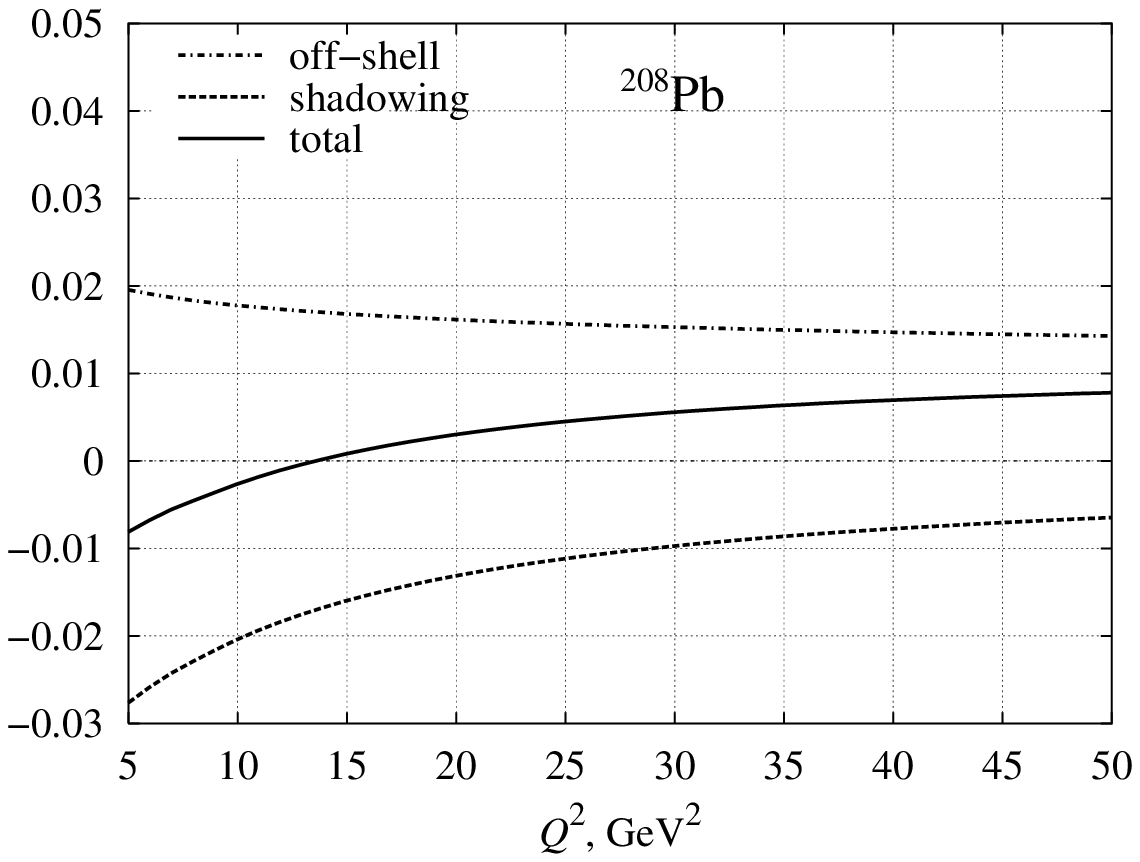,width=0.49\textwidth}
\caption{%
Relative off-shell ($\delta N_{\text{val}}^{\text{OS}}/3$) and
nuclear shadowing ($\delta N_{\text{val}}^{\text{NS}}/3$) corrections to
the normalization of the valence quark distribution for ${}^{56}$Fe
and ${}^{207}$Pb nuclei (left and right panel, respectively)
computed as described in Sec.\:\protect\ref{sec:valnorm}. The solid curve
shows the sum of the off-shell and the nuclear shadowing corrections. }
\label{fig:normval}\tightspace
\end{center}
\end{figure}

Let us first assume $\delta f_q(x) = \delta f_{\bar q}(x)$. From \eq{del:f2} 
we then also have $\delta f_2(x)=\delta f_q(x)$, that implies that we have a 
universal off-shell function for both quark (valence) and antiquark (sea) 
distributions. We evaluate $\delta N_{\text{val}}^{\rm OS}$ by 
\eq{del:val:off} as a function of $Q^2$ using the parameters of $\delta 
f(x)$ from Table~\ref{tab:fits} and the nucleon valence distribution of 
\cite{a02}.  The results for iron and lead are reported in 
Fig.~\ref{fig:normval}, indicating a positive off-shell correction of 
about $1.5-2\%$ that decreases with $Q^2$.
We then compute the shadowing correction $\delta N_{\text{val}}^{\text{NS}}$
by Eq.(\ref{sh:qmn}) using the effective cross section extracted from our
fits (see Sec.\:\ref{sec:xsec}). The results are shown in
Fig.~\ref{fig:normval}. It is important to observe that $\delta
N_{\text{val}}^{\text{NS}}$ is negative and there is large cancellation
between off-shell and shadowing effects in the normalization over a wide
range of $Q^2$.

Let us now test a different hypothesis, namely no off-shell effect in
the sea of bound nucleon $\delta f_{\bar q}=0$. From \eq{del:f2} we
then have for the isoscalar nucleon $(u+d)\delta
f_q=\tfrac{18}{5}F_2\delta f_2$.  In this case the off-shell
correction to valence quark number is dominated by the small $x$
region and even becomes divergent.%
\footnote{We obtain $\delta N_{\text{val}}^{\rm os}/3\approx -0.5$ for
iron if we cut off the contribution of the region
$x<10^{-5}$. Changing the lower limit to $x=10^{-6}$ increases the
magnitude of this correction by about factor of 2.}
Therefore, this assumption leads to unphysical results and we have to
rule out this case.

\subsection{Normalization constraints}
\label{sec:inter}

In the following we will favor the assumption of a single 
universal off-shell function $\delta f(x)$, according to the 
discussion of the previous Section. 
This is supported by the existing $\mathcal{R}_2$ data we used
to extract the phenomenological off-shell function. The universality
of $\delta f(x)$ should be further verified with both Drell-Yan data and
future precise neutrino data.
In our analysis we use the normalization condition in order to fix
parameters of the function $\delta f(x)$, in particular the parameter
$x_1$.
As explained in Sec.\:\ref{sec:fitres}, within all possible values of
$x_{1}$ providing comparable descriptions of data ($\chi^2$) we
selected the one minimizing the overall correction
$\delta N_{\text{val}}^{\text{OS}}+\delta N_{\text{val}}^{\text{NS}}$.

The function $\delta f(x)$ measures the change in the
quark-gluon structure of the nucleon in nuclear environment.
This function is
not accessible in experiments with isolated proton and/or neutron
but can generally be probed in nuclear reactions.
The results described in Sec.\:\ref{sec:fitres} demonstrate that
inclusive DIS data have a good sensitivity to off-shell 
effects, allowing a precise determination of this correction.

The phenomenological cross section in \eq{eq:effxsec} effectively
incorporates contributions to structure functions
due to all twists since it is extracted from data. Higher
twists are known to be important at low and intermediate $Q^2$ and for
this reason we should not expect an exact cancellation between $\delta
N_{\text{val}}^{\text{OS}}$ and $\delta N_{\text{val}}^{\text{NS}}$
calculated with phenomenological cross section. Nevertheless, we observe
from Fig.~\ref{fig:normval} that the cancellation becomes more accurate at
higher $Q^2$ indicating transition to the leading twist regime. In
particular, the exact cancellation takes place at $Q^2\approx 15\gevsq$.
We performed similar calculation for several nuclei and we thus verified
that this effect is independent of the choice of the nucleus.

\begin{figure}[htb]
\begin{center}
\tightspace  
\epsfig{file=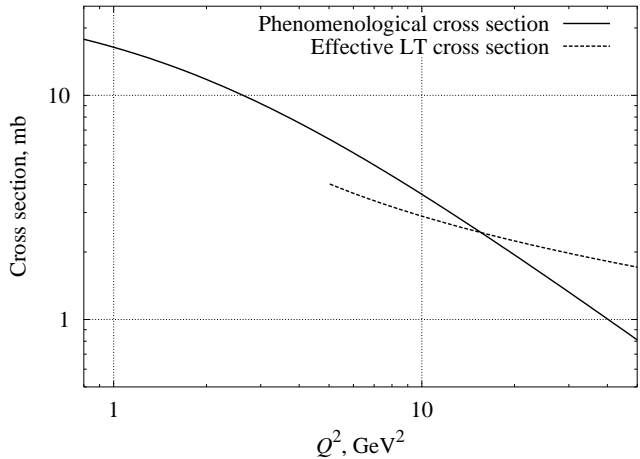,width=\figx}
\caption{%
Phenomenological cross section $\sigmabar_T$ extracted from
our fits (solid curve) and effective LT cross section (dashed curves)
computed for iron and lead nuclei as described in 
Sec.\:\protect\ref{sec:valnorm}. }
\label{fig:xsec}
\tightspace
\end{center}
\end{figure}

It should be noted that the nuclear data available in the shadowing region
are limited to relatively low $Q^2$ and for this reason the
phenomenological cross section (\ref{eq:effxsec}) is not constrained at
high $Q^2$. In this work we evaluate the effective cross section at
high $Q^2$ by treating the condition $\delta
N_{\text{val}}^{\text{OS}}+\delta N_{\text{val}}^{\text{NS}}=0$ as an
equation on the cross section. We solve this equation
numerically using the off-shell function $\delta f_2(x)$ from
Sec.\:\ref{sec:fitres}. The resulting cross section is presented in
Fig.~\ref{fig:xsec} together with phenomenological cross section extracted
from our fits. In this paper we use the following simple model for the 
effective cross section.
For $Q^2$ below the crossing point in Fig.~\ref{fig:xsec} we use
phenomenological cross section (\ref{eq:effxsec}) extracted form the fits,
and for higher $Q^2$ we use the cross section calculated from the
normalization condition. The difference between the two curves in
Fig.~\ref{fig:xsec} below the crossing point is attributed to high-twist
effects.

The function $\delta f(x)$ is positive for $x<x_{1}$ (see 
Fig.~\ref{fig:functions}). This implies a negative off-shell correction to 
the structure functions at small Bjorken $x$ because the offshellness $v$ 
of a bound nucleon is negative. Thus the off-shell correction at small $x$ 
appears as a leading twist shadowing correction. Therefore, in this region 
there is a certain interplay between nuclear effects due to coherent 
nuclear interactions and off-shell effect. 
In the region $x_1<x<x_0$ the function $\delta f(x)$ is negative that 
provides an enhancement of bound nucleon structure functions. Thus in our 
approach the antishadowing at $x\sim 0.1$ is linked to off-shell effects. 
It is important to note that for the valence distributions there is 
additional antishadowing mechanism due to coherent nuclear interactions. 
Indeed, the presence of substantial real part in the $C$-odd channel
($\alpha_\Delta=1$) results in the constructive interference of multiple 
scattering interactions at $x\sim 0.1$ for valence distributions
as will be discussed in more detail in Sec.\:\ref{sec:npdf}. 
%
%

\subsection{Off-shell effect and modification of the nucleon size in nuclei
}
\label{sec:radius}

From our analysis we obtain a positive off-shell correction at large $x>x_0$.
Since the virtuality $p^2-M^2$ of the bound nucleon is
negative this leads to the suppression of valence distribution in the
bound nucleon at large $x$. In order to give a qualitative interpretation 
to this result we consider a simple model of the valence
distribution in the nucleon and we argue that the behavior of $\delta
f_2$ at large $x$ observed in data can be related to
the increase of the nucleon core radius in nuclear environment.

Let us consider the valence quark distribution in terms of spectral
representation \eq{q:spec:off}. We will consider a simple model in which 
the spectrum of spectator states is approximated by a single mass $\bar s$ 
\cite{KPW94,Ku98}
\begin{align}
\label{q:spec:model}
D_{q/N} &=\delta(s-\bar s)\Phi(t,p^2),
\end{align}
where the function $\Phi(t,p^2)$ describes the distribution of valence
quarks over $t=k^2$ in the nucleon with the invariant mass $p^2$. 
For the on-shell
nucleon the distribution $\Phi(t)$ is characterized by a scale
$\Lambda_v^2$. In configuration space this scale should be related to
the size of the valence quark confinement region $r_c\sim \Lambda_v^{-1}$
(the nucleon core radius). From dimensional analysis one can write
$\Phi(t)=C_v\Lambda^{-2}_v\phi(t/\Lambda^2_v)$ where $\phi$ and $C_v$ are
dimensionless profile function and normalization constant.  We found
that a simple pole model $\phi(z)=(1-z)^{-n}$ results in a reasonable
description of the nucleon valence distribution at large $x$
and high $Q^2$. In particular, we obtain a reasonable fit to valence
distribution of \cite{a02} at $Q^2=15\gevsq$ and $x\ge
0.2$ by taking $\bar s=2.1\gevsq$, $\Lambda_v^2=1.2\gevsq$ and $n=4.4$.

In order to model off-shell dependence of parton distributions we
assume that the normalization constant $C_v$ and the scale $\Lambda_v$
become functions of $p^2$ while the profile function $\phi$ and
the average mass of spectator states $\bar s$ do not change off-shell.
We use \eq{q:spec:off} in order to calculate the off-shell
modification of the quark distribution $\delta f_q$.  The
$\delta$-function in \eq{q:spec:model} allows us to integrate over the
spectrum of residual system. Inspecting the resulting expression we observe
a relation between the derivatives of the quark distribution with
respect to $x$ and $p^2$.
After some algebra we obtain
\begin{align}
\label{del:fval:2}
\delta f_q &= c + \left[\ln q_{\rm val}(x)\right]' x(1-x)h(x),\\
h(x) &= \frac{(1-\lambda)(1-x)+\lambda\bar s/M^2}{(1-x)^2-\bar s/M^2},
\end{align}
where $c=\partial\ln C_v/\partial\ln p^2$ and
$\lambda=\partial\ln\Lambda_v^2/\partial\ln p^2$ taken at
$p^2=M^2$. It should be noted that \eq{del:fval:2} is independent
of the specific choice of the profile function $\phi$. 

\begin{figure}[htb]
\begin{center}
\tightspace 
\epsfig{file=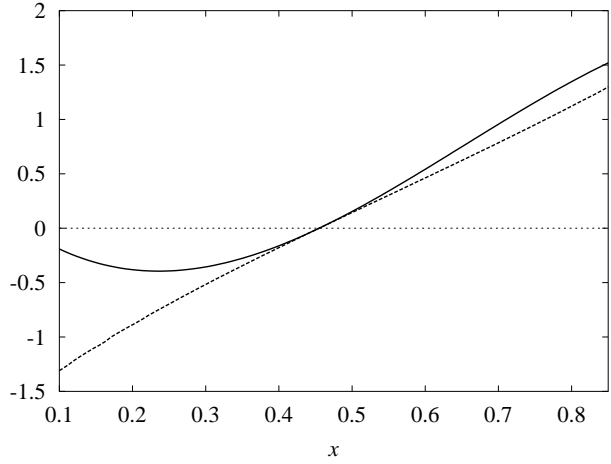,width=\figx}
\caption{%
Phenomenological off-shell function $\delta f_2(x)$ (solid) in comparison 
with $\delta f_q(x)$ (dashed) computed using \eq{del:fval:2} as 
described in text. }
\label{fig:delfmodel}
\tightspace
\end{center}
\end{figure}

We use \eq{del:fval:2} in order to reproduce phenomenological function
$\delta f_2$ at large $x$. In particular, we fix the parameters $c$
and $\lambda$ in order to reproduce the zero of $\delta f_2$ at large $x$
($x_0$) and the slope $\delta f_2'(x_0)$. Using $\bar s=2.1\gevsq$ we
obtain $\lambda= 1.03$ and $c=-2.31$. The function $\delta f_q(x)$ 
by \eq{del:fval:2} is shown in Fig.\ref{fig:delfmodel}
together with the phenomenological function $\delta f_2(x)$. One observes 
that this simple model agrees with phenomenology at large $x$
but not at small $x$ at which effect of the nucleon sea is important.

The positive sign of the parameter $\lambda$ suggests that the scale
parameter $\Lambda_v$ decreases in nuclear environment since $p^2<M^2$
for bound nucleon. This in turn indicates the increase in the nucleon
core $r_c$ in nuclear environment (``swelling'' of bound nucleon). In
order to quantitatively estimate this effect we consider the relative
change in the nucleon radius $\delta r_c/r_c$. We have $\delta
r_c/r_c\sim -\frac12 \delta\Lambda_v^2/\Lambda_v^2$. The relative
change in the scale $\Lambda_v$ can be estimated as 
$\delta\Lambda_v^2/\Lambda_v^2=\lambda \langle p^2-M^2\rangle/M^2$,
where averaging is taken over bound nucleons.
We evaluate this quantity 
using our model spectral function for iron and obtain $\sim 9\%$ increase 
in $r_c$.

To conclude this Section we remark that the swelling of bound nucleons was 
discussed in the context of quenching of nuclear longitudinal response 
function in \cite{noble}. The change of confinment scale in nuclei in 
terms of a different approach was discussed in the context of the EMC 
effect in \cite{close85,Frankfurt:nt}. The 
swelling effect was experimentally constrained to $<30\%$ from the analysis 
of Coulomb sum in \cite{chen}.
%
%
\begin{figure}
\begin{center}\tightspace
\epsfig{file=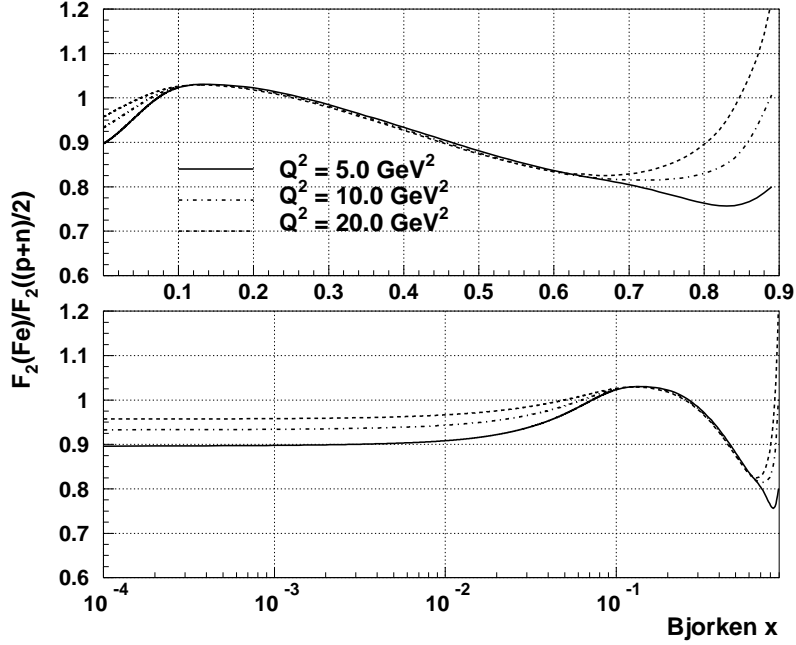,width=\figxx}
\caption{%
Our predictions for the ratio
of $^{56}$Fe and isoscalar nucleon structure functions calculated for
$Q^2=5,10,20~\gevsq$. The calculation takes into account the
non-isoscalarity correction for iron by \eq{nuke:FA}.
}
\label{fig:mf2fef2pn}\tightspace
\end{center}
\end{figure}

\subsection{The $Q^2$ and $A$ dependence}
\label{sec:qsq}

In order to illustrate the $Q^2$ dependence of $\mathcal{R}_2$ calculated in our approach, in
Fig.~\ref{fig:mf2fef2pn} we plot the ratio $\mathcal{R}_2({\rm Fe}/N)$, where $N$ is
isoscalar nucleon $(p+n)/2$, as a function of $x$ for a few fixed $Q^2$.
We observe from Fig.~\ref{fig:mf2fef2pn} significant variations of the
ratio $\mathcal{R}_2$ with $Q^2$ at small $x<0.1$ and large $x>0.65$. This $Q^2$
dependence can be attributed to several effects. In the nuclear shadowing
region at small $x$ the $Q^2$ dependence of the ratio $\mathcal{R}_2$ is due to the
corresponding dependence of effective cross section $\sigmabar_T$ (see
\eq{eq:effxsec} and Fig.~\ref{fig:functions}). It must be also noted that
in the region of $x$ between $0.01$ and $0.1$ the $Q^2$ dependence of
$\mathcal{R}_2$ is affected by the $Q^2$ dependence of longitudinal correlation
length $1/k_L$ (see Section \ref{smallx-nuke} and \Eqs{mult:D}{C2A}). For
$0.1<x<0.65$ the $Q^2$ dependence of $\mathcal{R}_2$ is negligible. At large $x$ the
$Q^2$ dependence is due to the target mass correction effect by \eq{TMC} in
convolution equations.
%
%
\begin{figure}[htp]
\begin{center}
\tightspace \vspace{-5mm}
\epsfig{file=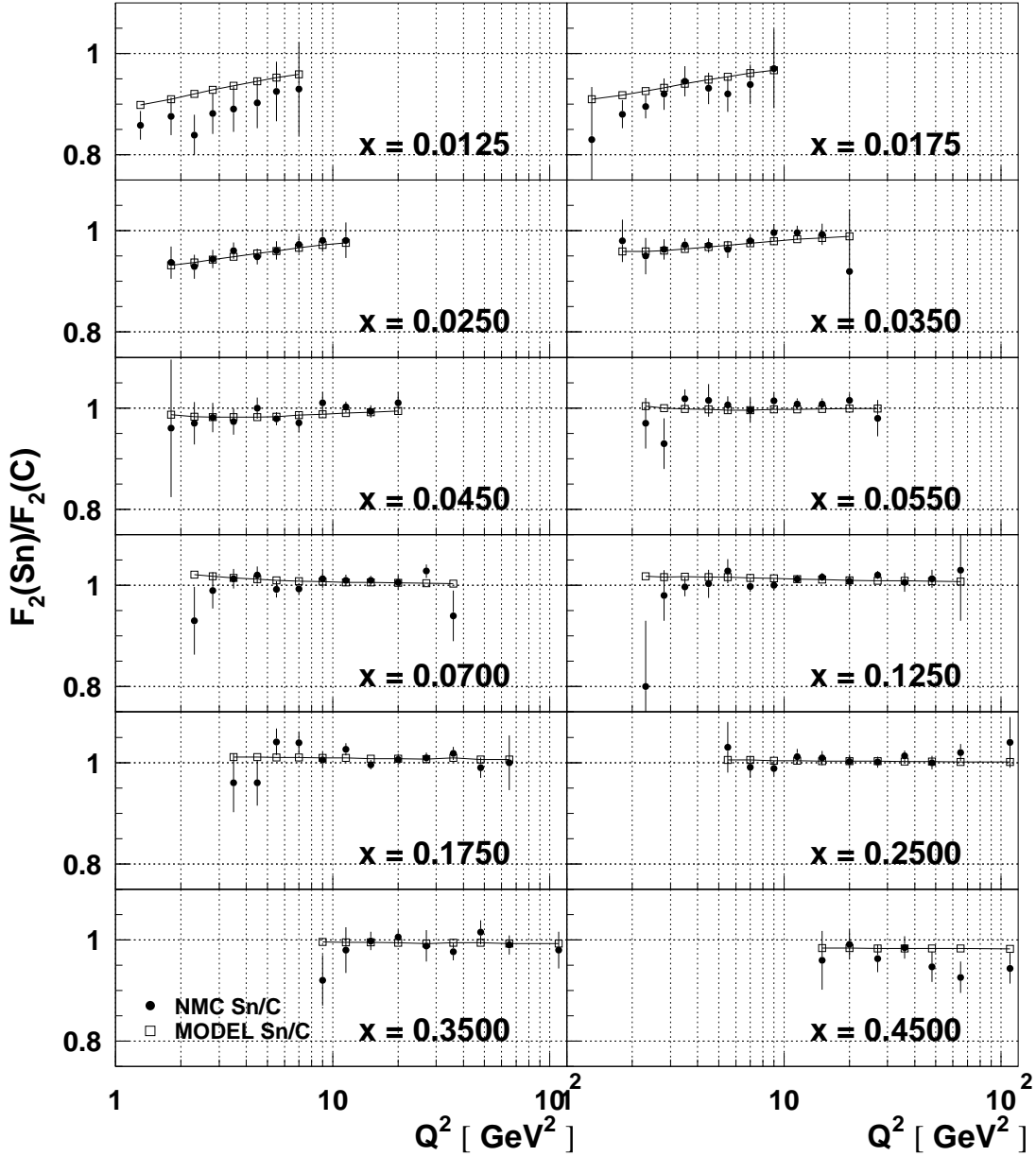,width=0.95\textwidth}
\caption{%
The $Q^2$ dependence of the ratio $\mathcal{R}_2(\mathrm{Sn/C})$ for
different values of $x$ as measured by the NMC \cite{NMCpb}. The curves with
open squares show the corresponding model calculations. For data points
the error bars correspond to the sum in quadrature of statistical and
systematic uncertainties, while the normalization uncertainty is not
shown.}
\label{fig:sncq2}
\tightspace
\end{center}
\end{figure}

In Fig.~\ref{fig:sncq2} we compare the NMC data on $Q^2$ dependence of the
ratio $\mathcal{R}_2({\rm Sn}/{\rm C})$ with our calculations. We observe an overall
good agreement between data and model calculations for all values of $x$
within available region of $Q^2$.
However, it should be remarked that available data on $Q^2$ dependence of nuclear effects
are still too scarce to make thorough phenomenological studies of this effect.
In particular, the correlation between $x$ and $Q^2$ for fixed
target experiments and the lack of information about the $Q^2$
distributions of data in each of the $x$ bins used (typically only the
average $Q^2$ is provided) can potentially bias the calculations where a
significant $Q^2$ dependence is expected.

\begin{figure}[htb]
\begin{center}
\tightspace
\epsfig{file=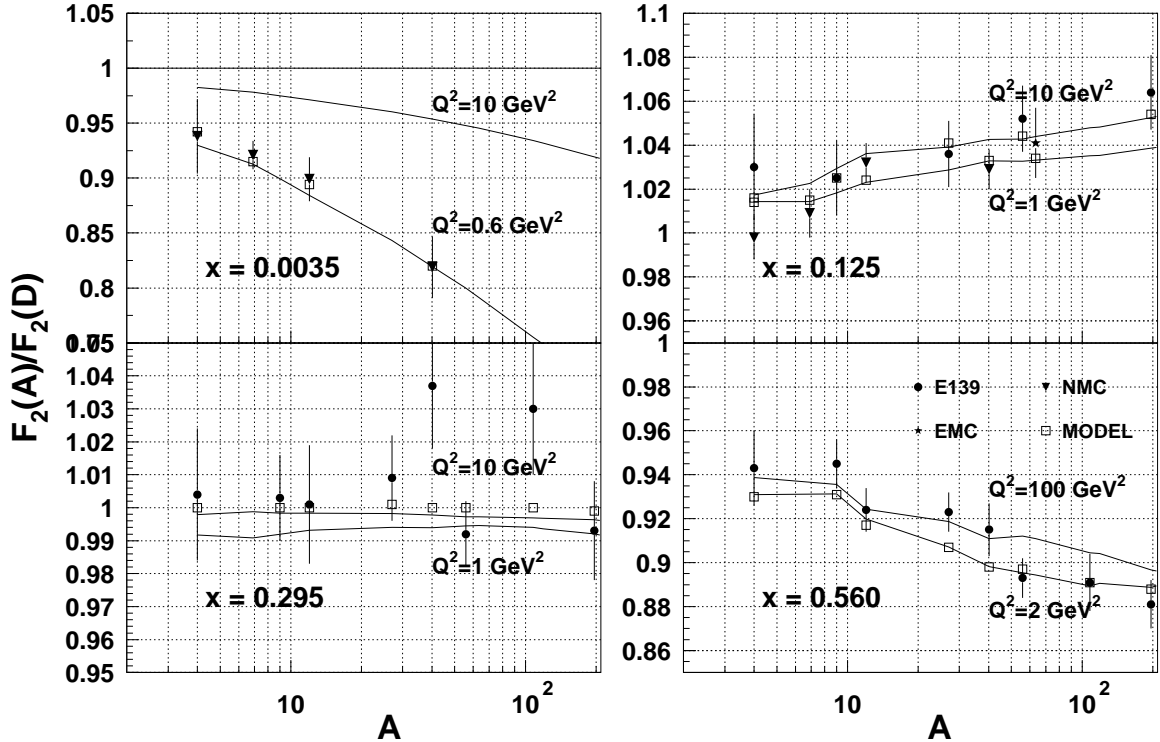,width=\textwidth}
\caption{%
The ratio $\mathcal{R}_2(\mathrm{A/D})$ as a function of $A$ for different
values of $x$. The open squares show the model calculations corresponding
to the average $Q^2$ of the data points. For data points the error bars
correspond to the sum in quadrature of statistical and systematic
uncertainties, while the normalization uncertainty is not shown. Two
curves calculated at constant $Q^2$ values are also shown for comparison.}
\label{fig:adep}
\tightspace
\end{center}
\end{figure}

Figures~\ref{fig:ratios1} and~\ref{fig:ratios2} show that the model
reproduces correctly the ratios $\mathcal{R}_2$ over a wide range of nuclei
and kinematic regions. The $A$ dependence of the ratio $\mathcal{R}_2$ is
illustrated in Fig.~\ref{fig:adep} for a few fixed values of $x$. At small
$x$ the A dependence is related to the multiple scattering coefficients in
\Eqs{C2A}{C3A}, through the nucleon number density distributions.
The increase in the nuclear shadowing effect with $A$ has "geometrical''
origin and can be attributed to the rising size of heavy nuclei. At large $x$ the
$A$ dependence of the ratio $\mathcal{R}_2$ is determined by the corresponding
dependence of parameters of nuclear spectral function. The slope of $\mathcal{R}_2$
as a function of $x$ at intermediate $x=0.5$--$0.6$ increases with $A$
because of the corresponding increase in the average separation and
kinetic energy of bound nucleons.
It is interesting to note that at
$x\approx 0.3$ the ratio $\mathcal{R}_2$ depends on neither $A$ nor $Q^2$.

%
%

\vspace{-5mm}
\subsection{Nuclear effects in Deuterium}
\label{sec:deuterium}

Understanding of nuclear effects in deuterium is an important issue since
deuterium data are often used as the source of information on the neutron
structure functions. As explained in Sec.\:\ref{sec:fits}, the determination
of $d$ and $u$ parton distributions is sensitive to nuclear corrections
to deuterium data (Section~\ref{phenom-sect}). In this Section we apply
our model with the parameters fixed from fit to data from heavy nuclei
(see Table~\ref{tab:fits})
in order to calculate nuclear
modifications in deuterium and compare our predictions with data.
We take into account nuclear binding, Fermi motion, off-shell,
nuclear pion and shadowing corrections as explained in Sec.\:\ref{nuke-sect}.
It should be emphasized that our approach does
not require any extrapolation from heavy nuclei to deuterium.

\begin{figure}[htb]
\begin{center}
\tightspace
\epsfig{file=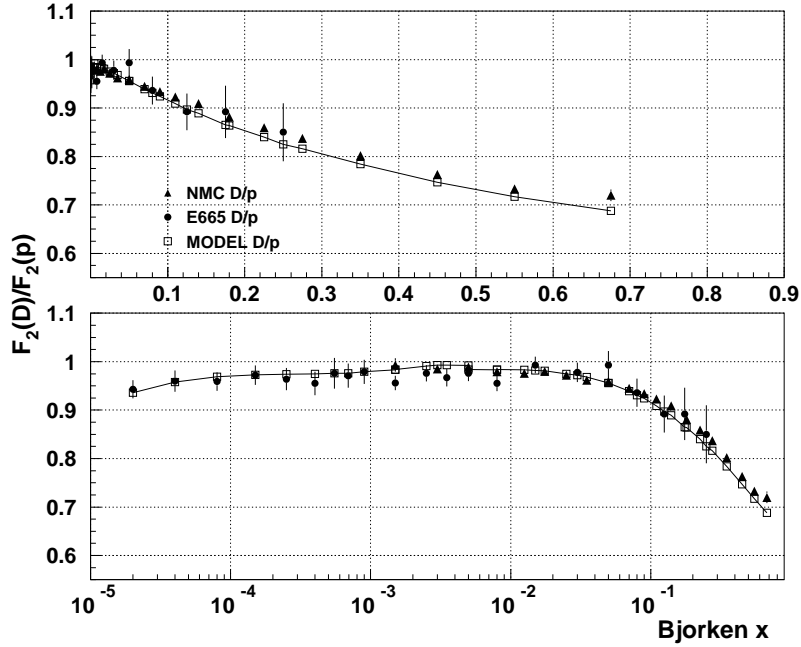,width=\figxx}
\caption{%
Comparison of E665 and NMC data to our calculations (curve with
open squares) for the ratio $\mathcal R_2(x,\mathrm{D}/p)$. For data 
points the error bars correspond to the sum in quadrature of statistical 
and systematic uncertainties, while the normalization uncertainty is not 
shown.
}
\label{fig:f2df2p}
\tightspace
\end{center}
\end{figure}

The ratio of the deuteron and the proton structure functions 
$\mathcal{R}_2(\mathrm{D}/p) = F_2^D/F_2^{p}$ was measured by the E665 and NMC 
collaborations \cite{NMCdp,E665dp} in a wide kinematical region of $x$ and 
$Q^2$. A comparison with these data provide a good test of the 
applicability of our model to D, since these data were not used in the 
fits described in Sec.\:\ref{sec:fits}. In Fig.~\ref{fig:f2df2p} we show 
the E665 and NMC data together with the results of our calculations. A 
good agreement is found between data and the model described in this 
paper. In particular, our prediction of a small shadowing effect in D 
seems to be supported by the measured values of $\mathcal{R}_2(\mathrm{D}/p)$ at 
small values of $x$. We note that the ratio $\mathcal{R}_2(\mathrm{D}/p)$ 
also provides a test of the parton distributions used in our calculation 
and in particular of the difference between $d$ and $u$ quark contents. 
This was not the case for all the remaining data listed in 
Table~\ref{tab:data} which were corrected by experiments for 
the neutron excess, thus providing an effective cancellation of PDFs in the 
ratios $\mathcal{R}_2(A^{\prime}/A)$ (Section~\ref{sec:fits}).
%
%
\begin{figure}[htb]
\begin{center}
\tightspace
\epsfig{file=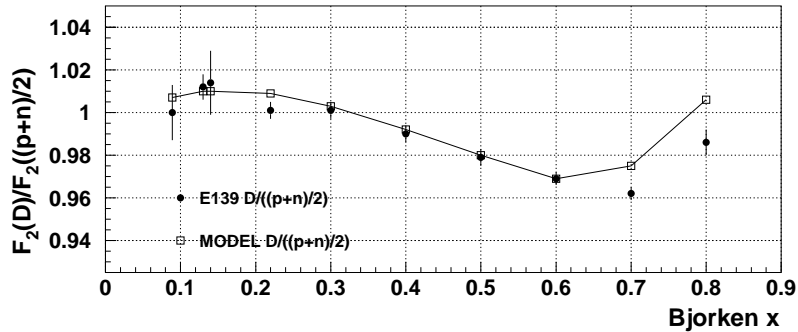,width=\figxx}
\caption{%
Comparison of the extrapolations of E139 nuclear data within
the nuclear density model of Ref.\cite{Gomez} to our calculations (curve
with open squares) for the ratio $\mathcal{R}_2(x,{\rm D}/{\rm N})$.
The error bars of the E139 data points correspond to the sum in quadrature of
statistical and systematic uncertainties. For each $x$ value, the model
calculation was performed at the average $Q^2$ of the experimental points
quoted in Ref.\cite{Gomez}.}
\label{fig:f2df2pn}
\tightspace
\end{center}
\end{figure}
\begin{figure}[htb]
\begin{center}
\tightspace
\epsfig{file=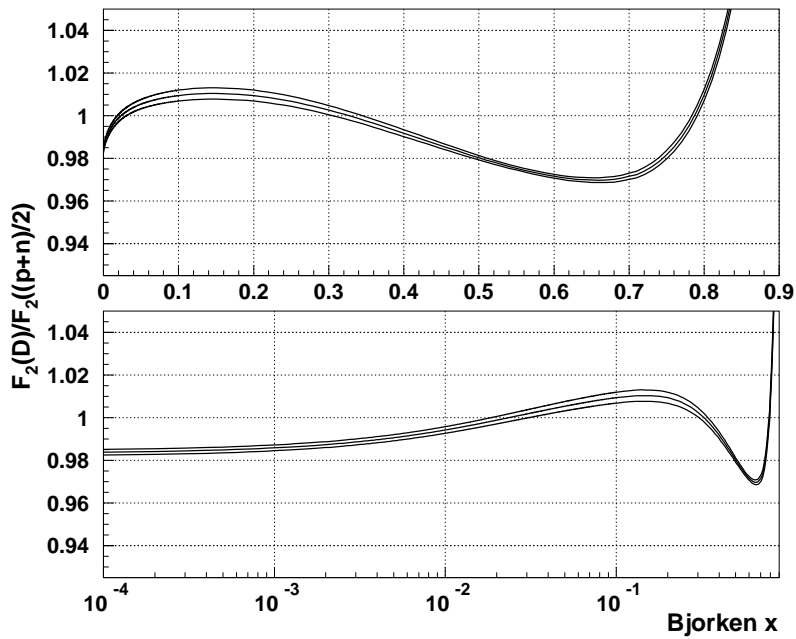,width=\figxx}
\caption{%
Our predictions for the ratio $\mathcal{R}_2(x,{\rm D}/{\rm N})$ of
deuterium to isoscalar nucleon at a fixed $Q^2=10~\gevsq$. The $\pm
1\sigma$ band is also given, including both statistical and systematic
uncertainties.}
\label{fig:mf2df2pn}
\tightspace
\end{center}
\end{figure}

Unlike the ratio $\mathcal{R}_2(\mathrm{D}/p)$ the ratio
$\mathcal{R}_2(\mathrm{D}/N) = F_2^D/F_2^{p+n}$ cannot be measured
directly because a free neutron target is not available.
The extraction of $\mathcal{R}_2(\mathrm{D}/N)$ from SLAC data
was discussed in Ref.\cite{Gomez} in terms of
a phenomenological model of the EMC effect in the deuterium.
In Ref.\cite{Gomez} the ratio
$\mathcal{R}_2(\mathrm{D}/N)$ was extracted by extrapolating the
measured ratios $\mathcal{R}_2(A/\mathrm{D})$ using the nuclear density model of
Ref.\cite{Frankfurt:nt}. The key assumption was made
that the quantity $\mathcal{R}_2(A/N)-1$ scales as
nuclear number density and it was also assumed that this ratio is
independent of $Q^2$. The values of $\mathcal{R}_2(\mathrm{D}/N)$ were
given in \cite{Gomez} for $x$ corresponding to the $x$ bins of SLAC
data. The results are shown in Fig.~\ref{fig:f2df2pn} together with
our calculation of the ratio $\mathcal{R}_2(\mathrm{D}/N)$ for the
same kinematics of the points presented in \cite{Gomez}.%
\protect\footnote{%
The theoretical uncertainties of such
extrapolation were not estimated in \cite{Gomez}. See also discussion
of these points in \cite{AKL03}.  }
In Fig.~\ref{fig:mf2df2pn} we show our prediction for the ratio
$\mathcal{R}_2(\mathrm{D}/N)$ at fixed $Q^2=10\gevsq$ and the corresponding
uncertainty band ($\pm 1\sigma$), including model systematics.


\section{Applications}
\label{sec:appl}

In this Section we apply our results to evaluate nuclear parton
distributions (Section~\ref{sec:npdf}) and make the predictions of nuclear
effects for neutrino structure functions (Section~\ref{sec:nuint}).

\subsection{Nuclear parton distributions}
\label{sec:npdf}

The parton distributions are process-independent characteristics of
the target in high-energy processes.
Different phenomenological approaches to the extraction of
nuclear PDFs (nPDFs) can be found in
Refs.~\cite{npdfs1,npdfs2,npdfs3,npdfs4}.
It should be remarked at this point that physics observables are the
cross sections and the structure functions, which include
contributions from all twists. The higher-twist terms are generally
process-dependent, can be essential in the region of relatively low
$Q^2$, and, furthermore, can be substantially affected by nuclear
environment. Therefore, the applicability of the leading twist
approximation must be considered in comparison with data as well as in
any attempt to extract nPDFs.
In our approach we derive nPDFs from the
analysis of nuclear structure functions
(Sections~\ref{sec:fitres} and~\ref{sec:inter}) that
allows us to determine both nPDFs and their uncertainties from
existing data. 
However, this paper is not aimed at the full nPDF analysis,
which will be published elsewhere~\cite{KP05}. We rather want to
discuss a few different effects which cause modifications of nuclear
quark distributions. The numerical results shown in this Section were
obtained using the NNLO proton and neutron PDFs
described in Sec.\:\ref{phenom-sect}.

\subsubsection{Nuclear convolution}
\label{sec:npdf:convol}

As discussed in Sec.\:\ref{nuke-sect}, in the region of high $Q^2$ and
large $x$ the nuclear structure functions can be approximated by
incoherent contributions from different nuclear constituents which can
be presented in a convolution form (see \Eqs{convol:N}{pion:conv}).
The convolution formulas look similar for all type of nuclear
structure functions suggesting that the convolution equations hold for
the parton distributions. We denote $q_{a/T}(x,Q^2)$ the distribution of
quarks of type $a$ in a target $T$. Then the quark distribution in a
nucleus can be written as
\begin{equation}
\label{npdf:conv}
q_{a/A}(x,Q^2) = \sum_{c=p,n,\pi}f_{c/A}\otimes q_{a/c},
\end{equation}
where the function $f_{c/A}(y,v)$ can be interpreted as the
distribution of particles of type $c$ in a nucleus over light-cone
momentum $y$ and invariant mass (virtuality) $v$ (for bound nucleons
and nuclear pions see \Eqs{ydist:N}{pion:f}, respectively).
The operation $f\otimes q$ denotes the convolution
\begin{equation}
\label{conv:def}
f \otimes q = \int_{x<y}\frac{\ud y\ud v}{y} f(y,v) q(x/y,Q^2,v).
\end{equation}
Equations similar to (\ref{npdf:conv}) can be written for antiquark and
gluon distributions in nuclei. Note also that the distribution functions
are independent of $Q^2$ and, therefore, the $Q^2$ evolution of nuclear
PDFs is goverened by the evolution of PDFs of nuclear constituents.

In view of applications to complex nuclei with different number of protons and neutrons,
it is usefull to sort out the contributions to the convolution equation
according to \emph{isospin}.
Let us consider the isoscalar and isovector quark distributions, $q_0=u+d$
and $q_1=u-d$. We first address the contributions from bound protons and
neutrons to nuclear quark distributions.
Assuming exact
isospin invariance of PDFs in the proton and neutron we have simple
relations between the isoscalar and the isovector distributions in the proton and
the neutron
\begin{subequations}
\label{q:01:np}
\begin{align}
q_{0/p}(x,Q^2) &= q_{0/n}(x,Q^2),\\
q_{1/p}(x,Q^2) &= -q_{1/n}(x,Q^2).
\end{align}
\end{subequations}
Using these relations we observe that the quark distributions with
different isospin decouple in the convolution equation. In
particular, for the isoscalar ($q_{0/A}$)
and the isovector ($q_{1/A}$) nuclear quark distributions we have
\begin{subequations}
\label{q:01:A}
\begin{align}
q_{0/A}(x,Q^2) &= A\,f_0 \otimes q_{0/p},\\
q_{1/A}(x,Q^2) &= (Z-N)\,f_1 \otimes q_{1/p},
\end{align}
\end{subequations}
where $f_0$ and $f_1$ are the isoscalar and the isovector nucleon
distributions in a nucleus. These distributions are given by
\eq{convol:N} with the spectral functions $\mathcal{P}_0$ and
$\mathcal{P}_1$ defined by
\eq{spfn:01}. Note that the distributions $f_{0}$ and $f_{1}$ are
normalized to unity.

Let us now discuss the pion contribution to \eq{npdf:conv}. Similar to the
nucleon case, we assume the isospin relations for quark distributions
in the pion: $q_{0/\pi^+} = q_{0/\pi^-} = q_{0/\pi^0}$ and
$q_{1/\pi^+} = -q_{1/\pi^-}$ and $q_{1/\pi^0}=0$.
Using these relations we have for the pion correction to
the isoscalar and isovector nuclear quark distributions
\begin{subequations}
\label{q:01:A:pi}
\begin{align}
q_{0/A}^\pi(x,Q^2) &= f_{\pi/A} \otimes q_{0/\pi},\\
q_{1/A}^\pi(x,Q^2) &= (f_{\pi^+/A}-f_{\pi^-/A}) \otimes q_{1/\pi}.
\end{align}
\end{subequations}
Here in the first equation $f_{\pi/A}$ is the sum of the
distributions over all pion states. It should be emphasized that in
\eqs{q:01:A:pi} the pion distributions refer to nuclear pion excess,
since scattering off virtual pions emitted and absorbed by same
nucleon (nucleon pion cloud) are accounted in the proton and neutron
PDFs. For the calculation of nuclear pion distributions in our model
see Sec.\:\ref{sec:pion:details}. 
\begin{figure}[htb]
\begin{center}
\epsfig{file=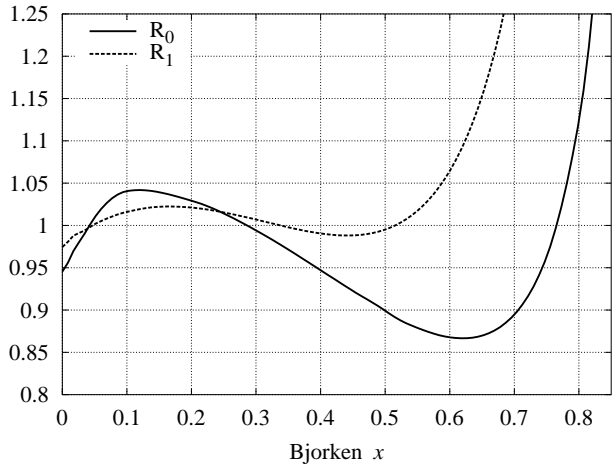,width=\figx}
\caption{%
Nuclear effects for the isoscalar and the isovector quark
distributions in ${}^{56}$Fe. 
The ratios $\mathcal{R}_0$ and $\mathcal{R}_1$ (see 
text) were calculated for the valence quark distributions at 
$Q^2=20\gevsq$. Nuclear shadowing and pion corrections are also included 
for the isoscalar distribution.  }
\label{fig:R01}
\tightspace
\end{center}
\end{figure}

The isovector component should vanish in isoscalar nuclei with $Z=N$.%
\footnote{It should be remarked that this statement applies to nuclear
states with the total nuclear isospin 0. If higher-isospin states are
present for a $Z=N$ nucleus, then the isovector distribution $q_{1/A}$
may be non-zero. The discussion of these issues is postponed for
future studies.  }
However, for a generic nucleus with different number of protons and
neutrons both the isoscalar and the isovector distributions are
present.  Heavy nuclei typically have a small excess of neutrons over
the protons and the distributions $f_0$ and $f_1$ are quite different
in such nuclei, as discussed in Section \ref{A-sect}. For this reason
nuclear effects in PDFs depend on isospin. In order to illustrate this
statement we calculate the ratios $\mathcal{R}_0=q_{0/A}(x)/(A\,q_{0/p}(x))$
and $\mathcal{R}_1=q_{1/A}(x)/[(Z-N)q_{1/p}(x)]$ for the iron nucleus using
the proton PDFs of Ref.\cite{a02}. The results are shown in
Fig.~\ref{fig:R01}. We note that the full nuclear correction
is shown in case of $q_{0/A}$, \ie the calculation includes
effect of nuclear spectral function, off-shell correction, nuclear
pion and shadowing effects. However, for the isovector quark
distribution $q_{1/A}$ we neglect possible nuclear pion and
shadowing effects.

\subsubsection{Nuclear shadowing}
\label{sec:npdf:sh}

In this Section we discuss coherent nuclear effects in the context of
parton distributions. To this end we want to apply the approach discusses
in Section \ref{sec:smallx:A}.
The multiple scattering effects are generally different for different
PDFs. We specify this statement by considering nuclear effects for quark
distributions of different $C$ parity, $q^{(\pm)}(x)=q(x)\pm \bar q(x)$.
In order to simplify discussion we consider the isoscalar
(anti)quark distributions, $q=u+d$ and $\bar q=\bar u+\bar d$.
The $C$-odd distribution is in fact the valence quark distribution in the
target $q^{(-)}=q_{\mathrm{val}}$. The $C$-even distribution at small $x$
describes the target quark sea.

In order to bridge between Sec.\:\ref{smallx-nuke} and the present
discussion we recall that the structure function $F_1$ in the LT
approximation is given by $C$-even distribution $q^{(+)}$.
The structure function $F_1$ is transverse helicity structure function
(to be more precize, the average over left- and
right-polarized transverse helicity structure function, see
\eq{SF:hel:nu}).
At small $x$, as discussed in Sec.\:\ref{smallx-nuke}, nuclear effects
are described by the propagation of hadronic component
of virtual boson with the proper helicity state in nuclear environment.
Equation (\ref{sh:T:A}) applies in the case of $q^{(+)}$.

Similarly, the structure function $F_3$ in the LT approximation is given 
by $C$-odd (valence) distribution $q^{(-)}$. In terms of helicity 
structure functions this is the asymmetry between left- and 
right-polarized states. Therefore, nuclear corrections to $q^{(-)}$ at 
small $x$ can be described by \eq{asym:A:2}.
We have for coherent nuclear corrections to $q^{(+)}$ and $q^{(-)}$ quark
distributions
\begin{subequations}\label{sh:pdf}
\begin{align}
\label{sh:qpl}
\delta \mathcal{R}^{(+)} &= \frac{\delta q_A^{(+)}(x)}{q_N^{(+)}(x)} =
\Re(a_T^2\mathcal{C}_2^A)/\Im a_T,\\
\label{sh:qmn}
\delta \mathcal{R}^{(-)} &= \frac{\delta q_A^{(-)}(x)}{q_N^{(-)}(x)} =
[2 \Re({\Delta}a\, a_T \mathcal{C}_2^A) - 
\Im({\Delta}a\, a_T^2\mathcal{C}_3^A)]/\Im {\Delta}a,
\end{align}
\end{subequations}
where $\mathcal{C}_2^A$ and $\mathcal{C}_3^A$ are given by
\Eqs{C2A}{C3A} with effective transverse scattering amplitude
$a_T=(i+\alpha_T)\sigmaeff/2$.
Equation (\ref{sh:qmn}) determines the nuclear shadowing effect for 
valence quark distribution $\delta \mathcal{R}_{\mathrm{val}}=\delta 
\mathcal{R}^{(-)}$. The amplitude ${\Delta}a$ describes the left-right 
asymmetry in the transverse amplitude. In other terms ${\Delta}a$ can be 
interpreted as the difference between $\bar qN$ and $qN$ scattering 
amplitudes \cite{Kulagin:1998wc}. As discussed in 
Sec.\:\ref{sec:smallx:A} the correction $\delta \mathcal{R}^{(-)}$ does 
not depend on the specific value of the cross section asymmetry 
$\Delta\sigma$ but does depend on $\alpha_{\Delta}=\Re {\Delta}a/\Im 
{\Delta}a$. The rate of nuclear effects for both $C$-even and $C$-odd 
distributions is determined by transverse amplitude $a_T$.
Nuclear shadowing effect for antiquark
distributions can readily be derived from \Eqs{sh:qpl}{sh:qmn}
and we have 
\begin{equation}
\delta \mathcal{R}_{\mathrm{sea}} = \frac{\delta\bar q_A(x)}{\bar q_N(x)}
= \delta \mathcal{R}^{(+)} + \frac{q_{\mathrm{val}/N}(x)}{2\bar q_N(x)}
        \left(\delta \mathcal{R}^{(+)} - \delta \mathcal{R}^{(-)}\right).
\label{sh:qbar}
\end{equation}

\begin{figure}[htb]
\begin{center}
\epsfig{file=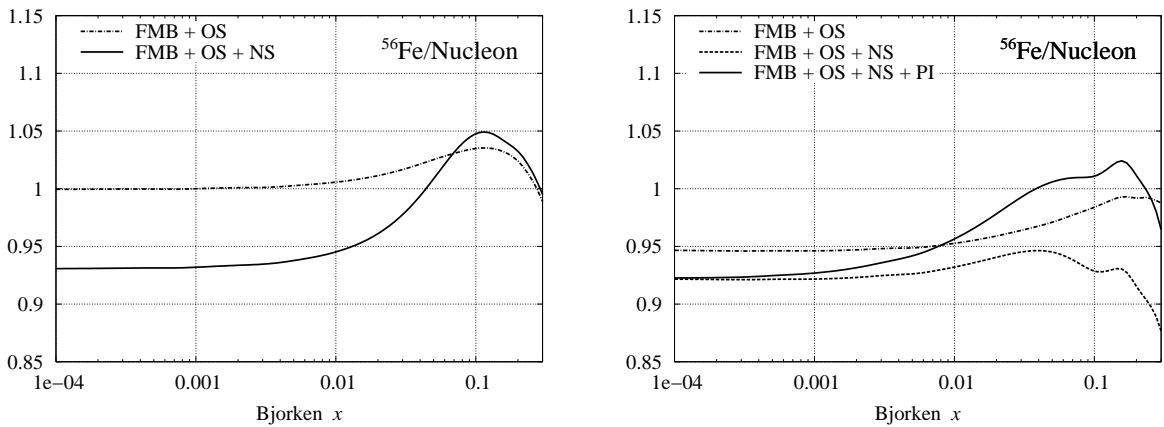,width=\textwidth}
\caption{
Nuclear effects for isoscalar valence and sea quark distributions
calculated for iron nucleus at $Q^2=20\gevsq$ (see text). The left
panel displays different contributions to $\mathcal{R}_{\text{val}}$:
the dot-dashed curve if only the effect of nuclear
spectral function (FMB) and off-shell (OS) corrections are taken into
account, the full curve is overall nuclear correction including
nuclear shadowing effect (NS). The right panel displays similar
contributions to $\mathcal{R}_{\text{sea}}$. The full curve also
includes the nuclear pion effect (PI), which is absent for the
valence quark distribution.
}
\label{fig:npdf}
\tightspace
\end{center}
\end{figure}

The results of calculation of nuclear effects for valence quark and
antiquark distributions are reported in Fig.~\ref{fig:npdf}. The
calculations account of the effects of smearing with nuclear spectral
function (FMB), off-shell corrections (OS), nuclear shadowing (NS),
and nuclear pion (PI) corrections. The FMB, OS, and PI corrections
have been computed as discussed in Sec.\:\ref{sec:npdf:convol} using
our model spectral function, pion distribution function
and off-shell
correction described in Sec.\:\ref{sec:model}. The NS correction for
valence and sea distributions are computed by \Eqs{sh:qmn}{sh:qbar}
using the parameters of effective scattering amplitude derived from
our fits (see Sections~\ref{sec:fitres} and \ref{sec:inter}).

A few remarks are in order.
At small $x<0.01$ the NS effect for valence quark distribution is
enhanced relative to that for nuclear sea. The underlying reason for
that is the enhancement of multiple scattering corrections for the
cross section asymmetry as discussed in Sec.\:\ref{sec:smallx:A}. If
we keep only the double scattering correction then the ratio $\delta
\mathcal{R}_{\mathrm{val}}/\delta \mathcal{R}_{\mathrm{sea}}$ is given by 
\eq{sh:3:D}.
The OS correction is negative in this region. However, the combined
effect of FMB and OS is somewhat different for valence and sea
distributions as displayed in Fig.~\ref{fig:npdf}. This is attributed
to different $x$ dependence of valence and sea in the nucleon which
affect the result of the averaging with nuclear spectral function.
Nevertheless, in spite of these differences, the overall nuclear
corrections are similar for valence and sea for $x<0.01$.%
\footnote{
Note that this discussion refers to a high $Q^2\sim 20\gevsq$.
At lower $Q^2$ the balance between different nuclear effects change.
}

One observes that nuclear corrections for valence and sea
distributions are different in the antishadowing region. The
antishadowing effect for valence (\ie  positive nuclear correction) is
a joint effect of two corrections both of which are positive: (1) the
FMB and OS corrections and (2) the constructive interference in the
multiple scattering effect which is due to a finite real part
$\alpha_{\Delta}$ of the effective scattering amplitude in the $C$-odd channel
(for this reason shadowing becomes antishadowing, see the left panel
of Fig.~\ref{fig:npdf}).
For sea-quark distribution in the antishadowing region we observe
a cancellation between different effects. In this respect we remark that
the contribution of the last term in \eq{sh:qbar} becomes increasingly
important at $x>0.05$, because of the ratio
$q_{\mathrm{val}/N}(x)/\bar q_N(x)$. This term is negative in this
region and cancels a positive nuclear pion contribution. As a result
the overall nuclear correction to antiquark distribution is small for
$0.02<x<0.2$. Note that this agrees with the results of E772
experiment, in which no enhancement of nuclear sea was observed in DY
nuclear processes \cite{Drell-Yan}.

It should be noted that the calculation of the relative nuclear correction for
valence quark distribution is stable with respect to the choice of the
PDF set for entire region of $x$ (see also Fig.~\ref{fig:R01} for 
nuclear correction to valence distributions). Nuclear effects for
sea quarks also depend weakly on the particular choice of PDF for small 
$x$. However, at high $x$ the calculation of nuclear effects for 
antiquark distributions has larger uncertainties and the result is 
sensitive to both the shape and the magnitude of the nucleon antiquark 
distribution (note the val/sea ratio in \eq{sh:qbar}).

\subsection{Neutrino interactions with nuclei}
\label{sec:nuint}

\begin{sidewaysfigure}[p]
\begin{center}
\epsfig{file=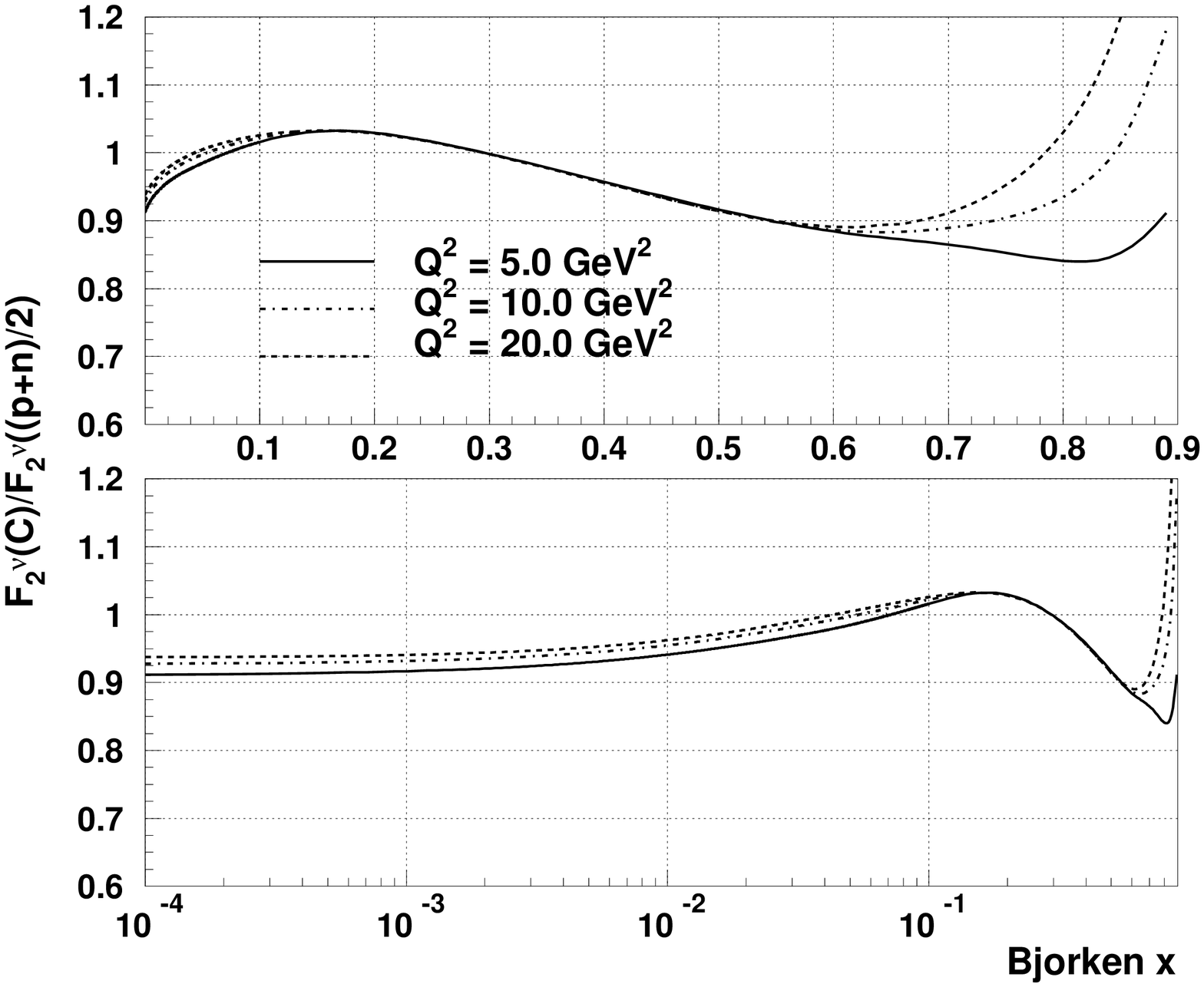,width=10.0cm}%
\epsfig{file=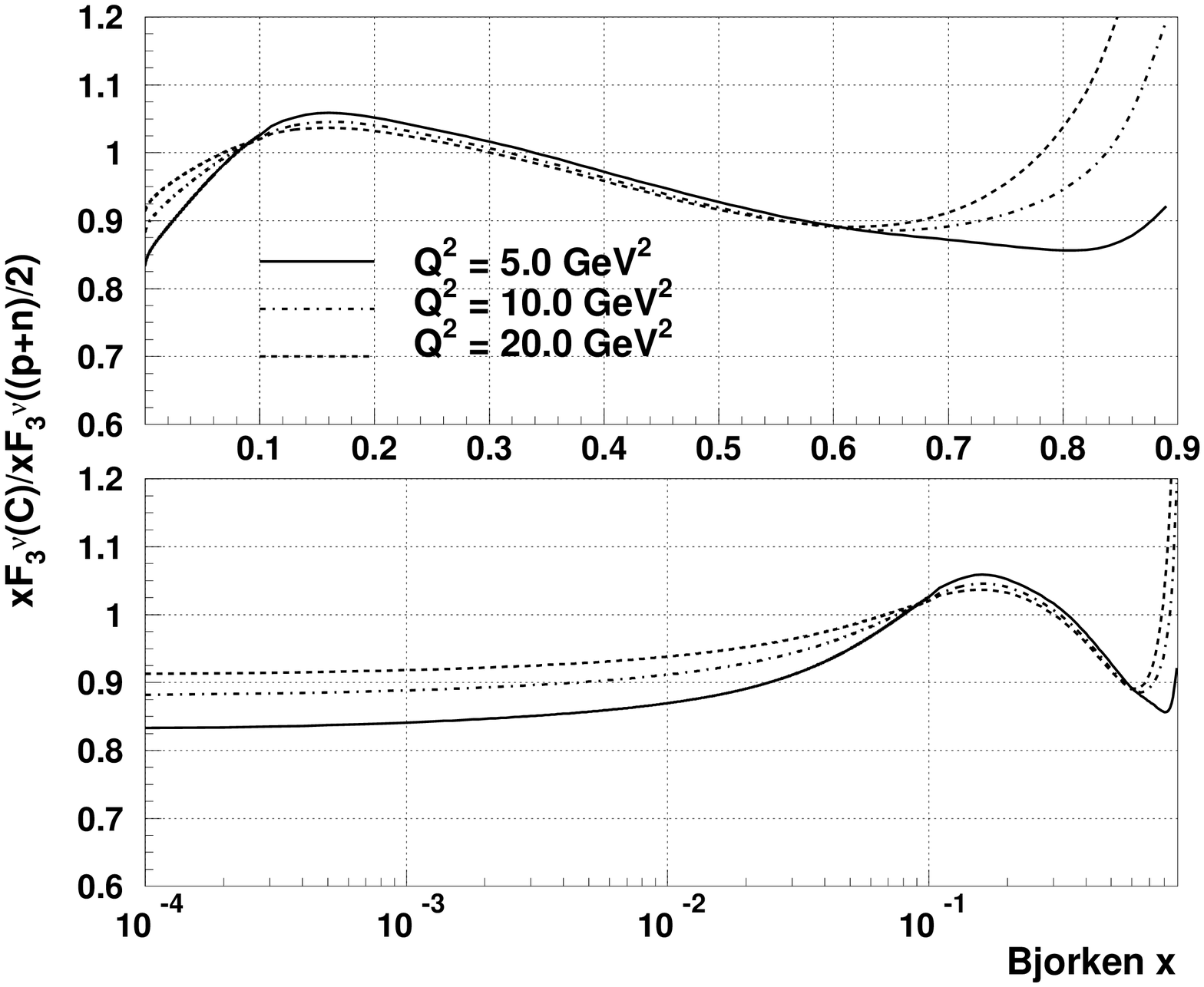,width=10.0cm}
\epsfig{file=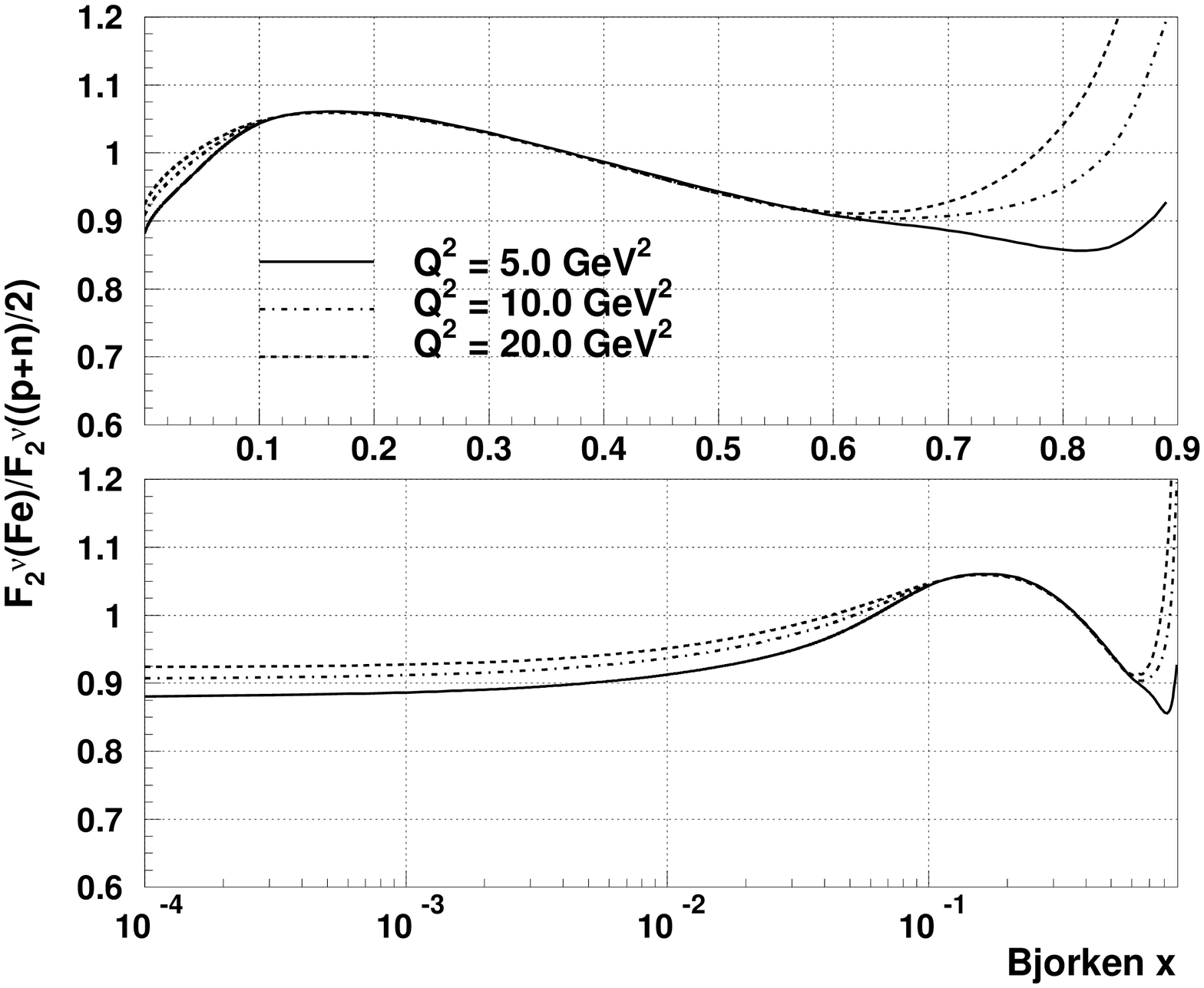,width=10.0cm}%
\epsfig{file=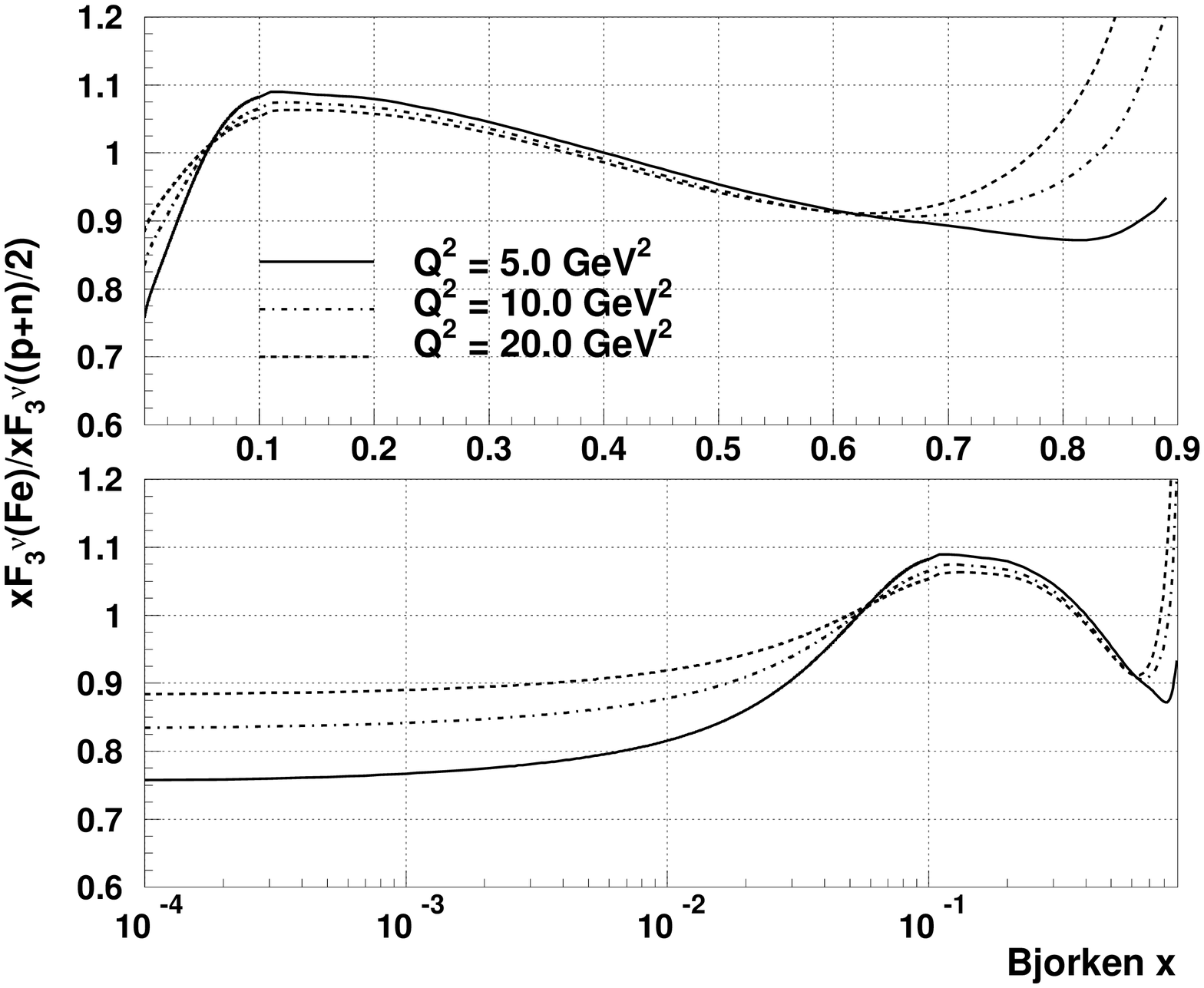,width=10.0cm}
\caption{Our predictions for ratios of $F_{2}$ (left plots) and $xF_{3}$
(right plots) for neutrino scattering on ${}^{12}$C and ${}^{56}$Fe and
the corresponding values on isoscalar nucleon (p+n)/2.
The curves are drawn for $Q^2=5,10,20~\gevsq$ and take
into account the non-isoscalarity correction.}
\label{fig:nuratios1}
\end{center}
\end{sidewaysfigure}
\begin{sidewaysfigure}[p]
\begin{center}
\epsfig{file=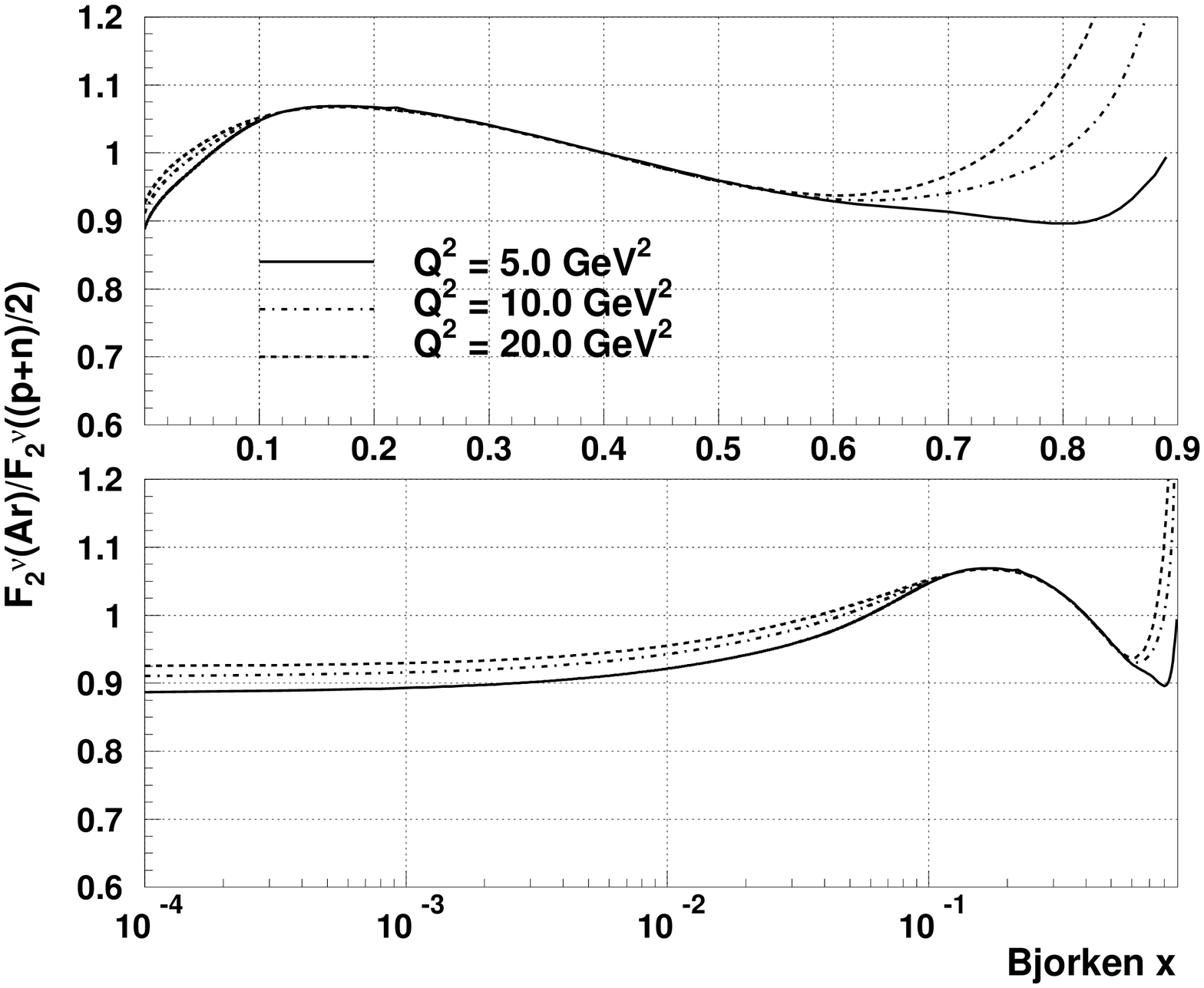,width=10.0cm}%
\epsfig{file=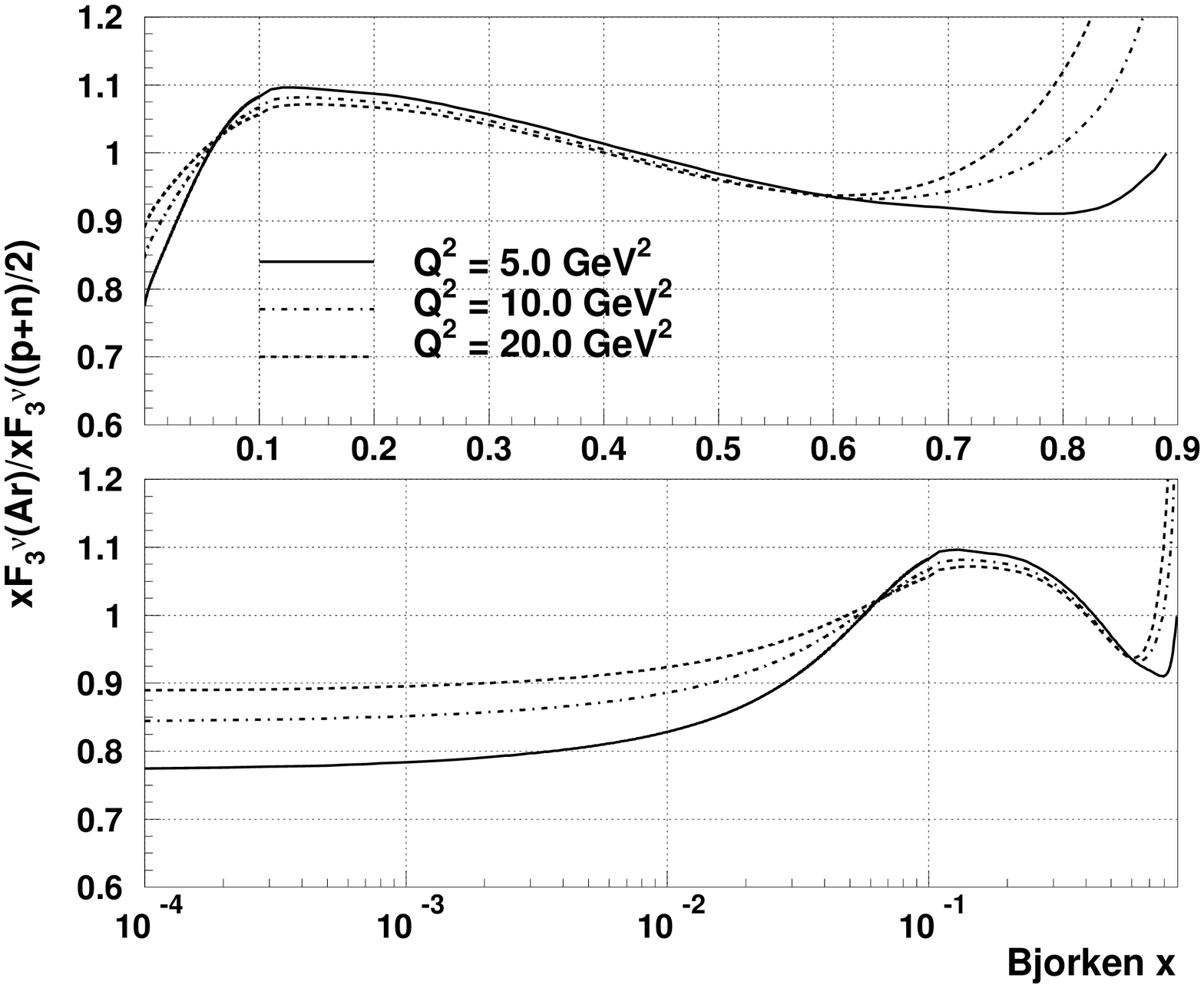,width=10.0cm}
\epsfig{file=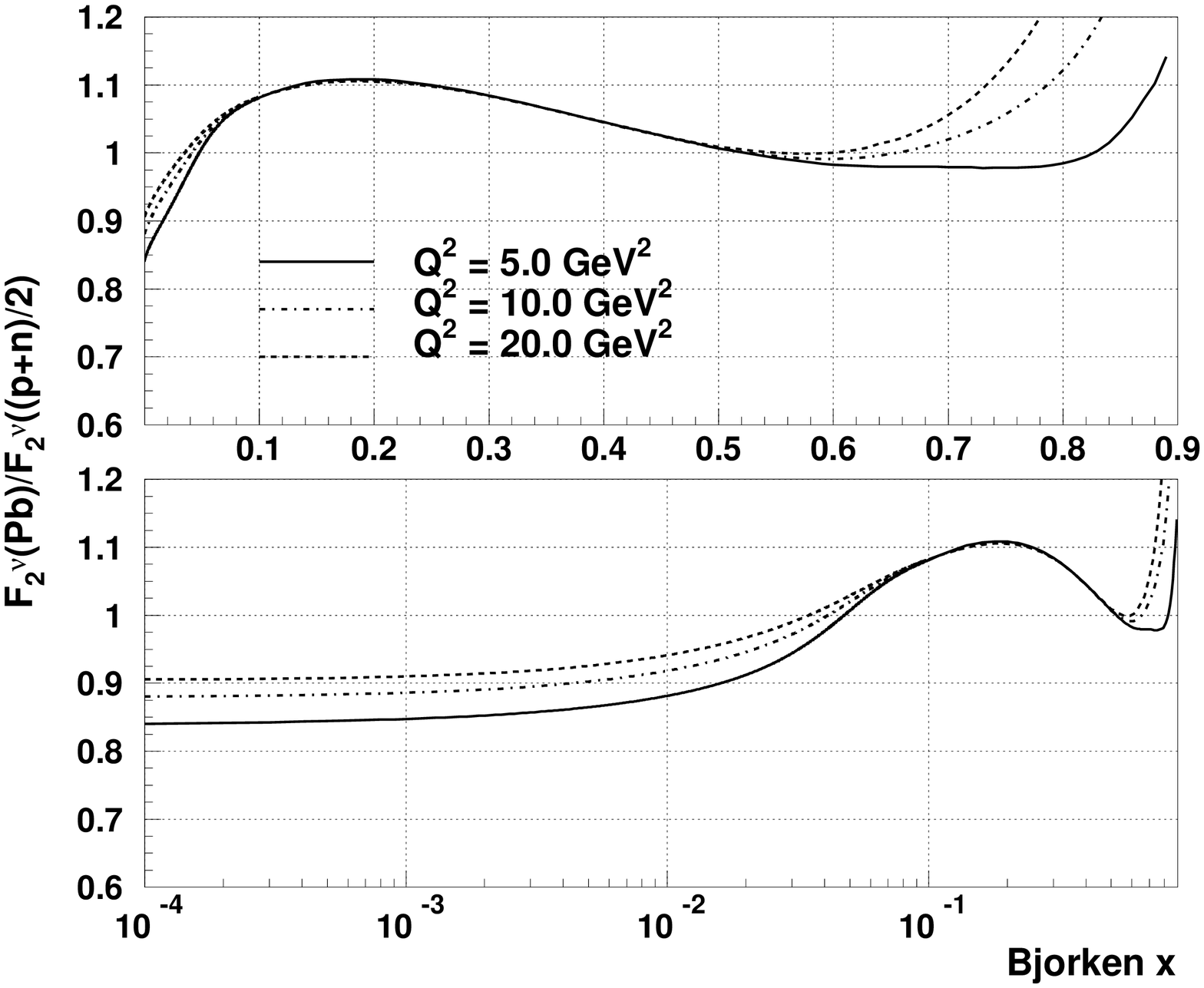,width=10.0cm}%
\epsfig{file=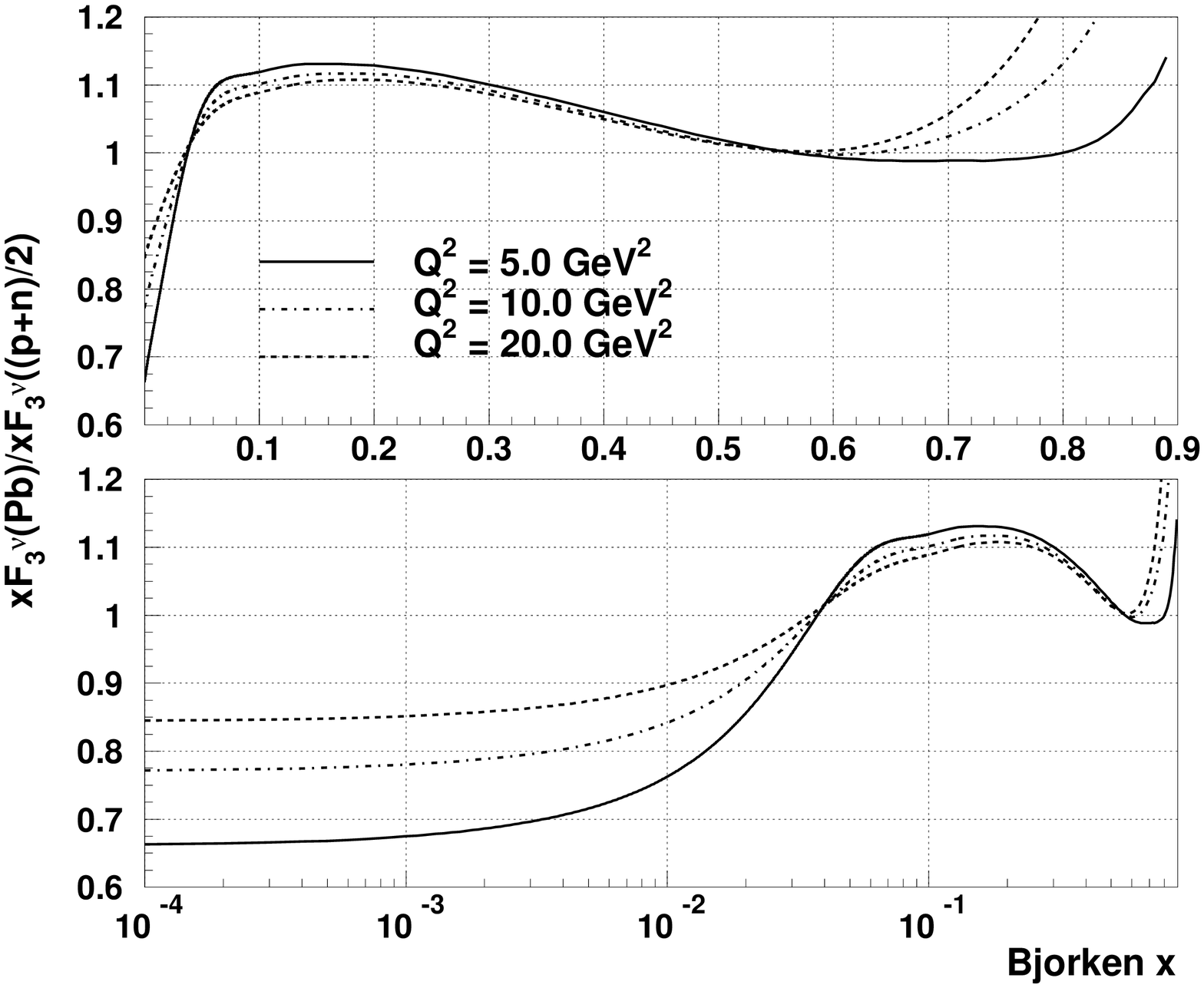,width=10.0cm}
\caption{Our predictions for ratios of $F_{2}$ (left plots) and $xF_{3}$
(right plots) for neutrino scattering on ${}^{40}$Ar and ${}^{207}$Pb
and the corresponding values on isoscalar nucleon (p+n)/2. The curves
are drawn for $Q^2=5,10,20~\gevsq$ and take into account the
non-isoscalarity
correction.}
\label{fig:nuratios2}
\end{center}
\end{sidewaysfigure}

In this Section we calculate nuclear effects for neutrino
charged-current structure functions using the
approach developed in the previous sections.
The study of neutrino interactions is particularly
interesting to this end since they are flavour sensitive
and they are strongly influenced by the structure function
$F_{3}$, which is not present in the electromagnetic
case. We also note that due to the
low interaction probability in practice the detection
of neutrinos always requires heavy nuclear targets.
Therefore, the knowledge of nuclear effects is crucial for
understanding of neutrino cross-sections.

In order to compute corrections to $F_2^\nu$ and $F_3^\nu$
related to the averaging with nuclear spectral function (FMB and OS effects)
we apply \Eqs{FA-2}{FA-3} and use the off-shell
function $\delta f_2$ extracted from the analysis of
Sec.\:\ref{sec:fitres} for both $F_2$ and $F_3$.
Nuclear shadowing/antishadowing corrections are computed as discussed
in Sec.\:\ref{smallx-nuke}.

We focus here on the region of relatively high momentum transfer
$Q^2>5\gevsq$ and assume that coherent nuclear interactions driven by
axial current are similar to those of vector current at large $Q^2$
and that they can be described by the effective amplitude extracted
from the analysis of Sections~\ref{sec:xsec}
to \ref{sec:syst}.%
\footnote{Note that the interactions of the axial-vector current at low $Q^2$ are
essentially different from those of the vector current. This region requires
a special analysis which goes beyond the scope of the present paper.}
A detailed study of nuclear effects in (anti)neutrino interactions including
the low $Q^2$ region will be the subject of a future publication~\cite{KP05}.

We calculate the ratios $\mathcal{R}_2^{\nu}=F_2^{\nu A}/(A F_2^{\nu N})$ and
$\mathcal{R}_3^\nu=xF_3^{\nu A}/(A xF_3^{\nu N})$, where $N$ denotes the isoscalar
nucleon (averaged over proton and neutron), for the most common nuclear
targets used by recent neutrino experiments: ${}^{12}$C
(NOMAD~\cite{nomad}), ${}^{56}$Fe (NuTeV~\cite{nutev},
MINOS~\cite{minos}), ${}^{40}$Ar (ICARUS~\cite{icarus}) and ${}^{207}$Pb
(OPERA~\cite{opera}, CHORUS~\cite{chorus}).
Our results for both $F_2$ and $xF_3$ are shown in
Figures~\ref{fig:nuratios1} and~\ref{fig:nuratios2} for different values
of $Q^2$. We briefly comment on the main features that distinguish the
nuclear corrections in neutrino DIS from the ones in charged-lepton DIS
($F_2^\mu$). By comparing Fig.~\ref{fig:mf2fef2pn} and \ref{fig:nuratios1}
we observe that nuclear effects for $F_2^\nu$ and $F_2^\mu$ in the
coherent region are similar (note that we restrict the present discussion
to relatively high $Q^2$). However, at large $x$ nuclear effects for
$F_2^\nu$ and $F_2^\mu$ are somewhat different. In particular, we note
that $\mathcal{R}_2^\nu > \mathcal{R}_2^\mu$ in the dip region of $x\sim 0.6-0.8$. 
This is because the neutron excess correction is positive for $F_2^\nu$, 
while it is negative for $F_2^\mu$.
From Fig.~\ref{fig:nuratios1} and \ref{fig:nuratios2} one can also observe
that nuclear effects at large $x$ are similar for neutrino $F_2$ and
$xF_3$. However, at small $x$ the nuclear shadowing effect for $xF_3$ is
systematically larger as follows from \eq{sh:3:D}.

\begin{figure}[htb]
\begin{center}
\tightspace
\epsfig{file=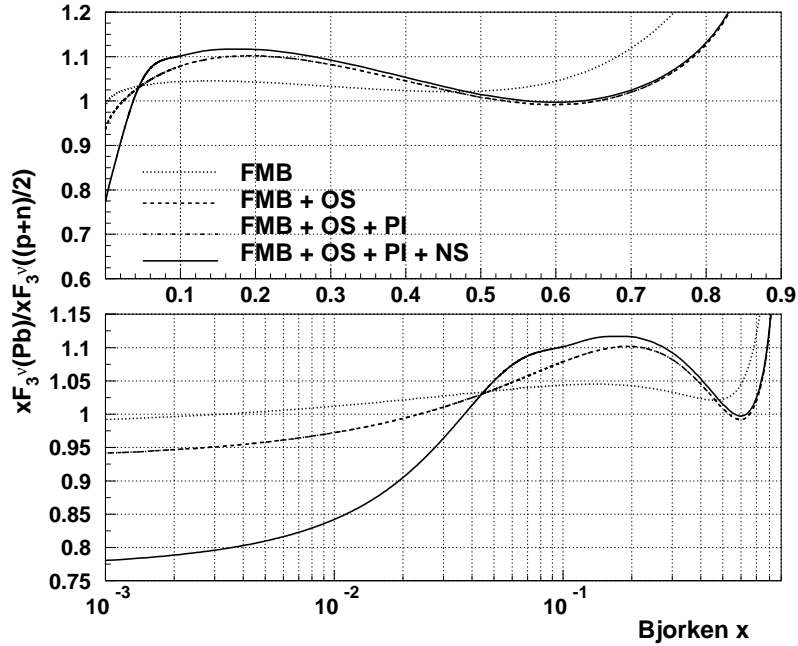,width=\figxx}
\caption{%
Comparison of different nuclear effects calculated for neutrino $xF_{3}$
at $Q^2=10\gevsq$ for ${}^{207}$Pb target. The labels on the curves
correspond to the effects included in turn: the averaging with nuclear
spectral function (FMB), off-shell correction (OS), nuclear pion
excess (PI) and coherent multiple-scattering correction (NS). The
calculation takes into account the target mass and the neutron excess
corrections. Note that $xF_{3}$ is not corrected for pion excess effect
(overlapping dashed and dashed-dotted curves).
}
\label{fig:xF3contsPb}
\tightspace
\end{center}
\end{figure}

Figure~\ref{fig:xF3contsPb} illustrates different nuclear corrections
to $xF_3^\nu$ for ${}^{207}$Pb target computed at fixed $Q^2$.  The
enhancement at intermediate $x$ values is a joined effect of all
considered nuclear correction (see also Fig.~\ref{fig:npdf} and
discussion in Sec.\:\ref{sec:npdf:sh}). In the case of $F_2$ the
``antishadowing'' at $x\sim 0.1$ is due to off-shell and nuclear pion
corrections.%
\protect\footnote{%
See also Fig.~\ref{fig:F2contsAu} for $F_2^\mu$. Note, however, that
the neutron excess correction has a different sign for $F_2^\mu$ and
$F_2^\nu$ that explains the differences between the magnitude of
nuclear effects in Figs.~\ref{fig:F2contsAu} and \ref{fig:nuratios2}.
}
Note that the nuclear pion excess effect can be neglected in the case of
$xF_3$, in contrast to the case of $F_2$. Indeed, in the isoscalar nucleus 
the pion correction depends on pion structure functions averaged 
over different pion states and $F_3^\pi$ vanishes after such averaging. A 
small isovector correction which is proportional to $\pi^+ - \pi^-$ 
asymmetry in the nuclear pion distribution functions (see \eq{q:01:A:pi}) 
is also neglected.

The study of $xF_{3}$ is particularly important since it allows to
test the normalization of nuclear valence quark number.
As discussed in
Sections~\ref{sec:valnorm} and~\ref{sec:inter} the conservation
of the nuclear valence quark number was used in our analysis in order
to test the balance between nuclear shadowing and off-shell effects.
The valence quark (baryon) number of the target is related to the
integral of neutrino and antineutrino averaged $F_3$, the
Gross--Llewellyn-Smith (GLS) sum rule~\cite{Gross:1969jf}. We remark,
however, that in QCD this relation is not exact and only holds in the
leading twist and the leading order in $\as$ and is corrected by both
the radiative \cite{gls:rad} and the higher-twist
effects. It would be interesting to experimentally address the
question of nuclear modification of the GLS sum rule.  New measurements
of $xF_{3}$ from neutrino and antineutrino scattering off different
nuclei would help to clarify this issue, provided they can reach a
precision comparable to the size of the effects we observe in our
analysis (typically 1\%, see Fig.~\ref{fig:normval}).

We conclude this Section by remarking that
in spite of the major interest of neutrinos as a probe for nuclear effects,
virtually no precise experimental information is available so far in DIS
region. The only direct measurements of nuclear effects on neutrino DIS
cross-sections were performed by BEBC~\cite{BEBC} (${}^{20}$Ne/D) and
CDHSW~\cite{CDHSW} (${}^{56}$Fe/p). However, these results are affected by
large statistical and systematic uncertainties. It should be emphasized
that neutrino DIS provides information complementary to that of the
charged-lepton scattering and, therefore, the completion of new
high-statistics measurements would have a large impact on our
understanding of nuclear effects. 
The NOMAD experiment~\cite{nomad} collected large neutrino samples on
${}^{12}$C, ${}^{27}$Al and ${}^{56}$Fe targets allowing a study of nuclear 
effects from ${}^{27}$Al/${}^{12}$C and ${}^{56}$Fe/${}^{12}$C 
ratios~\cite{ichep04}. The recent NuTeV cross-section data~\cite{nutev-xsec} 
also provide information on nuclear effects in ${}^{56}$Fe. In addition, the CHORUS 
experiment~\cite{chorus} is extracting neutrino cross-sections  
from the interactions collected on ${}^{207}$Pb.  
Table~\ref{tab:nudata} summarizes the various (anti)neutrino data samples. 

\begin{table}[htb]
\begin{center}
\begin{tabular}{l|c|c|c|c}
Experiment & Targets & CC statistics ($\times 10^6$) & Data taking & Reference \\ \hline\hline 
NOMAD & ${}^{12}{\rm C}$  & $1.3(0.06) \;\nu (\bar{\nu})$  & 1995-1998 & \cite{nomad} \\ 
      & ${}^{27}{\rm Al}$ & $1.5(0.07) \;\nu (\bar{\nu})$  &  &  \\ 
      & ${}^{56}{\rm Fe}$ & $12.5(0.6) \;\nu (\bar{\nu})$  &  &  \\ \hline 
CHORUS & ${}^{207}{\rm Pb}$ & $1.3(0.3) \;\nu (\bar{\nu})$  & 1998 & \cite{chorus} \\ \hline 
NuTeV & ${}^{56}{\rm Fe}$ & $2.7(1.2) \;\nu (\bar{\nu})$  & 1996-1997 & \cite{nutev} \\ \hline 
\end{tabular}
\caption{
List of recent nuclear data which can be used to study nuclear effects on  
neutrino structure functions.  
\label{tab:nudata}
}
\end{center}
\tunespace
\end{table}


\section{Summary}
\label{sec:concl}

We presented a detailed phenomenological study of unpolarized 
nuclear structure functions for a wide kinematical region of $x$ and 
$Q^2$. A general approach was developed which, on one side, 
includes the main nuclear corrections and, on the other side, 
provides a good description of data on nuclear structure functions.
We take into account the QCD treatment of  
the nucleon structure functions and address a number of nuclear 
effects including nuclear shadowing, Fermi motion and nuclear binding, nuclear 
pions and off-shell corrections to bound nucleon structure functions.

Starting from a relativistic approach in the description of nuclear
DIS we then exploited the fact that characteristic
energy and momentum of bound nucleon are small compared to
the nucleon mass. This allowed us to compute nuclear corrections in
terms of nonrelativistic nuclear spectral function, the quantity which
is well constrained by data at low- and intermediate-energy
regions.
Our analysis suggested that data cannot be quantitatively explained
in impulse approximation by applying ``standard'' Fermi motion and
nuclear binding corrections even at large $x$.
This motivated us to address the off-shell effect in bound nucleon
structure functions.
This correction was parametrized in terms of a few parameters which 
were extracted from data, together with their uncertainties.
The effective scattering amplitude which determines the magnitude of 
nuclear shadowing effect was also addressed phenomenologically.
 
It should be emphasized that the phenomenological parameters of our  
model refer to the nucleon structure and for this reason they are
common to all nuclei. We verified this hypothesis by extracting them 
from different subsets of nuclei. Overall, we obtained an excellent
agreement between our calculations and data by using only three
independent parameters.

Our results show that inclusive nuclear DIS data have a good
sensitivity to off-shell effects, allowing a precise determination of
this correction. We also note that the study of semi-inclusive
nuclear DIS in which the kinematics of the active nucleon can be
controlled by selecting certain final states would provide 
additional information on the off-shell effect.
The off-shell effect is related to the modification of the nucleon
structure in nuclear environment. This relation was discussed in terms
of a simple model in which the off-shell effect at large $x$ was
linked to the modification of the bound nucleon core radius. We found
that the off-shell correction derived from our analysis favours the
increase in the nucleon core radius in nuclear environment.

We studied in detail the $Q^2$ and $A$ dependencies of nuclear corrections.
One important application was the calculation of nuclear effects for 
deuterium, which is of primary interest for the problem of the 
extraction of the neutron structure functions.
We also applied our model to study nuclear valence and 
sea quark distributions, as well as the flavour (isospin) dependence of 
nuclear effects.

Another important application was the calculation of nuclear structure 
functions for neutrino scattering. In the present paper we evaluated 
nuclear corrections for charged-current neutrino structure functions for 
relatively high $Q^2$, which are relevant for the analysis of 
existing DIS neutrino data.
More detailed studies of neutrino and antineutrino interactions for both 
charged-current and neutral-current scattering are planned in future 
publications.

\section*{Acknowledgments}

We would like to thank M. Arneodo for useful information on NMC data,
S. Alekhin, A. Butkevich and A. Kataev for fruitful discussions 
on different stages of this work.
The work of S.K. was partially supported by the Russian Foundation for 
Basic Research grant 03-02-17177 and by INTAS project 03-51-4007.


\appendix
\section{Integration in nuclear convolution}
\label{apndx:conv}

The integration in convolution formulas is constrained by the requirement
that the invariant mass of bound nucleon and the virtual photon is
high enough for producing physical final states. In particular, 
for the region of invariant masses of final states larger than a given mass
$M_X$ the required relation is
\begin{equation}
W^2 \ge M_X^2,
\label{ineq:0}
\end{equation}
where $W^2=(p+q)^2$ and $p$ the four-momentum of the bound nucleon. The 
threshold of inelastic channels corresponds to $M_X=M+m_\pi$ and by setting 
$M_X=M$ we take into account the elastic channel. Here we 
discuss in detail the constraints on the integration region in the 
convolution formulas due to \eq{ineq:0}. Note that \eq{ineq:0} is 
equivalent to
\begin{equation}
p_0+q_0 \ge E_X,
\label{ineq:1}
\end{equation}
where $E_X=\left(M_X^2+(\bm{p}+\bm{q})^2\right)^{1/2}$ and $p_0=M+\ceps$.
Using this equation we can write the integral over the bound nucleon
four-momentum in convolution formulas as
\begin{equation}
\int\ud^4 p\: \theta(W^2-M_X^2)=
\int{\ud^3\bm{p}}
\int\limits_{\lefteqn{\scriptstyle E_X-q_0-M}}{\ud\ceps}.
\label{int:1}
\end{equation}
This equation should be applied together with the nuclear spectral function
and other functions which enter the convolution formulas. The energy
integration in \eq{int:1} corresponds to the integration over the excitation
energies of the residual nucleus.

We first consider the spectral function
\begin{align}
\label{model:spfn}
\mathcal{P}(\ceps,\bm{p}) &=2\pi\delta(\ceps-\ceps_{\bm{p}}) n(\bm{p}),
\\
\ceps_{\bm{p}} &=\ceps_0 - \bm{p}^2/(2m_0).
\label{model:spectrum}
\end{align}
This is the relevant case for the deuterium, for which $\ceps_0=\ceps_D$ 
and $m_0=M$ (see \eq{spfn:D}), and also for the model spectral function 
$\mathcal{P}_{\mathrm{MF}}$ with $m_0=M_{A-1}$ the mass of the 
residual nucleus and $\ceps_0=-E^{(1)}$ the nucleon separation energy 
averaged over mean-field configurations of the residual 
nucleus (see \eq{model:PMF} and the discussion thereafter).

The energy integration in Eq.(\ref{int:1}) can easily be performed and
inequality (\ref{ineq:1}) then becomes
\begin{equation}
q_0+M+\ceps_{\bm{p}} \ge E_X.
\label{ineq:2}
\end{equation}
This inequality provides the constraints on the momentum
space in \eq{int:1}. In order to solve it explicitly we chose the
coordinate system such that the momentum transfer has only longitudinal
component $q=(q_0,\bm{0}_\perp,-|\bm{q}|)$.
Then after some algebra (\ref{ineq:2}) can be written as (we retain only
the terms linear in $\ceps_{\bm{p}}$)
\begin{equation}
\label{ineq:3}
{\bm{p}^2}/{(2m_*)} - p_z - p_* \le 0,
\end{equation}
where the notations are
\begin{subequations}\label{par:star}
\begin{align}
\gamma p_* &= M\left[1-x\left(1+\frac{\Delta}{Q^2}\right)
+\frac{\ceps_0\gamma_2}{M} \right],
\\
{m_*} &= \frac{m_0|\bm{q}|}{m_0+q_0+M},
\end{align}
\end{subequations}
and $\Delta=M_X^2-M^2$, $\gamma=|\bm{q}|/q_0$,
and $\gamma_2=1+M/q_0$.

Inequality (\ref{ineq:3}) is most easily solved in terms of
\emph{longitudinal} and \emph{transverse coordinates}, $\bm{p}=(\bm{p}_\perp,
p_z)$. In this case, the solution to (\ref{ineq:3}) can be written as
\begin{eqnarray}
\label{sol:tz}
\left\{
\begin{array}{rclcl}
        p_z^- &\le& p_z &\le& p_z^+, \\
        0 &\le& \bm{p}_\perp^2 &\le& T^2, \\
\end{array}
        \right.
\end{eqnarray}
where $T^2=m_*^2+2m_*p_*$ is maximum transverse
momentum squared of the bound nucleon (for the given kinematical
conditions) and $p_z^\pm$ correspond to those longitudinal momenta
at which the left side of (\ref{ineq:3}) is 0,
\begin{equation}
\label{pz:bounds}
p_z^\pm = m_* \pm \left(T^2-\bm{p}_\perp^2\right)^{1/2}
\end{equation}
The momentum integral in Eq.(\ref{int:1}) in terms of these variables is
\begin{eqnarray}
\label{int:tz}
\int\limits_{\lefteqn{\vphantom{\bigl(}\scriptstyle{
        W^2 \ge M_X^2}}} \ud^3\bm{p} =
\pi\,\theta(T)
\int_0^{T^2} \ud p_\perp^2 
\int_{p_{z}^-}^{p_{z}^+} \ud p_z.
\end{eqnarray}
The reqirement $T^2\ge 0$ gives the constraint on possible $x$ and
$Q^2$ in inelastic scattering off bound nucleon.
\protect\footnote{%
The equation $T^2=0$ determines the maximum possible $x$ which can be
achieved in DIS from bound nucleon. In application to the deuteron this
gives $x=3/2$ (neglecting $Q^{-2}$ terms and $\varepsilon_d/M$ corrections
in \eq{par:star}). This is different from the kinematical
maximum $x=M_D/M\approx 2$, which corresponds to elastic scattering from the
deuteron as a whole. We comment that the limit $x=3/2$ was
derived keeping linear terms in $\ceps/M$. However, 
the events with such large $x$ are due to high-momentum configurations 
$p\sim M$ in the wave function and, therefore, require fully relativistic 
description.
}

In \emph{spherical coordinates} we introduce the azimuthal angle
$\theta$ between the $z$-axis and the direction of the momentum
$p_z=p\cos\theta$ (here $p=|\bm{p}|$).
The solution to (\ref{ineq:3}) splits into two different regions with 
respect to the sign of $p_*$ and 
the momentum integral in
convolution formula can be written as
\begin{equation}
\label{int:sph}
\int
\limits_{\lefteqn{\vphantom{\bigl(}\scriptstyle{
        W^2 \ge M_X^2}}} \ud^3\bm{p} =
\left\{\
        \begin{array}{ll}
   \displaystyle{2\pi 
        \int \limits_{-1}^1
		\ud \cos\theta
        \int \limits_0^{\lefteqn{\vphantom{(}\scriptstyle{
		p_+(\cos\theta)} }}
                \ud p\:p^2, } 
& \quad\mathrm{if}\ p_* > 0, 
\\
   \displaystyle{2\pi
        \int \limits_{\lefteqn{\scriptstyle{c_*}}}^1
	\ud \cos\theta
        \int \limits_{\lefteqn{\scriptstyle{
	p_-(\cos\theta)}}}^{\lefteqn{\vphantom{(}\scriptstyle{
	p_+(\cos\theta)} }}
        \ud p\:p^2,}
& \quad\mathrm{if}\ {-}\dfrac{m_*}{2} \le p_* \le 0,
	\end{array}
\right.
\end{equation}
where $p_\pm(\cos\theta)$ are the values of $p$ at which the left side of
(\ref{ineq:3}) is 0
\begin{equation}
\label{p:bounds}
p_\pm(\cos\theta)=m_*\cos\theta \pm
\sqrt{(m_*\cos\theta)^2 + 2m_* p_*}
\end{equation}
and $c_*=(2|p_*|/m_*)^{1/2}$.
The first case in (\ref{int:sph}) applies if $x<1$ as can be readily seen 
from Eqs.(\ref{par:star}), while the last case concerns 
the regions $x\sim 1$ and $x>1$.

We now consider generic spectrum in Eq.(\ref{int:1}). 
The upper limit of energy 
integration is determined by the threshold separation energy $\eth$. We 
recall that in our notations the
separation energy $\ceps=E^A_0-E^{A-1}$, where $E^A_0$ is the ground state 
energy of the target nucleus and $E^{A-1}$ is the energy of the residual 
nucleus including the recoil energy. Therefore, for the given recoil 
momentum $\eth=\ceps_0-\bm{p}^2/(2m_0)$, where $\ceps_0=E^A_0-E^{A-1}_0$ is 
the difference of the ground state energies of the target and the residual 
nucleus and $m_0=M_{A-1}$ is the mass of the residual nucleus.
The constraints on the momentum space in \eq{int:1} directly follow from 
\begin{equation}
E_X-q_0-M \le \eth.
\label{ineq:4}
\end{equation}
This inequality, written in terms of $\ceps_0$ and $m_0$, is equivalent to 
(\ref{ineq:2}). Therefore, the discussion of (\ref{ineq:2}) can be taken over 
(\ref{ineq:4}). In particular, the solutions to (\ref{ineq:4}) in terms of 
the longitudinal and transverse momentum are given by Eqs.(\ref{sol:tz}) and 
(\ref{pz:bounds}). 
The integration in (\ref{int:1}) in terms of these variables can
be written as
\begin{equation}
\label{int1:tz}
\int\limits_{\lefteqn{\vphantom{\bigl(}\scriptstyle{
        W^2 \ge M_X^2}}} \ud^4{p} =
\pi\,\theta(T^2)
\int_0^{T^2} \ud p_\perp^2
\int_{p_{z}^-}^{p_{z}^+} \ud p_z
\int_{\lefteqn{\scriptstyle E_X-q_0-M}}^{\eth}\ud\ceps,
\end{equation}
where the limits of integration are similar to those in \eq{int:tz} and 
given by (\ref{par:star},\ref{sol:tz},\ref{pz:bounds}).

The integration in spherical coordinates is
\begin{equation}
\label{int1:sph}
\int\limits_{\lefteqn{\vphantom{\bigl(}\scriptstyle{
        W^2 \ge M_X^2}}} \ud^4{p} =
\left\{\
        \begin{array}{ll}
   \displaystyle{2\pi 
        \int \limits_{-1}^1
		\ud \cos\theta
        \int \limits_0^{\lefteqn{\vphantom{(}\scriptstyle{
		p_+(\cos\theta)} }}
                \ud p\:p^2
        \int \limits_{\lefteqn{\scriptstyle E_X-q_0-M}}^%
        {\lefteqn{\vphantom{(}\scriptstyle \eth}} \ud\ceps, } 
& \quad\mathrm{if}\ p_* > 0, 
\\
   \displaystyle{2\pi
        \int \limits_{\lefteqn{\scriptstyle{c_*}}}^1
	\ud \cos\theta
        \int \limits_{\lefteqn{\scriptstyle{
	p_-(\cos\theta)}}}^{\lefteqn{\vphantom{(}\scriptstyle{
	p_+(\cos\theta)} }}
        \ud p\:p^2
        \int \limits_{\lefteqn{\scriptstyle E_X-q_0-M}}^%
        {\lefteqn{\vphantom{(}\scriptstyle \eth}} \ud\ceps, }
& \quad\mathrm{if}\ {-}\dfrac{m_*}{2} \le p_* \le 0,
	\end{array}
\right.
\end{equation}
where the notations are similar to those in \eq{int:sph}.

\section{Multiple scattering coefficients for uniform nuclear density}
\label{apndx:ms}

The magnitude of coherent nuclear effects in the $C$-even and $C$-odd 
structure functions $F_2$ and $F_3$ is determined by the terms 
$\mathcal{C}_2^A$ and $\mathcal{C}_2^A$ (see \Eqs{C2A}{C3A}). These 
quantities can be computed analytically for uniform density distribution 
with a sharp edge (square well model), 
$\rho_A(\bm{r})=\rho_0\theta(R_A-|\bm{r}|)$, which is a reasonable 
approximation for large nuclei \cite{deshalit}. The nuclear radius $R_A$ in 
this model is related to the r.m.s. nuclear radius as 
$R_A^2=\frac53\langle r^2\rangle$ and the central nuclear density is 
$\rho_0=A/(\frac{4\pi}{3}R_A^3)$ with $A$ the number of nucleons. The 
coefficients $\mathcal{C}_2^A$ and $\mathcal{C}_2^A$  are
\begin{subequations}
\begin{align}
\mathcal{C}_2^A &= A \rho_0 R_A\, \varphi_2^{\rm SW}(y),\\
\mathcal{C}_3^A &= A \left(\rho_0 R_A\right)^2 \varphi_3^{\rm SW}(y),
\end{align}
\end{subequations}
where $\varphi_{2,3}^{\rm SW}$ are the functions of dimensionless and complex 
variable $y=2i(\rho_0 f - k_L)R_A$
\begin{subequations}
\begin{align}
\varphi_2^{\rm SW}(y) &= \left[6-3y^2-2y^3+6(y-1)\exp(y)\right]/y^4,\\
\varphi_3^{\rm SW}(y) &= 12\left[-4+y^2+y^3/3+(2-y)^2\exp(y)\right]/y^5.
\end{align}
\end{subequations}
If the real part of the amplitude $f$ and $k_L$ can be neglected 
(which is a reasonable approximation for $x\ll 0.1$), then $y=R_A/l_f$ with 
$l_f=(\rho_0\sigma)^{-1}$ the mean free path of the particle in a nucleus.


\end{document}